\documentclass[useAMS,usenatbib]{mnras}

\usepackage{graphicx,picture,calc} \usepackage{amssymb}
\usepackage{mathrsfs} \usepackage{amsfonts}
\usepackage{amsmath}
\usepackage[usenames,dvipsnames]{xcolor}
\usepackage{url} 

\newcommand{\aea}{A\&A}

\title[The Herschel-ATLAS Data Release 1]{The \textit{Herschel}-ATLAS Data 
Release 1 Paper I: Maps, Catalogues and Number Counts\thanks{{\it Herschel} is an ESA space 
observatory with science instruments provided by European-led Principal Investigator
 consortia and with important participation from NASA.}}

\author[Valiante et al.]{E. Valiante$^{1}$\thanks{E-mail: Elisabetta.Valiante@astro.cf.ac.uk},
M.W.L. Smith$^{1}$, S. Eales$^{1}$, S.J. Maddox$^{1,2}$, E. Ibar$^{3}$, R. Hopwood$^4$, \newauthor
L. Dunne$^{1,2}$, P.J. Cigan$^{1}$, S. Dye$^5$, E. Pascale$^1$, E.E. Rigby$^{6}$, N. Bourne$^2$,  \newauthor 
C. Furlanetto$^{5,7}$, R.J. Ivison$^{2,8}$\\
$^1$ School of Physics and Astronomy, Cardiff University, The Parade, Cardiff CF24 3AA, UK\\
$^2$ Institute for Astronomy, The University of Edinburgh, Royal Observatory, Blackford Hill,
Edinburgh, EH9 3HJ, UK\\
$^3$ Instituto de F\'isica y Astronom\'ia, Universidad de Valpara\'iso, Avda. Gran Breta\~na 1111,
Valpara\'iso, Chile\\
$^4$ Department of Physics, Imperial College London, Prince Consort Road, London SW7 2AZ, UK\\
$^5$ School of Physics and Astronomy, University of Nottingham, University Park, Nottingham
NG7 2RD, UK\\
$^6$ Leiden Observatory, Leiden University, PO Box 9513, 2300 RA, Leiden, The Netherlands \\
$^7$ CAPES Foundation, Ministry of Education of Brazil, Bras\'ilia/DF, 70040-020, Brazil\\
$^8$ European Southern Observatory, Karl-Schwarzschild-Stra{\ss}e 2, D-85748
Garching bei M\"unchen, Germany\\ 
}

\begin{document}

\date{Accepted 2016 July 21. Received 2016 July 4; in original form 2015 December 22}

\pagerange{\pageref{firstpage}--\pageref{lastpage}} \pubyear{xxx}

\maketitle

\label{firstpage}

\begin{abstract} 

We present the first major data release of the largest single key-project in area carried out in open time with the {\it Herschel} Space Observatory. The {\it Herschel} Astrophysical Terahertz Large Area Survey
(H-ATLAS) is a 
survey of 600\,deg$^2$ in five photometric bands - 100, 160, 250, 350 and 500\,$\mu$m
 - with the PACS and SPIRE cameras.
In this paper and a companion paper \citep{bourne16} we present the survey of 
three fields on the celestial equator, 
covering a total area of 161.6 deg$^2$ and previously observed in the Galaxy and Mass Assembly (GAMA)
spectroscopic survey. This paper describes
the {\it Herschel} images and catalogues of the sources detected on the SPIRE 250\,\micron\ images. 
The 1$\sigma$ noise for source detection, including
both confusion and instrumental noise,
is 7.4, 9.4 and 10.2\,mJy at 250, 350
and 500\,$\mu$m.
Our catalogue includes 120230 sources in total, with 113995, 46209 and 11011 sources detected at $>4\sigma$ at
250, 350 and 500\,$\mu$m. 
The catalogue contains detections at $>3\sigma$ at
100 and 160\,$\mu$m for 4650 and 5685 sources,
and the typical noise at these wavelengths is
44 and 49\,mJy. 
We include estimates of the completeness of the survey and of the effects
of flux bias and also describe a novel method
for determining the true source counts.
The H-ATLAS source counts are very similar to the source
counts from the deeper HerMES survey at 250 and 350\,$\mu$m, with a small
difference at 500\,$\mu$m. Appendix A provides a quick start  in using the released 
datasets, including
instructions and cautions on how to use them.

\end{abstract}


\defcitealias{bourne16}{Paper~II}

\section{Introduction}\label{sect:intro}
We describe the 
first major data release (hereafter the ``Data Release 1'') of the largest single
key-project in area carried out in open time with the
\textit{Herschel} Space Observatory \citep{pilbratt10}. The {\it Herschel} Astrophysical Terahertz 
Large Area Survey (the \textit{Herschel}-ATLAS or H-ATLAS) was allocated 600 hrs 
of observing time and is a survey of $600\,\deg^2$ of sky in five photometric bands:
100, 160, 250, 350 and 500\,$\mu$m \citep{eales10}. 
The chief original goal of the survey was to provide a relatively shallow {\it Herschel}
survey over a very large area of sky, with the specific aims of
providing measurements of the dust masses and dust-obscured star formation 
rate for tens of thousands of nearby galaxies \citep{dunne11}, 
complementing the large optical
spectroscopic surveys of the nearby Universe, such as the Sloan Digital Sky
Survey (SDSS; \citealt{ab09}) and the Galaxy and Mass Assembly project 
(GAMA\footnote{\url{http://www.gama-survey.org}}; \citealt{driver09,liske15}). However, the exceptional sensitivity of {\it Herschel}, aided by the
large and negative $k-$correction at submillimeter wavelengths \citep{blain93},
has meant that a significant
fraction of the sources in H-ATLAS
actually lie at high redshift \citep{amblard10,lapi11,gonzalez12,pearson13}. The H-ATLAS survey is therefore
useful both to astronomers studying the
nearby Universe and to 
those studying the early ($z > 1$) Universe. The large area of the survey
means that there are also potential uses for it in Galactic astronomy \citep{eales10}, 
with one practical example being a search for debris disks
around stars \citep{thompson10}.

We selected the H-ATLAS fields to avoid bright continuum emission from 
dust in the Galaxy,
as seen on the IRAS 100\,$\mu$m image, which shows well the cool low surface brightness dust in the interstellar medium. 
We also chose the fields to provide the
maximum amount of data in other wavebands. In the latter respect,
a high priority
was to choose fields that had already been surveyed in one of the major
optical spectroscopic surveys of the nearby Universe. We selected five
fields: a large ($\sim170$\,deg$^2$) field close to the North Galactic Pole,
which was observed in the SDSS; a second large field ($\sim285$\,deg$^2$) close to
the South Galactic Pole, which was observed in the 2-Degree-Field Galaxy Redshift
Survey (2dF; \citealt{colless01}); and three fields on the celestial equator, which are
each about 54\,deg$^2$ in size, which were observed in the 
SDSS, 2dF and GAMA redshift
surveys. The existence of the GAMA data is
particularly important because this survey 
is much deeper than the SDSS (more than two magnitudes in the $r$ band), yielding a surface-density of
galaxies with redshifts approximately four times greater.
The positions of these latter equatorial fields are given in Table~\ref{tab:gama}. The fields
are at right ascensions of approximately 9 hours, 12 hours and 15 hours
and hereafter we will call these fields GAMA9, GAMA12 and GAMA15.

Apart from the photometric data in the five H-ATLAS fields, there
is a large amount of imaging data in many other wavebands.
In particular, these fields have been imaged with the Galaxy
Evolution Explorer in the FUV and NUV filters (GALEX; \citealt{martin05}) 
and with the Wide-field Infrared Survey Explorer in the {\it W1, W2, W3, W4} filters (WISE; \citealt{wright10})
and as part
of the UK Infrared Deep Sky Survey Large Area Survey
(UKIDSS-LAS; \citealt{lawrence07}) and the VISTA Kilo-Degree Infrared Galaxy Survey in the {\it Z, Y, J, H, $K_S$} filters
(VIKING; \citealt{edge13}) and the VST Kilo-Degree Survey in the {\it u, g, r, i} filters (KIDS; \citealt{dejong13}).

This is the first of two papers describing the public release of the H-ATLAS data for
the entire GAMA fields. We intend to release the data for the other fields over the following few months. 
The second paper describing the data release for the GAMA fields is \citet{bourne16} (hereafter \citetalias{bourne16}) which describes the 
properties of the optical counterparts of the H-ATLAS sources. 
We released part of the data for the GAMA9 field at the end of the 
{\it Herschel} Science Demonstration Phase (hereafter SDP).
This dataset was described
by \citet{pascale11}, \citet{ibar10}, \citet{rigby11} and \citet{smith11}. The data we release now supersedes this
earlier dataset. 
A comparison between the two datasets is described in \S~\ref{sect:SDPcomp}.

This paper describes the {\it Herschel} images of the fields and the
catalogues of sources that we have constructed based on the images. 
We have tried to describe the dataset in a way that
is easily comprehensible to astronomers not expert in 
submillimetre data, referring those who are
interested in some of the more technical details to other publications. 
Nevertheless, there are some unavoidable technical issues
that do need to be understood if the data are to be used in a reliable way.
The most important
of these is the importance of the noise in the images produced by the blending together
of faint sources - so-called ``source confusion''. 
One result of this is that there are two kinds of noise on submillimetre images: instrumental noise, which is
usually uncorrelated between image
pixels, and confusion noise which is highly correlated between pixels.
One practical result of this is that the optimum filter for detecting unresolved
sources differs from
the point spread function of the telescope (\S~\ref{sect:mapfiltering}), which is the optimum filter
for maximising the signal-to-noise of unresolved sources if the noise is uncorrelated between pixels
\citep{chapin11}.

A second technical issue is
the statistical bias 
created by the effect of random errors if the distribution of the property being
measured is not uniform. 
\citet{eddington13} recognised this effect for the
first time, showing that there was a significant bias in the measured
number counts of stars 
even when the errors on the flux densities of the stars have the usual Gaussian
distribution. The cause of this bias is that there are more faint stars
to be scattered to bright flux densities by errors than bright stars to be
scattered to faint flux densities by errors, and thus any sample of stars
in a narrow range of flux density will contain more stars 
whose true flux densities are lower than this range than stars
whose true flux densities are brighter than this range.
This effect is present in optical catalogues but is much more
severe in submillimetre catalogues because of the steepness of
the submillimetre source counts \citep{clements10,oliver10,lapi11}.
This effect creates systematic errors in both the number counts, the phenomenon spotted
by Eddington, and also the
flux densities of individual sources.
In this paper we will refer to the first phenomenon as ``Eddington bias'' and the 
second phenomenon as ``flux bias''\footnote{Arguably both should be called
``Eddington bias'' because although \citet{eddington13} did not explicity refer
to the bias on individual flux densities, it was implicit in the phenomenon
he did describe. Nevertheless, in this paper we keep the distinction, partly because it
has become a common distinction in the astronomical community and partly
because, as we show in this paper, it is possible to correct
for the statistical (Eddington) bias on the distribution of flux densities
without correcting
the flux densities of the individual sources.}.
The second effect is often called
`flux boosting', but we prefer our name because, in principle, the bias
could be negative if there were fewer faint sources than bright sources.

We have included in this paper extensive modelling
of the effect of Eddington and flux bias, as well as other
simulations and modelling that
will allow the astronomical community to use the H-ATLAS images and
catalogues in a reliable way.
\citetalias{bourne16} describes the catalogue of the galaxies
from the SDSS r-band images that we have identified as the counterparts to the
{\it Herschel} sources and the multi-wavelength data for each of these galaxies.
All of the data described in these two papers can be found on \url{http://www.h-atlas.org}.

The layout of this paper is as follows. Section~\ref{sect:observations} contains a description of the
{\it Herschel} observations. Section~\ref{sect:spire} describes the reduction and properties of the
images made with one of the two cameras on {\it Herschel}, the Spectral and Photometric
Imaging Receiver (SPIRE; \citealt{griffin10}). Section~\ref{sect:pacs} describes the data
reduction and properties of the images made with {\it Herschel}'s other camera,
the Photoconductor Array Camera and Spectrometer (PACS; \citealt{poglitsch10}).
Section~\ref{sect:phot} describes the detection and photometry of the sources. 
Section~\ref{sect:simulations} describes the simulations we have carried out to investigate the
effects of source confusion and instrumental noise on the catalogues. 
Section~\ref{sect:counts250}
describes our determination of the 250\,$\mu$m source counts, corrected for the
effect of Eddington Bias
using the results of the simulations. Section~\ref{sect:corrections} describes our models
of the completeness of the survey and of the flux bias 
of individual sources at 250, 350 and 500\,$\mu$m.
Section~\ref{sect:comparisons} describes a comparison of our flux-density measurements with
previous measurements at these wavelengths.
Section~\ref{sect:catalogue} describes the catalogue of sources.
The main results from the paper are 
summarised in Section~\ref{sect:summary}.

For those who wish to use the data as soon as possible without wading through
a lengthy technical paper, we have provided a quick start in the form of
Appendix A. In this appendix we describe the basic datasets
we have released, with instructions and cautions on how to use them and references
back to the full paper. Appendix B provides a list of the asteroids 
detected in our maps, since these are not included in the catalogues of sources
provided in the data release.

\section{The \textit{Herschel} Observations}\label{sect:observations}
The GAMA fields
were observed with {\it Herschel}
in parallel mode, with the two {\it Herschel} instruments, PACS \citep{poglitsch10} 
and SPIRE \citep{griffin10}, operating simultaneously. 
The {\it Herschel} survey of each GAMA field consists of four overlapping rectangular regions, 
strung along the celestial equator. With one exception\footnote{The third quadrant
in GAMA12 (in order of increasing right ascension) was observed three times
because of the failure, part of the way through, of 
one of the observations. The combination of these three observations gives 
two orthogonal scans, the same
as for the other quadrants.},
each quadrant was observed twice by the telescope,
each observation lasting about nine hours. 
The telescope scanned along great circles on the sky at a constant angular speed
of
$60\,{\rm arcsec}\,{\rm s}^{-1}$, with the scan directions of the two observations
being roughly orthogonal
(about $ 85\,^{\circ}$ apart).
By making maps out of two datasets made with orthogonal scan directions, it is
possible, with the correct map-making algorithm, to correct for any large-scale 
artefacts that would result from the
gradual changes in the response of the
camera, which was a major
concern before launch \citep{waskett07}.
In operation, SPIRE, although not PACS, proved sufficiently stable that
only a very simple map-making algorithm was required and
maps made from observations made with a single scan direction were usually free
of artefacts from responsivity changes.

The time delay between the two observations made it 
possible to look for variations in the submillimetre sky by comparing maps made
from the two observations of each field. 
We used the differences of these maps to look for variable sources
but only found asteroids (see App.~\ref{app:asteroids}). We
also used these maps
for the astrometric registration of the data (\S~\ref{sect:dataproc}). We have not included 
these maps in this data release (maps made from the individual H-ATLAS
observations can be found in the {\it Herschel} archive).

During the observations, the telescope scans at a constant velocity
along a great circle across the field.
At the end of each ``scan leg'', the telescope decelerates to rest, moves a constant distance
orthogonal to the scan leg, and then scans backwards across the field.
A single observation is then made up of a large number of scan legs during which the
telescope is moving at a constant speed. 
Data are still being taken during the sections between scan legs when the telescope
is accelerating, but this ``turn-around'' data is not included in the final maps.
In parallel mode, the scan lines were
separated by $155\,{\rm arcsec}$ in order to achieve a good coverage 
with both PACS and SPIRE. More details
can be found in the SPIRE and PACS
Observers' Manuals, which are available at http://herschel.esac.esa.int. 

\begin{table} 
\centering 
\caption{Fields coordinates and areas.} 
\begin{minipage}{80mm} 
\centering
\label{tab:gama}
\begin{tabular}{@{}lccc@{}} 
\hline 
             &                   RA    &                  DEC   & area \\ 
             & \multicolumn{2}{c}{[J2000 $\deg$]}&[$\deg^2$] \\
\hline 
GAMA9  &    127.5...142.0           &       +3.1...-2.0           &    53.43              \\                    
GAMA12&    172.5...186.9           &       +2.0...-3.0           &     53.56               \\
GAMA15&    210.2...224.8           &       +3.1...-2.2           &     54.56               \\ 
\hline
\end{tabular} 
\end{minipage} 
\end{table}

\section{The SPIRE Maps}\label{sect:spire}

H-ATLAS imaged the sky with the SPIRE camera simultaneously
in three submillimetre bands centred at 250, 350 and
500\,$\mu$m. Each band is approximately 30 per cent wide in $\Delta \lambda / \lambda$.
More technical detail of the camera is given in \citet{pascale11} and
a full description of the camera is in \citet{griffin10}.

\subsection{Map Making}\label{sect:dataproc}
Several map-making software packages are capable of working with time-ordered data
pre-processed by the standard SPIRE pipeline. Some of these have been developed directly for
SPIRE and others were adapted from different instruments. The description and a detailed comparison
of the performance of some of the most used map-makers is presented in \citet{xu14}.

We processed the data with a very similar
method to one we used for the data we released at the end of the
SDP \citep{pascale11}.
In this paper we give a basic description of the method, emphasising any differences
from the method we used during SDP.
Some additional technical details can be found in \citet{pascale11}.

We processed the data within the {\it Herschel} Interactive Pipeline Environment (HIPE; \citealt{ott10}).
We used HIPE version 8. Our fundamental flux calibration is based
on the value of Neptune assumed in SPIRE Calibration Tree Version 8.
Note that our
SDP results \citep{pascale11,rigby11} were based on Version 5 of the
SPIRE Calibration Tree. Between Version~5 and Version~8 there was no change in the
calibration at 250 and 500\,$\mu$m and a change of only 1.0067 at 350\,$\mu$m, in the sense
that the flux densities in the current data release are lower by this factor than they would
have been if we had assumed the calibration we used in the SDP.

The {\it Herschel} Level-1 data consists of fully calibrated time-line data: files of flux density
versus time
for each bolometer. We produced the Level-1 data using the standard SPIRE data-reduction pipeline
with a few exceptions. The first exception was the way we corrected the time-line
data for glitches produced by cosmic rays. 
We used the {\sc SigmaKappaDeglitcher} module instead of the default {\sc WaveletDeglitcher} module
as we found the former performed better in masking glitches for our fast-scan parallel-mode observations. 

The second exception is the way we corrected for gradual changes in the
bolometer signals created by gradual changes in the temperature of the instrument.
SPIRE contains thermistors which monitored this temperature change, and we
designed our own procedure to use the thermistor results to correct
the bolometer signals. We first visually inspected the thermistor signals for sudden
jumps, which are spurious and do not represent temperature changes, and we removed these
jumps by adding an appropriate constant to
the thermistor timeline immediately after each jump. 
In the standard pipeline, each scan leg is corrected separately for the
drift in the bolometer signal caused by the temperature changes,
and the temperature information in the turn-around sections (\S~\ref{sect:observations}) is
not used at all. However, to make full use of all available information,
we used the thermistor data for the entire observation at once, including the
data from the turn-around sections.
We fitted the following function to the relationship between the temperature
measured by a suitable thermistor ($T$) and the
signal for the i'th bolometer ($S_{\rm bolom,i}$) for the
entire nine hours of data: $S_{\rm bolom,i} = a \times T + c$.
We then subtracted this relationship from the bolometer signals, effectively removing the
effect of the gradual temperature changes. There were some parts 
of the time-lines where a linear relation
clearly did not fit. These almost always occurred six hours after a cycle of the SPIRE cooling system
and became known in the trade as ``cooler burps''. We fitted the bolometer and
thermistor data in these regions with a fifth-order polynomial rather than a linear relation, and used this
function to correct the data. This polynomial
fitting stage is the only difference from the method we used for our SDP data \citep{pascale11}.

To remove any residual gradual drift in signal caused by thermal or other effects, we then
applied a high-pass filter to the time-line data after first masking bright astronomical sources.
The high-pass filter corresponds to a scale on the sky of 4.2$^{\circ}$, and thus
our images will not contain structure larger than this scale. This value is
the same as the angular size of the image made from a single nine-hour observation
and was chosen  
in order to minimise the $1/f$ noise of the maps \citep{pascale11}.

As described above (\S~\ref{sect:observations}), each field was mapped twice, with the scan direction of the
telescope in roughly orthogonal directions. After processing the time-line data as above,
we made initial maps using the data from the separate observations. We visually inspected the maps 
to spot any missed jumps in bolometer or thermistor signals, 
which we then fixed in the time-line data. We could also
see linear features on the maps in the scan direction which were caused by the residual
effect of bright cosmic rays. We masked these ``glitch tails'' in the time-line data.

We also used these initial maps to perform an astrometric calibration of the data, as done by \citet{smith11}.
We first produced initial source catalogues for each map using
our source-detection method (\S~\ref{sect:madx}). We then produced histograms 
of the separations in RA and Dec
between the sources and all objects on the
SDSS DR7 r-band images \citep{ab09}
within 50\,arcsec of each source.
We
then fitted these distributions using a Gaussian
model for the SPIRE positional errors and allowing for the effects of
galaxy clustering in the SDSS data. More details of this procedure
are given in \citet{smith11}.
This procedure allowed us to measure the average difference in positions
in both RA and Dec for each dataset between the {\it Herschel} positions and
the SDSS positions with a precision of $\sim0.05$\,arcsec in each
direction. The shifts
we found range from less than an arcsec to few arcsec, in agreement with the
1$\sigma$ pointing uncertainty of $\sim2$ arcsec given for {\it Herschel} 
\citep{pilbratt10}.
We used these shifts to correct the astrometry for each {\it Herschel} observation, 
so that the  
effective astrometric calibration of our maps and catalogues should be the same
as that of the SDSS.

We corrected the astrometry of the time-lines and reprocessed them,
removing the remaining thermistor or bolometer
jumps and masking the glitch tails. We then made the final maps using all the
time-line datasets for each GAMA field. We made these images
using the standard ``naive'' map-maker
in HIPE. In this map maker, the flux density in each pixel of the final image 
is calculated by taking the mean
of the values for the samples of the time-line data that have positions that fall within this
image pixel.
We did not use any turn-around data to make the final images
or data from regions for which there were data in only one scan direction.
We did not use the default pixel size of the standard pipeline (6, 10 and 14\,arcsec for 
250, 350 and 500\,$\mu$m, respectively), but instead 
6, 8 and 12\,arcsec at 250, 350 and 500\,$\mu$m, respectively.
We made this choice
because this corresponds approximately to 1/3 of the full-width half-maximum
of the point spread function (PSF, see \S~\ref{sect:psf}) in all bands and provides a good
compromise  between having a sufficient number of samples in each image
pixel to avoid high shot noise and 
producing a good sampling of the PSF.
Note that this choice differs from the one we used for our SDP data release, in which the
maps have pixel scales of 5, 10 and 10 arcsec at 250, 350 and 500\,$\mu$m, respectively
\citep{pascale11}.

\begin{figure*}
\centering
\includegraphics[trim=1cm 0.8cm 1cm 1cm,clip,width=190mm]{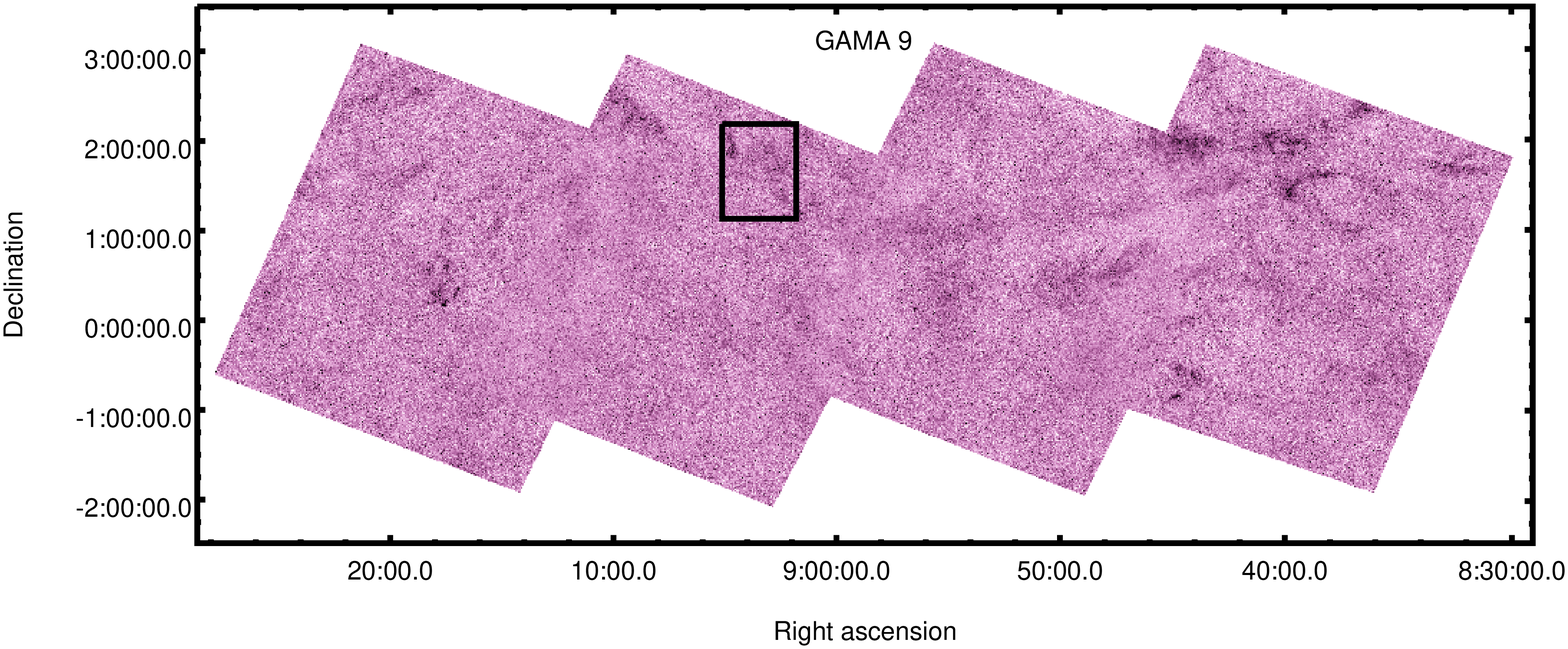}
\includegraphics[width=170mm,keepaspectratio]{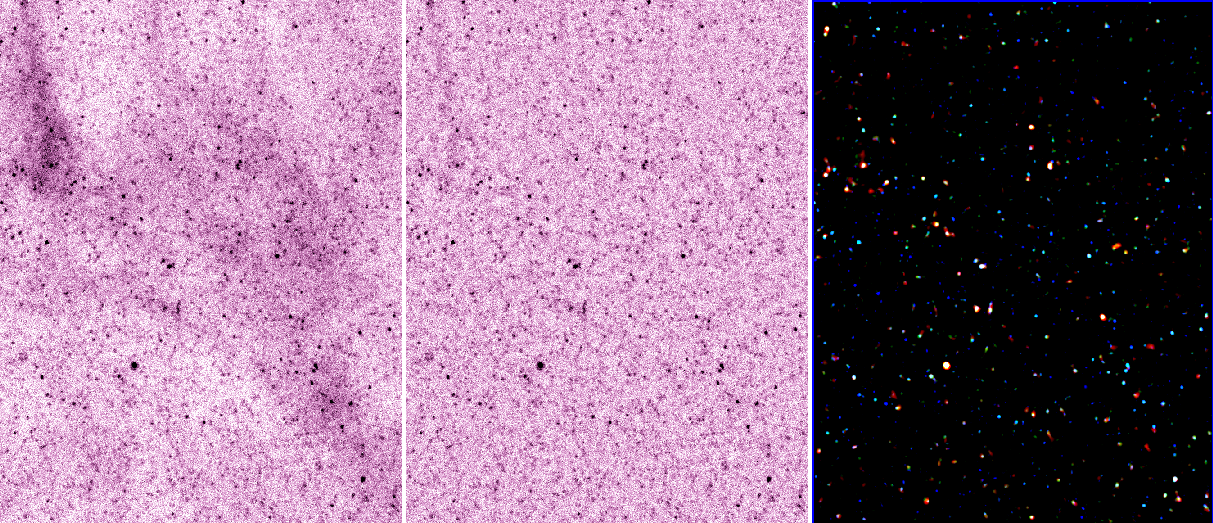}
\caption{\textit{Top}: raw image of the whole GAMA9 field at $250\,\micron$, with strong contamination by Galactic dust emission. 
The field we observed during the SDP is 
the second rectangular region from the left. \textit{Bottom}: images 
of a small part of the GAMA9 field. The left-hand panel shows the raw 250\,$\mu$m image and the middle panel
shows the same part of the 250\,$\mu$m image after both
{\it Nebuliser} (\S~\ref{sect:dataproc}) and the matched filter (\S~\ref{sect:mapfiltering}) have been applied. The
right-hand panel shows a pseudo-colour image of this region made by combining the 250, 350 and 500\,$\mu$m
images. On this image, red represents a source that is particularly bright at 500\,$\mu$m and blue one that is
particularly bright at 250\,$\mu$m. Red sources are generally ones which contain cold dust or at high redshifts
\citep{amblard10}. Images realised using {\sc cubehelix} \citep{green11}.
}
\label{fig:backsub}
\end{figure*}

\begin{figure*}
\centering
\includegraphics[trim=1cm 0cm 5cm 1cm,clip,width=170mm]{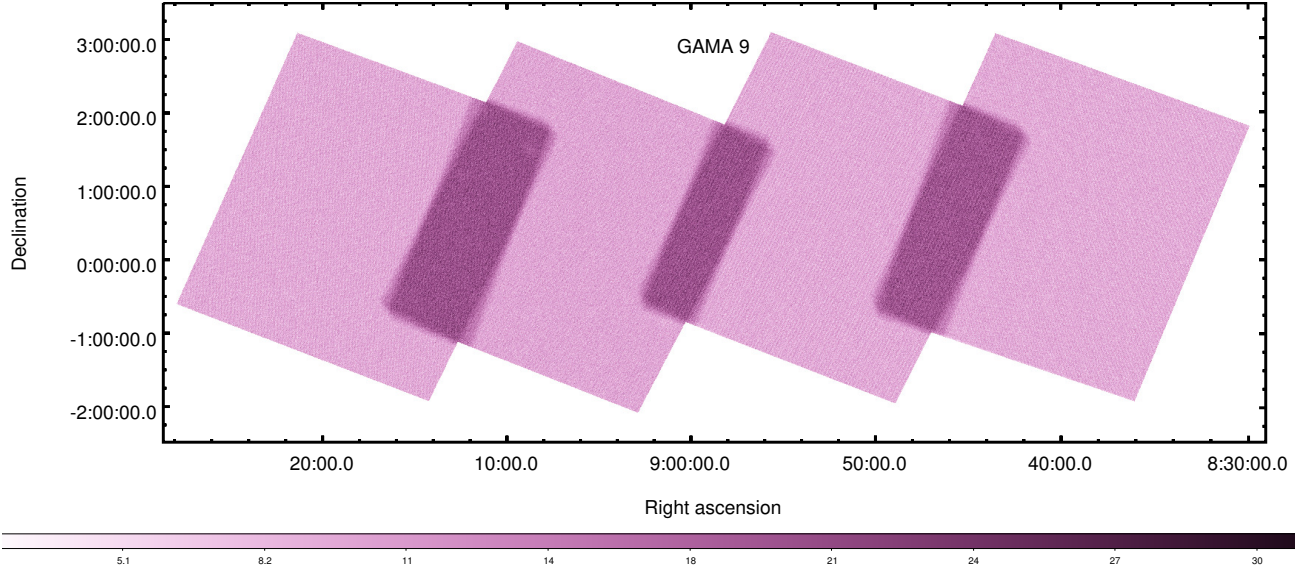}
\caption{Coverage map of the whole GAMA9 field at $250\,\micron$. The maps show the number of samples
per pixel. The coverage,
and thus the instrumental noise, is quite uniform through the whole map, with the exception of
the parts where the rectangular regions overlap. Image realised using {\sc cubehelix} \citep{green11}.
}
\label{fig:coverage}
\end{figure*}

We provide these images in the data release (App.~\ref{app:spire}).
Table~\ref{tab:gama} gives the precise areas covered by each map.
Figure~\ref{fig:backsub} ({\it top}) shows the basic 250\,$\mu$m images of GAMA9 and Figure~\ref{fig:coverage}
shows the coverage map for the same field. 
The field we observed during the SDP is the second quadrant from
the left.
The large-scale features
visible in the GAMA9 field are emission from dust in the Galaxy (``cirrus emission''). This was
the one field where we had to accept some significant cirrus emission as the price
of some high-quality multi-wavelength data. 

In an H-ATLAS SPIRE image made from an observation made with a single scan-direction,
it should be possible to reconstruct 
the submillimetre structure of the sky in the scan direction on all scales
up to $\sim4.2^{\circ}$, the angular scale of the high-pass filter. A single
observation, however, contains little information about the structure of the
sky in a direction orthogonal to the scan direction. Nevertheless, by using 
the observations from both orthogonal scan directions and a Maximum Likelihood 
estimator \citep{pascale11}, it is possible to  produce an accurate image of the sky up to 
the maximum scale set by the high pass filter. However, in this data release our maps have been made 
with the ``naive'' map-maker, and as a result the 
Fourier modes with scales greater than 20\,arcmin may be attenuated.
Therefore, our images will be 
adequate for measuring the submillimetre flux densities of even large nearby galaxies but should not be used 
to investigate structure on larger scales.

All {\it Herschel} images contain emission from dust in the Galaxy. Before trying to detect extragalactic
sources, it is preferable to try and remove this emission. 
We used the {\it Nebuliser} 
algorithm\footnote{\url{http://casu.ast.cam.ac.uk/surveys-projects/software-release/background-filtering}},
developed by the Cambridge Astronomical Survey Unit, to attempt to remove the emission from
Galactic dust from the SPIRE images in all the three wavebands.
The algorithm is fairly simple. In each pixel, the median of the 
intensity values in a square of ${\rm N} \times{\rm N}$ pixels
centered on the pixel is used to estimate the background at that position. The background map is 
then smoothed using a box-car mean filter with box size chosen to be ${\rm N}/2\times {\rm N}/2$. The user can also set positive 
and negative signal-to-noise thresholds relative to the filtered map, outside of which the pixels are not used 
to compute the median. We used lower  and upper thresholds of $-10\sigma$ and $+3\sigma$, and carried out three iterations 
of pixel rejection. We tested different values of N and concluded that with ${\rm N} = 30$ 
(corresponding to 3\,arcmin), there was no significant effect on the flux densities of point sources. As 
discussed in \S~\ref{sect:pacsmaps}, some flux is lost from sources whose size is comparable to the median 
box size, but the effect is negligible for sources with radius $\lesssim 1$\, arcmin.
We have provided these background-subtracted images as part of the
data release (see App.~\ref{app:spire}). These images do not contain extended emission
from the Galaxy and any kind of extra-galactic signal with spatial scale larger than $\sim 3$\,arcmin.
Figure~\ref{fig:backsub} ({\it bottom}) shows the difference between the GAMA9 map at 250\,$\mu$m
and the same map after {\it Nebuliser} has been applied; the application of {\it Nebuliser} has clearly
efficiently removed the extended emission from the Galactic dust. Nevertheless,
very compact Galactic dust regions and cirrus knots are still present in the map,
and our final catalogues will contain compact Galactic sources,
such as debris disks
\citep{thompson10}.

\subsection{The Point Spread Function}\label{sect:psf}

To characterise the point spread function (PSF) in our images, we created images of Neptune \citep{griffin13}, 
using the same pixel scale as for the H-ATLAS images. From these images we 
made azimuthally-averaged and normalised PSF profiles to use in
the source extraction (see \S~\ref{sect:madx}).
Although the actual SPIRE PSF is slightly elongated (see SPIRE Handbook\footnote{\url{http://herschel.esac.esa.int/Docs/SPIRE/spire_handbook.pdf}}), 
our images are made from multiple observations with different scan directions,
so the approximation that the PSF is circularly symmetric should be a good one.
Besides it is the approximation made in the calibration of the
SPIRE flux densities (Griffin, private communication). 
We measured the full-width half-maximum of the azimuthally-averaged PSF to be
17.8, 24.0 and 35.2\,arcsec at 250, 350 and 500\,$\mu$m, respectively. We provide this 
azimuthally-averaged normalised PSF profile
as part of the data release (see App.~\ref{app:spire}).

\subsection{The Instrumental Noise}\label{sect:noise_from_jn}

The uncertainty maps produced by HIPE, which come as extensions to the standard images, 
are obtained by calculating the variance of all the time-line samples that contribute
to each pixel.
However, this method does not produce an accurate estimate of the
instrumental noise in each pixel for two reasons.
First, the small
number of samples in each pixel means the estimate of the variance
itself has a significant uncertainty, leading
to two pixels with the same number of samples from equally sensitive bolometers
having in general different uncertainties even though the instrumental noise
should be the same.
Second, if the pixel coincides with a bright
source
the variance will be higher 
because each time-line sample will have been measured at a slightly different position
on the source. 

We therefore adopted a different technique 
for estimating the instrumental noise in each pixel.
We used the statistics of the
of the difference between
the two maps, $m_1$ and $m_2$, of the same region made with the different scan directions. 
This jackknife map only contains instrumental noise because any real astronomical
structure will be removed by subtracting the images (apart from a small number of asteroids - App.~\ref{app:asteroids}). 
\citet{pascale11} have shown that the histogram of pixel intensities from the jackknife
map closely follows a Gaussian distribution.
If $\sigma_1$ is the instrumental noise from one sample, the variance in each pixel
of the jackknife map will be

\begin{equation}
{\rm var} (m_1-m_2) = \sigma_1^2 (1/c_1+1/c_2),
\end{equation}

\noindent in which $c_1$ and $c_2$ are the numbers of samples in that pixel in each 
of the two maps.
Our estimate of the instrumental noise per sample, $\sigma_1$, is then given by

\begin{equation}
\sigma_1^2 = \frac{ \sum_i^{N_{pix}} (m_{1i} - m_{2i} )^2}{\sum_i^{N_{pix}} 1/c_{1i} + 1/c_{2i}}
\end{equation}

\noindent in which $i$ refers to the i'th pixel and the sum is carried out over all the
$N_{\rm pix}$ pixels in the image. 
We have estimated $\sigma_1$ for each GAMA field and then calculated
the average for the three fields as 30.1, 30.9 and 36.1\,mJy/beam per time-line sample
at 250, 350 and 500\,$\mu$m, respectively.
These values are quite similar to the values obtained by 
\citet{pascale11} from the data of the Science Demonstration Phase and also broadly consistent with the 
values measured by \citet{nguyen10} from data taken as part
of the HerMES survey \citep{oliver12}, after allowing for the fact
that the bolometer sampling during 
the H-ATLAS parallel-mode observations ($10\,{\rm Hz}$) was less frequent than during the
HerMES observations ($18.2\,{\rm Hz}$). 

The instrumental noise in each pixel of our final images is given by 
$\sigma_1 / \sqrt{c_i}$, where $c_i$ is
the number of data samples in that pixel. The instrumental noise in our fields is quite 
uniform,
because most of the survey
area has been observed by the telescope for
the same exposure time (see \S~\ref{sect:observations}). 
The exceptions are the overlapping regions between two adjacent quadrants, which 
had twice the exposure time of the rest of the survey
(see Figure~\ref{fig:coverage}).
We have provided maps of this instrumental noise as part of the data release
(App.~\ref{app:spire}) and also estimated the typical instrumental noise in each image from the following
equation:

\begin{equation}\label{eq:sigma_inst}
\sigma_{\rm inst} = \sqrt{ {1 \over N_{\rm pix}} \sum_i {\sigma_1^2 \over c_i} }
\end{equation}

\noindent in which the sum is carried out over all the pixels in the image.
We give these estimates in Table~\ref{tab:noisevar} (top two panels).
We also tried an alternative to Eq.~\ref{eq:sigma_inst} to estimate
the average instrumental noise on a map\footnote{In this alternative method, 
we started with the noise map produced
by the jackknife technique, in which the
noise in the i'th pixel is given by $\sigma_i = \sigma_1/\sqrt{c_i}$),
in which $c_i$ is the number of time-line pixels contributing
to that pixel. 
We then used this noise map to
create a Monte-Carlo
realisation of 
an image with no real astronomical sources, on the assumption that
the noise has a Gaussian distribution and with the standard deviation
of the Gaussian function used to generate the intensity values
in each pixel given by $\sigma_i$. We then plotted a histogram of
the intensities in all the pixels of the Monte-Carlo image, and then
fitted a Gaussian function to this distribution, using the standard
deviation of the Gaussian as our estimate of the noise on the image.}.
These values are given in Table~\ref{tab:noisegauss}.
The good agreement between the estimates
of the instrumental noise in Tables~\ref{tab:noisevar} and \ref{tab:noisegauss} 
gives us confidence that our estimates of the instrumental noise are reliable.

\begin{table*}
\centering
\begin{minipage}{140mm}
\centering
\caption{Noise from the variance analysis: 
$\sigma_{\rm tot\_Var}$ is calculated using 
Eq.~\ref{eq:sigma_tot_var} and $F_{\rm lim}=200$\,mJy, 
$\sigma_{\rm inst}$ is calculated using Eq.~\ref{eq:sigma_inst} 
and $\sigma_{\rm conf\_Var}$ is calculated from the first two using Eq.~\ref{eq:confnoise}. 
These values of the confusion noise and total noise 
are effectively upper limits on the actual confusion noise at the position of a source
and on the $1\sigma$ uncertainty in the flux density of the source 
(see \S~\ref{sect:conf_noise} and \ref{sect:fluxunc} for more details).}
\label{tab:noisevar}
\begin{tabular}{@{}lccccccccc@{}}
\hline
             & \multicolumn{3}{c}{$\sigma_{\rm inst}$ [mJy/beam]}    & \multicolumn{3}{c}{$\sigma_{\rm conf\_Var}$ [mJy/beam]}   & \multicolumn{3}{c}{$\sigma_{\rm tot\_Var}$ [mJy/beam]} \\
             &$250\,\micron$&$350\,\micron$&$500\,\micron$&$250\,\micron$&$350\,\micron$&$500\,\micron$&$250\,\micron$&$350\,\micron$&$500\,\micron$\\
\hline
            & \multicolumn{9}{c}{Raw maps} \\
\hline
GAMA9  &    9.4            &        9.2           &        10.7                     &    7.5            &        7.9           &        7.5                         &    12.0            &        12.1           &        13.0\\
GAMA12&    9.4           &         9.2           &        10.6                     &    6.7            &        7.2           &        6.9                         &    11.6            &        11.7           &        12.6\\
GAMA15&    9.3             &       9.1           &        10.5                     &    6.9            &        7.4           &        7.1                         &    11.6            &        11.7           &        12.7\\
\hline
            & \multicolumn{9}{c}{Background-subtracted} \\
\hline
GAMA9  &    9.4            &        9.2           &        10.7                     &    6.3            &        6.9           &        6.5                         &    11.4            &        11.5           &        12.5\\
GAMA12&    9.4            &        9.2           &        10.6                     &    6.3            &        6.8           &        6.4                         &    11.4            &        11.5           &        12.4\\
GAMA15&    9.3            &        9.1           &        10.5                     &    6.5            &        7.0           &        6.6                         &    11.4            &        11.5           &        12.4\\
\hline
            & \multicolumn{9}{c}{Background-subtracted and smoothed with the matched filter} \\
\hline
GAMA9  &    4.6            &        4.5           &        5.3                     &    7.2            &        7.6           &        7.3                         &    8.6            &        8.8           &        9.0\\
GAMA12&    4.6            &        4.5           &        5.3                     &    7.2            &        7.5           &        7.1                         &    8.5            &        8.7           &        8.9\\
GAMA15&    4.5            &        4.5           &        5.3                     &    7.4            &        7.7           &        7.3                         &    8.7            &        8.9           &        9.0\\
\hline
            & \multicolumn{9}{c}{Background-subtracted and smoothed with the point spread function} \\
\hline
GAMA9  &    4.2            &        4.0           &        4.7                     &    9.7            &        10.4           &        10.2                         &   10.6            &      11.1           &        11.2\\
GAMA12&    4.2            &        4.0          &         4.7                     &    9.5            &        10.1           &        9.8                           &   10.3            &        10.9           &        10.9\\
GAMA15&    4.1            &        3.9           &        4.7                     &    9.7            &        10.4           &        10.1                         &   10.6            &       11.1           &        11.1\\
\hline
\end{tabular}
\end{minipage}
\end{table*}

\begin{table*} 
\centering 
\begin{minipage}{140mm} 
\centering
\caption{Noise analysis from Gaussian fits: $\sigma_{\rm tot\_Gauss}$ is 
calculated by fitting a Gaussian to the negative side of the histograms of pixel intensities (see
Figure~\ref{fig:noisegauss}), $\sigma_{\rm inst}$ is 
calculated from Monte-Carlo realisations of the noise 
(see \S~\ref{sect:noise_from_jn}) and $\sigma_{\rm conf\_Gauss}$ is 
calculated from the first two using Eq.~\ref{eq:confnoise}.These values of the confusion noise
and the total noise are lower limits on the actual confusion noise
at the position of a source and on the $1\sigma$ uncertainty in the
flux density of the source (see \S~\ref{sect:conf_noise} and \ref{sect:fluxunc} for more details).}
\label{tab:noisegauss}
\begin{tabular}{@{}lccccccccc@{}} 
\hline
             & \multicolumn{3}{c}{$\sigma_{\rm inst}$ [mJy/beam]}    & \multicolumn{3}{c}{$\sigma_{\rm conf\_Gauss}$ [mJy/beam]}   & \multicolumn{3}{c}{$\sigma_{\rm tot\_Gauss}$ [mJy/beam]} \\ 
             &$250\,\micron$&$350\,\micron$&$500\,\micron$&$250\,\micron$&$350\,\micron$&$500\,\micron$&$250\,\micron$&$350\,\micron$&$500\,\micron$\\
\hline
	    & \multicolumn{9}{c}{Background-subtracted maps} \\  
\hline
GAMA9  &    9.7            &        9.5           &        10.9                     &    3.3            &        4.2           &        4.8                         &    10.2            &        10.4           &        11.9\\                    
GAMA12&    9.9            &        9.6           &        11.0                     &    4.2            &        4.4           &        5.4                         &    10.7            &        10.6           &        12.2\\
GAMA15&    9.7            &        9.4           &        10.9                     &    3.9            &        4.3           &        5.1                         &    10.4            &        10.3           &        12.0\\ 
\hline
	    & \multicolumn{9}{c}{Background-subtracted and smoothed with the matched filter} \\ 
\hline 
GAMA9  &    4.7            &        4.6           &        5.4                     &    2.0            &        3.1           &        4.2                         &    5.1            &        5.6           &        6.8\\                    
GAMA12&    4.8            &        4.7           &        5.4                     &    2.0            &        3.1           &        4.0                         &    5.2            &        5.6           &        6.8\\
GAMA15&    4.7            &        4.6           &        5.4                     &    2.1            &        3.1           &        3.8                         &    5.2            &        5.5           &        6.6\\
\hline
	    & \multicolumn{9}{c}{Background-subtracted and smoothed with the point spread function} \\ 
\hline 
GAMA9  &    4.3            &        4.1           &        4.8                     &    3.6            &        5.1           &        7.0                         &    5.6            &        6.5           &        8.5\\                    
GAMA12&    4.4            &        4.2           &        4.9                     &    3.3            &        4.8           &        6.0                         &    5.5            &        6.4           &        7.7\\
GAMA15&    4.3            &        4.1           &        4.8                     &    3.6            &        4.8           &        5.8                         &    5.6            &        6.3           &        7.5\\
\hline
\end{tabular} 
\end{minipage} 
\end{table*}

\subsection{The Confusion Noise}\label{sect:conf_noise}

\begin{figure}
\sbox0{\includegraphics[trim=2cm 0cm 1cm 1cm,clip,width=84mm]{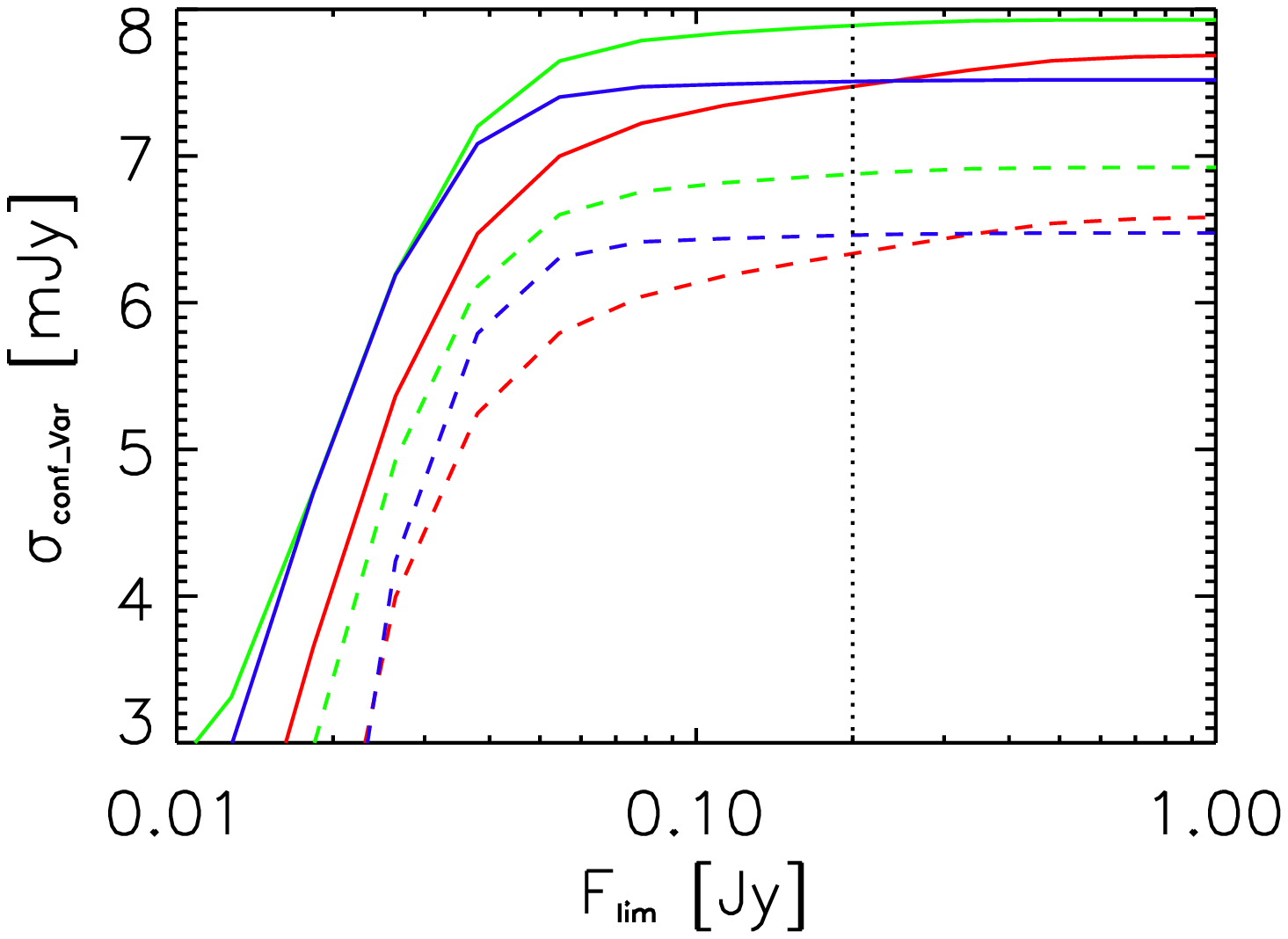}}
\sbox2{\includegraphics[trim=2cm 0cm 1cm 1cm,clip,width=34mm]{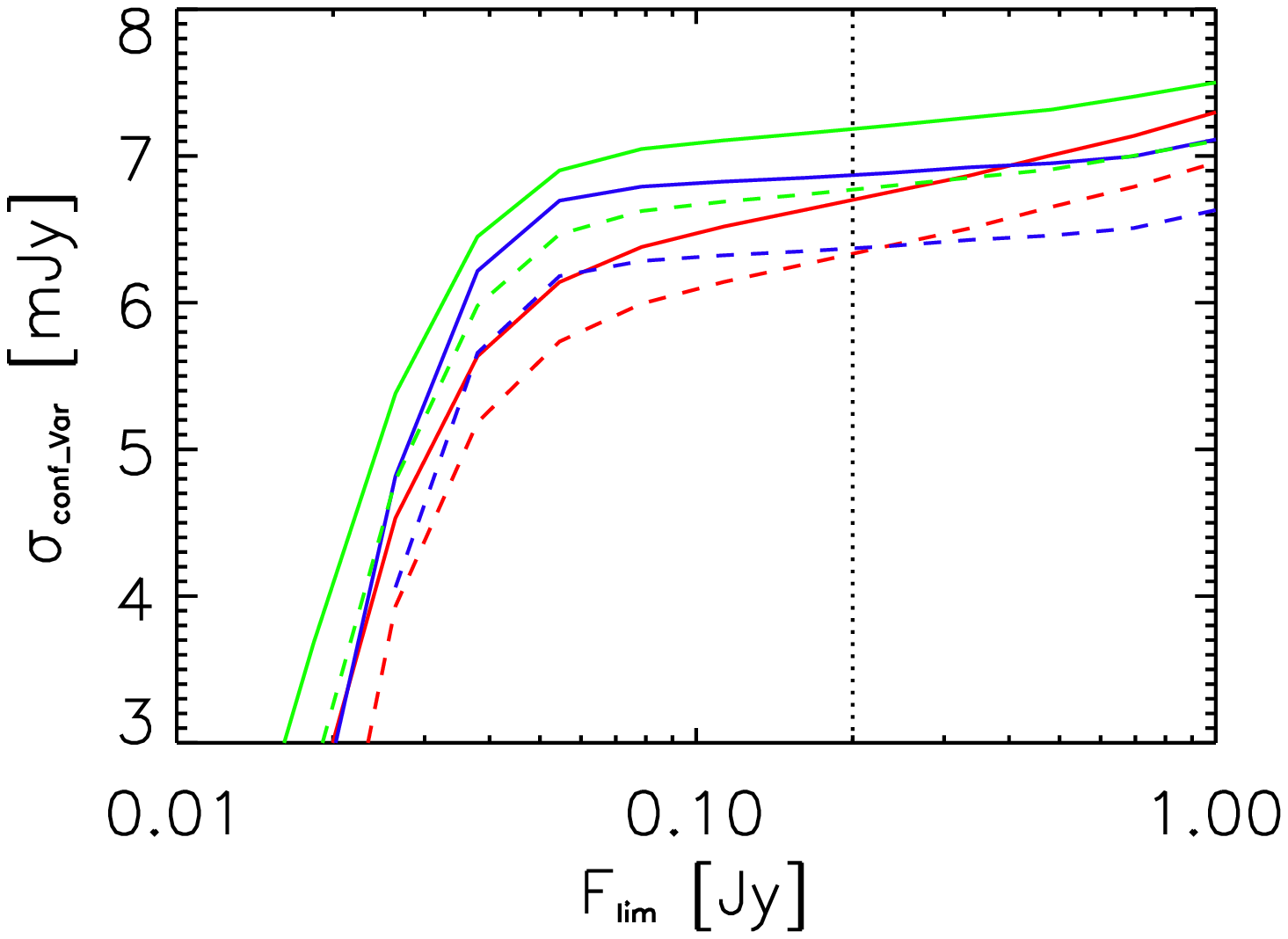}}

\begin{picture}(\wd0,\ht0)
\put(0,0){\usebox0}
\put(\wd0 - \wd2 - 0.3cm,\ht0 - \ht2 - 2.4cm){\usebox{2}}
\end{picture}

\caption{ 
Plot of the confusion noise, $\sigma_{\rm conf\_Var}$, plotted against the upper limit of the flux density, $F_{\rm lim}$, 
in the GAMA9 ({\it main plot}) and in the GAMA12 ({\it inset}) fields (see \S~\ref{sect:conf_noise}). 
The blue, green and red lines show the measurements
for 250, 350, 500\,$\mu$m, respectively, with the solid lines showing the measurements from the raw maps
and the dashed lines showing the measurements from the images that have had emission from Galactic dust 
removed using \textit{Nebuliser}. The vertical dotted line corresponds to $F_{\rm lim}=200$\,mJy, the limit used t
o calculate the values reported in Table~\ref{tab:noisevar}. Note the monotonic increase of $\sigma_{\rm conf\_Var}$: 
at 500\,\micron\, this effect is due to gravitationally lensed sources, which are particularly numerous in the GAMA9 field
(see main figure), while at shorter wavelengths is due to bright nearby galaxies, which populate the GAMA12 field 
(see inset figure). 
}
\label{fig:noisevar} 
\end{figure}

\begin{figure}
\center
\includegraphics[trim=0mm 5mm 5mm 10mm,clip,width=84mm]{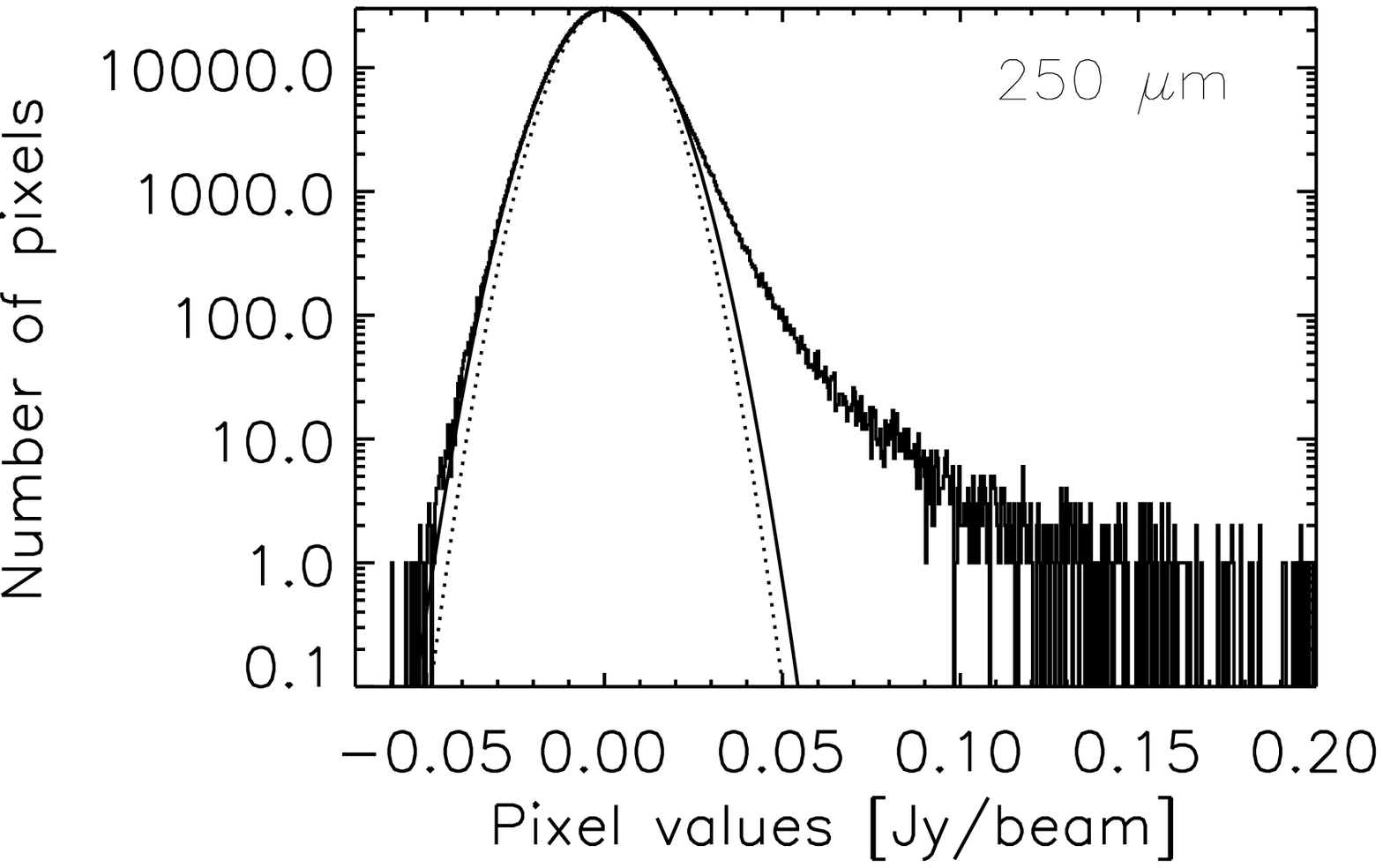}
\includegraphics[trim=0mm 5mm 5mm 10mm,clip,width=84mm]{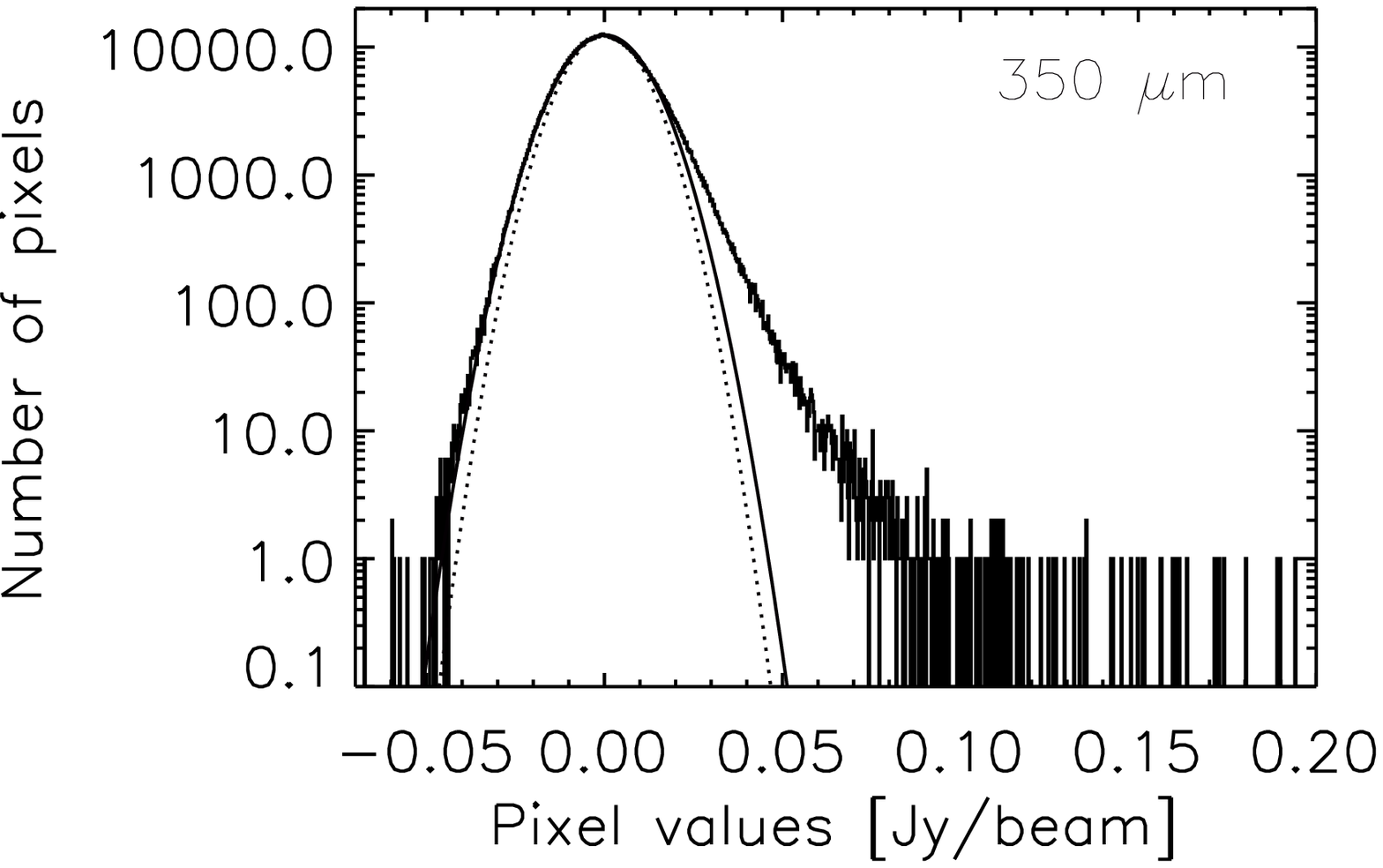}
\includegraphics[trim=2mm 5mm 5mm 10mm,clip,width=84mm]{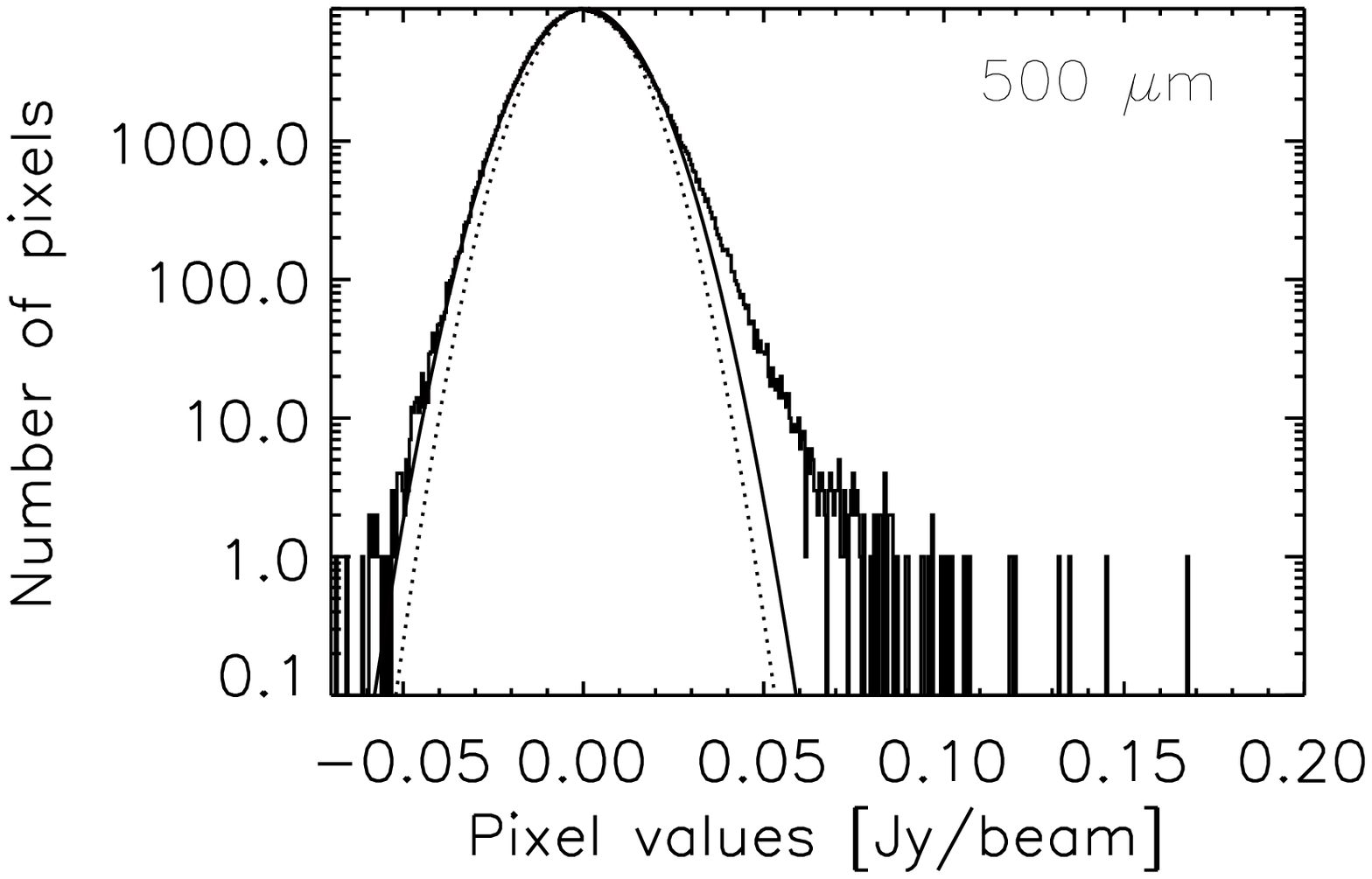} 
\caption{ 
Histograms of 
flux density for the pixels 
in, from top to bottom, the GAMA12 250, 350 and 500\,$\mu$m images.
The same region, in which the noise is approximately uniform, has been
used in all the images.
The histograms are fitted well by a Gaussian with an 
excess at positive flux densities due to bright sources on the images. 
The continuous line is the Gaussian that fits the negative side of the 
histogram and represents the total noise in the map. The dotted line shows the distribution
for instrumental noise predicted from the jackknife results (\S~\ref{sect:noise_from_jn}).
The difference between the total and instrumental noise is the result of the noise caused
by the confusion of faint sources. 
Our estimates of instrumental noise from the jackknife analysis and of
the total noise and confusion noise from these fits are given in Table ~\ref{tab:noisegauss}.}
\label{fig:noisegauss} 
\end{figure}

The best way to detect sources on submillimetre images depends on the
relative proportions of instrumental noise and noise due to the confusion of
faint sources. For example, in an image with instrumental noise and containing
only one point source
(and therefore no confusion noise), the maximum signal-to-noise is obtained by convolving the image 
with the point spread function (PSF) of the telescope \citep{north43}. 
On the other hand, in an image of confused point sources with no instrumental noise, the optimal
way to find the sources would be to take the Fourier Transform of the image, divide this
by the Fourier Transform of the PSF, and then take the inverse Fourier Transform of the
result,
a procedure which results in a perfect deconvolution of the original image.
Our H-ATLAS images, with their mixture of confusion noise and instrumental noise,
are somewhere between these two extremes.

\citet{chapin11} have shown that for images
in this midway regime it is possible to calculate 
a
convolving function or `matched filter' that will produce the maximum signal-to-noise
for unresolved sources for any ratio of confusion to
instrumental noise. It is therefore important to measure this ratio, and in this
section we describe a detailed analysis of the properties of the noise on the
H-ATLAS images. 

There is not a unique definition of confusion noise and this makes this matter, ironically, quite 
confusing. One definition of confusion is that it is the root-mean-square value of the fluctuations
in an image due to the emission of the astrophysical components of the sky background. 
In our maps, the extended Galactic emission has been removed (or at least most of it, see \S~\ref{sect:dataproc}). 
The only source of confusion noise will then be the galaxies we are trying to detect, in particular the ones fainter 
than the detection limit.
We tried two different methods to estimate the confusion noise on our images.
In both methods, we first estimated the total noise on the maps and
then calculated the confusion noise from:

\begin{equation}\label{eq:confnoise}
\sigma_{\rm conf} = \sqrt{ \sigma_{\rm tot}^2 - \sigma_{\rm inst}^2}
\end{equation}

\noindent in which $\sigma_{\rm inst}$ is the instrumental noise 
measured using the jackknife analysis described in \S~\ref{sect:noise_from_jn}.

In the first method, we calculated the total noise from the variance of the map:

\begin{equation}\label{eq:sigma_tot_var}
\sigma_{\rm tot\_Var} = \sqrt{\frac{1}{N_{\rm pix}} \sum_i^{N_{\rm pix}} (F_i - <F_{\rm map}>)^2}\ \ \ F_i < F_{\rm lim}
\end{equation}

\noindent
in which $F_i$ is the flux density in the i'th pixel, $N_{\rm pix}$ is the number of pixels in the image, $<F_{\rm map}>$ is the 
mean value of the pixels in the map and $F_{\rm lim}$ is 
an upper limit to the flux densities that are used in the calculation. 
We used an upper limit to the flux density because the effect of confusion
on a source is only produced by signals fainter than the flux density of this
source. In other words, in a pixel with flux density equal to $F_{\rm lim}$, the confusion
noise is calculated from the variance of all pixels with flux density fainter than $F_{\rm lim}$. If a brighter
pixel contributed to the noise, we would measure a brighter flux in that pixel.

The solid lines in
Figure~\ref{fig:noisevar} shows the relationship between $\sigma_{\rm conf\_Var}$ 
and $F_{\rm lim}$ for the GAMA9 and GAMA12 fields.
The figure shows that
the confusion value does not asymptotically approach some fixed value
as $F_{\rm lim}$ increases. 
It is easy enough to see why this should be. If the differential source
counts are given by ${\rm d}N/{\rm d}S \propto S^{\alpha}$, 
a value of $\alpha>-3$ leads to a monotonic increase of the
confusion noise with increasing
$F_{\rm lim}$.
This is likely to be the case for the number counts of submillimetre sources, which at bright flux
densities are dominated by a mixture of nearby galaxies, with $\alpha \simeq -2.5$, and gravitationally
lensed sources \citep{blain98,negrello10}. 
In fact, as shown by Figure~\ref{fig:noisevar}, this effect is more evident in the GAMA12 and GAMA15
fields (the latter is not shown, but it is very similar to GAMA12), where we find big local bright galaxies. It is
noticeable only at 500\,\micron\ in the GAMA9 (see inset of Figure~\ref{fig:noisevar}), where no particularly big 
galaxies are present, but where we detected a conspicuous number of gravitational lensed objects (Negrello et al., in prep).
The maximum value of the confusion noise is set by the brightest source in the survey, because
in that case we would calculate the variance of all pixels in the map. This limit will tend to be higher 
for surveys covering larger areas of sky.

We therefore have to be careful when comparing our estimates of the confusion
noise with those from other surveys.
In the HerMES survey \citep{oliver12}, which was deeper but covered a smaller
area of sky,
\citet{nguyen10} used a method very similar to ours to measure the
confusion noise, finding values for
$\sigma_{\rm conf\_Var}$ of
5.8, 6.3 and
6.8\,mJy/beam at 250, 350 and 500\,$\mu$m, respectively.
Figure~5 in \citet{nguyen10} shows that the
confusion values measured in the HerMES fields are changing very slowly with $F_{\rm lim}$
at $F_{\rm lim}=200$\,mJy, and so to make a fair comparison we calculated $\sigma_{\rm tot\_Var}$ and
$\sigma_{\rm conf\_Var}$ with the same limit.
The upper panel of Table~\ref{tab:noisevar} gives our measurements for all three GAMA fields.
The range of values for the three GAMA fields is 
$\sim0.5$\,mJy at all wavelengths, and the mean values
are slightly higher
than those from HerMES. However, the second panel in Table~\ref{tab:noisevar} shows our 
measurements from the images that have had the Galactic dust emission removed with {\it Nebuliser}   
(the dashed lines in Figure~\ref{fig:noisevar} shows how $\sigma_{\rm conf\_Var}$ depends
on $F_{\rm lim}$ for these images).
The values for the confusion noise in Table~\ref{tab:noisevar} are now very similar for the 
three GAMA fields, with a range of 0.2 mJy, with mean
values of 6.4, 6.9 and 6.5\,mJy/beam, fairly similar to 
the HerMES values.
This comparison shows that the emission from Galactic dust makes a significant
contribution to the confusion noise, and so in comparing confusion estimates
one must take into account both the areas covered by the different surveys and also the
emission from Galactic dust.

In the second method, we estimated
the noise by 
fitting a Gaussian to the negative side of the histograms
of pixel intensities.
Figure~\ref{fig:noisegauss} shows that the histograms
roughly follow a Gaussian at negative and small positive flux densities, but 
deviate strongly from a Gaussian at bright flux densities.
The natural interpretation of this distribution is 
that the non-Gaussian tail is produced by individual bright sources,
whereas the Gaussian distribution
is the
result of the combination of the instrumental noise and the population of very faint 
sources, which have such a high surface-density that the confusion noise they produce 
follows a Gaussian distribution, as a result of the Central Limit Theorem.
 
Table~\ref{tab:noisegauss} gives the results of this analysis.
As expected, the values for the
total noise and the confusion noise given in this table are lower than the values 
in Table~\ref{tab:noisevar} because we are now not including the contribution of the strongly
non-Gaussian tail that is seen at positive flux densities (Figure~\ref{fig:noisegauss}).

From a practical point of view, 
the crucial thing we need to know 
is the uncertainty in the flux density of each source in the survey.
If a source is close to the detection limit of the survey ($\sim 30-40$\,mJy), the
noise measurements in Table~\ref{tab:noisevar} will be larger than the true error
in the source's flux density, because
these measurements contain a confusion contribution from pixels brighter than this
limit. On the other hand, the values in Table~\ref{tab:noisegauss} will be too low because they
do not include a contribution from the positive tail of bright sources.
We return to this question in 
\S~\ref{sect:fluxunc}, in which we use the results of a simulation to
estimate the true errors in the flux densities.

\subsection{Filtering the Maps}\label{sect:mapfiltering}

\begin{figure}
\center
\includegraphics[trim=10mm 5mm 5mm 10mm,clip,width=84mm]{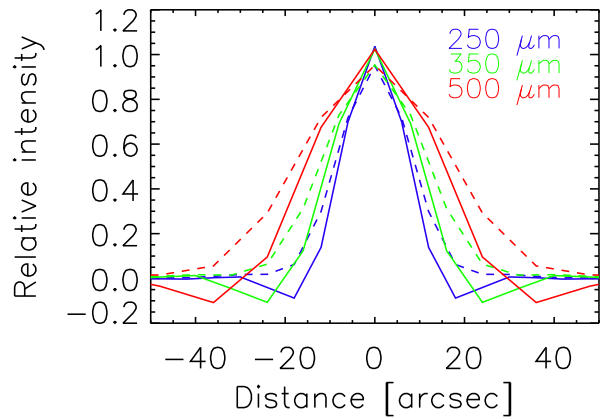}
\caption{Flux density verses radius in arcsec for the point spread function (dashed line) and
the matched filter (solid line) at the three wavelengths.}
\label{fig:profile}
\end{figure}

\citet{chapin11} show how to
calculate a  ``matched-filter'' that maximises the signal-to-noise for unresolved sources
from the ratio of the confusion noise and instrumental noise on the map.
We have used this method to calculate matched-filters for the
SPIRE images, using the
values for the instrumental and confusion noise estimated from the Gaussian fits to the 
histograms of the background-subtracted maps (top panel of Table~\ref{tab:noisegauss}). 
This is because we want to maximise the detection sensitivity of the survey, so we optimise 
the filter using the noise appropriate for sources near the detection limit.
Figure~\ref{fig:profile} shows the PSFs and the matched-filters. The matched-filters have the
negative lobes that are generated by this method
(see Figure~A1 of \citealt{chapin11}), which is the feature which makes
it possible to reduce the effect of confusing sources by
resolving sources that would otherwise be blended together.

To investigate whether this method gave an improvement 
in the signal-to-noise over using the PSF,
we produced two sets of convolved H-ATLAS images, one
set convolved with the matched-filter and one set with the PSF.
We then applied the same two methods of measuring the noise statistics that
we used for the raw images (see \S~\ref{sect:conf_noise}). Table~\ref{tab:noisevar} gives 
the results from the variance method and Table~\ref{tab:noisegauss} gives the
results from the Gaussian-fitting method.

First, consider Table~\ref{tab:noisevar}, which lists the values from the variance
analysis.
The reduction of the total noise due to the convolution with the PSF
results in an increase in signal-to-noise of 1.09, 1.04 and 1.12 at
250, 350 and 500 $\mu$m, respectively.
The gain in the signal-to-noise is only very small
because of the
large component of the variance that
comes from bright sources that are not confused, which
does not change when the images are convolved. 
The
reduction of the total noise due to the convolution with the
matched-filter results in an increase in the signal-to-noise
of 1.33, 1.30 and 1.39 at 250, 350 and 500 $\mu$m, respectively.
The gain in signal-to-noise from using the matched-filter rather than the PSF
is therefore 1.22, 1.25 and 1.24 at the three wavelengths.

Now consider Table~\ref{tab:noisegauss}, which lists the values from the Gaussian fitting.
The reduction of the total noise due to the convolution with the PSF
results in an increase in signal-to-noise of 1.87, 1.62 and 1.52 at
250, 350 and 500 $\mu$m, respectively.
The reduction of the total noise due to the convolution with the
matched-filter results in an increase in the signal-to-noise
of 2.01, 1.87 and 1.79 at 250, 350 and 500 $\mu$m, respectively.
The gain in signal-to-noise from using the matched-filter rather than the PSF
is therefore 1.07, 1.15 and 1.18 at the three wavelengths.

Therefore, whichever way we assess the noise, the use of the matched-filter
rather than the PSF when convolving the images results in a small but useful
gain in signal-to-noise.
\citet{chapin11}
found a similar useful 
but modest improvement in signal-to-noise when they applied their technique to data from the 
Balloon-borne Large Aperture Submillimeter Telescope 
(see their Table~1).

\section{The PACS Maps}\label{sect:pacs}

\subsection{Making the Images}\label{sect:pacsmaps}

We observed the sky simultaneously in two bands with the PACS camera, at 100 and 160\,$\mu$m. 
The two bands were designed to cover the wavelength ranges 85-130\, $\mu$m and 130-210\,$\mu$m 
\citep{poglitsch10}, although in practice the bands are slightly different than this. The detailed
spectral response of PACS in the two bands is shown in the PACS Observer's Manual
\footnote{\url{http://herschel.esac.esa.int/Docs/PACS/html/pacs_om.html}}.
In parallel mode, the SPIRE and PACS observations were
made simultaneously, with the two cameras offset on the
sky by $\simeq 21$\,arcmin. The final SPIRE and PACS images
of the GAMA fields therefore have a small
offset.

Processing the PACS data, which was taken in fast-scan parallel
mode, presented a number of challenges: (1) 
the huge datasets (the data for each GAMA field consist of
eight or nine observations
of 8-9 hours each) limited
the data processing to relatively powerful machines with at
least 64\,GB of random access memory; (2) 
the noise power on PACS images has a weak dependence on 
spatial frequency ($\propto 1/f^{\alpha}$ with
$\alpha\simeq 0.5$), which makes it difficult to reduce the noise by
spatial filtering
without affecting the properties of extended sources;
(3) the length of the observations meant that the observations were
affected by rare strong ``glitches'' that are missed by the standard
data-reduction software;
(4) the
on-board averaging of the PACS data that was required in parallel mode to make
it possible to transmit the data to the Earth
made it difficult to
disentangle glitches from real sources; (5)
the on-board averaging significantly affects the point
spread function (PSF), making it more elongated along the scan
direction (especially at 100\,$\mu$m).

The standard data-reduction pipeline provided by the PACS
Instrument Control Centre (ICC) working within the {\it Herschel}
Interactive Processing Environment ({\sc HIPE} v10.2747;
\citealt{ott10}) was not optimised for the H-ATLAS data and so
we made some modifications to it. 
The data reduction procedure we adopted 
is largely the one described by
\citet{ibar10}, but we made two major changes to the
procedure described in that paper.

\begin{figure*}
  \centering
  \includegraphics[width=175mm,keepaspectratio,angle=0]{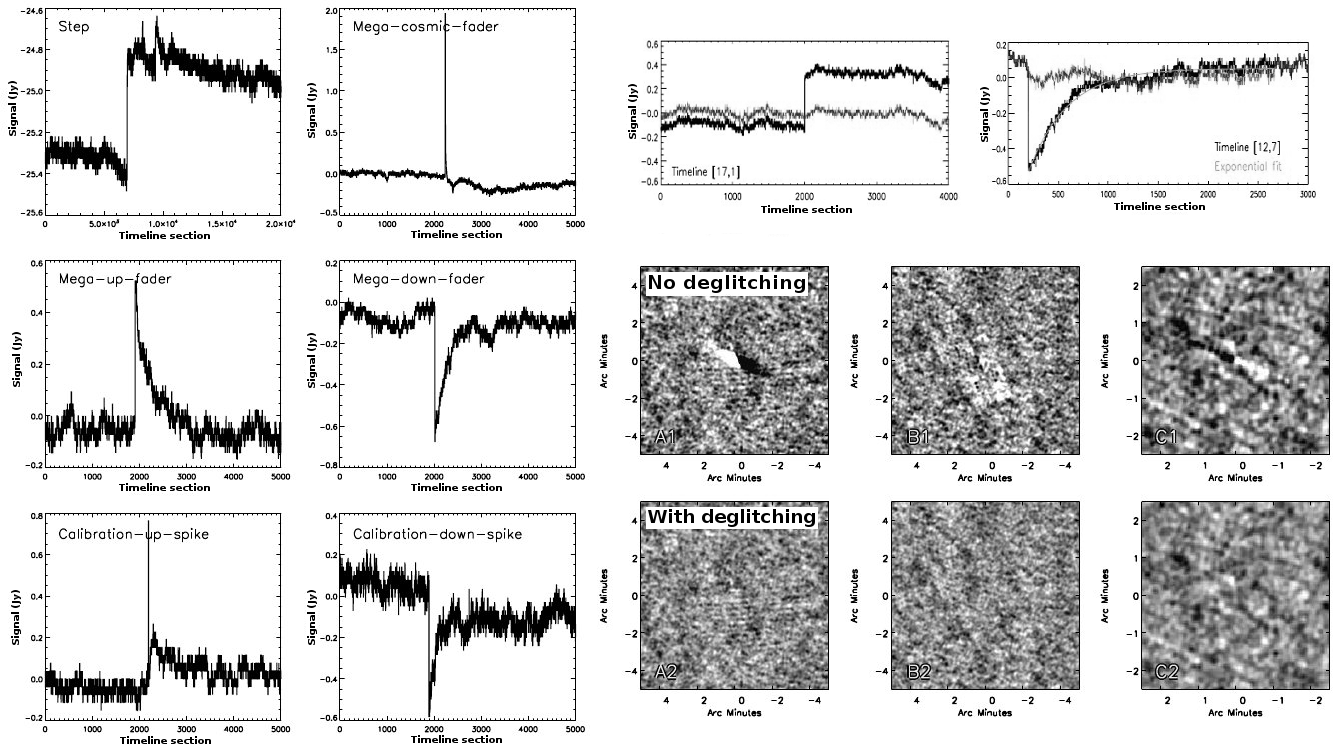}
  \caption{The first two columns at the left show six different
types of glitches in the PACS timelines that
were
missed by the standard de-glitching
approaches in {\sc HIPE}. 
The two panels at the top right show parts of the timelines
around two glitches before (black line) and after (grey line)
the application of our de-glitching program.
The six panels at the bottom right show images at the positions
of three different glitches, with the panels at the top showing the images
if no correction is made for the glitches and the panels at
the bottom showing the images after the application of 
our de-glitching program.}
\label{fig:glitches}
\end{figure*}

We processed the Level 1 data (the ``timelines'' - signal verses time for each
bolometer) largely using the method described in \citet{ibar10} with one major
modification, which is the way we dealt with spurious
jumps in timelines (``glitches'').
Our very long observations meant that our data
included rare types of glitch. In particular,
we found that there were several types of
strong glitch that occurred every 1-2 hours that
were
missed
by the de-glitching
methods used in the standard pipeline 
(Figure~\ref{fig:glitches}).
We wrote a program specifically designed for the H-ATLAS data
to find and correct them.
We corrected these glitches by fitting them with
either a Heaviside or an exponential function and subtracting this function
from the timeline.
If none of these functions provided a good fit to the glitch, we
simply masked the part of the timeline containing it.
Figure~\ref{fig:glitches} shows the effect on the final images of
detecting and removing these glitches.

The second major modification we made to the old procedure was
how we dealt with the problem that the noise on PACS images only
depends very weakly on spatial frequency (see above).
In producing our SDP images, \citet{ibar10} tried to reduce
the noise on the images by 
first filtering of the time-line data with a high-pass filter. We then 
made ``naive'' images from the timeline data, in which the sky brightness
in each image pixel is the average of the values for the time-line samples that fall within this pixel.
The filtering did reduce the noise, but at the price of removing 
extended structure and also reducing the
flux densities of the point sources, an effect we attempted to correct for \citep{ibar10}.

The naive images
do not use the information that there are at least two observations of each field, with
roughly orthogonal scan directions. A number of algorithms, such as 
the {\it Scanamorphos} algorithm \citep{roussel13}, use this information
to separate genuine extended emission on the sky from noise fluctuations in the
time-line data with low spatial frequency. 

Our approach for the Data Release 1 was therefore {\it not} to apply high-pass filtering
to the Level-1 data but instead to use one of these algorithms
to distinguish the noise fluctuations with low spatial frequency from genuine
large-scale emission.

As the first step, to calibrate the astrometry, we
made naive images from the data for each individual nine-hour observation. 
In principle, we should have been able to apply the
astrometric corrections for SPIRE maps (see \S~\ref{sect:dataproc}) to
the PACS data.
However, because the new PACS images were created using
a different model of where Herschel was pointing
(the ``pointing product'' provided 
within {\sc HIPE} v10.2747) to that used
for making the SPIRE images, we could not
simply assume that the sources detected by PACS were at exactly
the same positions as those detected by SPIRE.

To measure any offsets between the SPIRE and PACS positions,
we selected sources detected on the 160\,$\mu$m images
with peak flux densities
greater
than 65\,mJy/pixel and measured their positions
by fitting each source with a 
2D-Gaussian function. All of these sources
were detected on the SPIRE 250\,$\mu$m images and, for each individual
PACS image, we calculated
the mean differences between the positions of the sources
measured from the PACS and SPIRE images.
These mean differences were usually $\simeq 1$\,arcsec in both
RA and Declination.
We made our final PACS images after applying an astrometric
correction to each individual PACS dataset, which was the
sum of the astrometric correction obtained from making the
comparison between SPIRE and SDSS (see \S~\ref{sect:dataproc}) and the offset
obtained from comparing the PACS and SPIRE positions.

We tested three different algorithms for making the
final PACS images: the standard implementation of the
{\it Scanamorphos} algorithm \citep{roussel13}, the {\it JScanamorphos}
version of the algorithm provided as part of {\sc HIPE} \citep{graciacarpio15},
and {\it Unimap} \citep{piazzo15}.
We tested the images made using the different algorithms in
two different ways. First, we measured the noise on the
images by measuring flux densities through
a large number of apertures placed at random positions on the
images (\S~\ref{sect:pacsphot}). Second, we measured the point spread function
(PSF)
for each set of images using a stacking technique that is described
in the next section.
We found that the noise on the {\it Scanamorphos} images
was lower than on the other two sets of images, but that
the PSF on these images had negative sidelobes  
extending $\sim 1$\,arcmin from the peak of the PSF
in each scan direction. 
We are not sure of the reason for the sidelobes, but their existence 
suggests
that some spatial filtering was occurring in the version of
the algorithm we tested, which would also explain the lower noise.

There was no strong reason from our tests
to prefer one of the other two algorithms but for practical
reasons we decided on {\it JScanamorphos}.

We made the final images from the Level-1 data, after
making the astrometric corrections, using {\it JScanamorphos}.
We made the images using the ``parallel'' and ``nonthermal'' flags.
Since the PACS dataset for each GAMA field is too large for it to be practical to run 
{\it JScanamorphos} on the entire dataset, we made images
separately for each of the four rectangular regions (see Figure~\ref{fig:backsub} to visualise this).
We used a pixel scale of 3\,arcsec at 100\,$\mu$m and 4\,arcsec at 160\,$\mu$m,
which is approximately 1/3 of the full-width half-maximum at each wavelength,
which is how we selected the pixel scale for the SPIRE maps (\S~\ref{sect:dataproc}).

We then removed any residual large-scale artefacts from the
images by applying the {\it Nebuliser} algorithm,
which removes large-scale emission by using a combination of mean and median filters.
There is obviously a danger in this of removing genuine emission from extended
sources, such as nearby galaxies. We tested this effect by injecting
artificial galaxies on to the images, applying {\it Nebuliser} and then
measuring the flux densities of the galaxies. 
The galaxies were given exponential disks truncated at 5 scale lengths, and we tested a range of 
diameters from $12\arcsec$ to $96\arcsec$ and fluxes from 20\,mJy to 1\,Jy. This showed that a significant fraction 
of flux is lost from faint extended sources when the source radius is $>1/3$ of the filter scale. We chose to 
use a median filter scale of $300\arcsec$, and mean filter scale $150\arcsec$, which is means that galaxies 
smaller than $100\arcsec$ in radius will not be affected by the filter.

Finally, we masked parts of each
image for which there was data from only one scan direction, and then combined the
images for each GAMA field
using the {\sc SWARP} algorithm\footnote{\url{http://astromatic.iap.fr/software/swarp}} \citep{bertin02}.

The layout of the PACS mosaics is the same as that of the SPIRE mosaics
shown in Figure~\ref{fig:coverage}: four overlapping rectangular regions.
Each
quadrant has been made from two nine-hour observations with
orthogonal scan directions.
In the overlapping regions
the images have been made from four observations, and so the noise
level in these regions is approximately $\sqrt{2}$ less
than in the other areas.
In contrast to the SPIRE images, 
the PACS images are entirely dominated by
instrumental noise, which is much greater than the
confusion noise expected at these wavelengths
\citep{berta11}.

\subsection{The Point Spread Function}\label{sect:pacspsf}

The PACS team has carried out a detailed study of the PACS PSF \citep{lutz15}.
Although the PACS team did not have available
observations of a point source in our 
observing mode - parallel mode with a scan speed of 60\,arcsec\,s$^{-1}$ - it
was possible to estimate the PSF for this observing mode in the following way.
The crucial feature of the PSF in parallel mode is that 
it is elongated in the scan direction, especially at
70 and 100\,$\mu$m, because of
the
on-board averaging of the PACS data necessary to transmit both the
PACS and SPIRE data to the Earth.
Using observations with PACS alone of the asteroid Vesta and Mars
made with a scan speed of 60\,arcsec\,s$^{-1}$ and with the same
scan angles used in parallel mode, the PACS team simulated in software
the on-board averaging
used in parallel mode to produce an estimate of the PSF 
out to a radius
of 1000 arcsec.
However, although this should provide a reasonable estimate of our
PSF, the real PSF also depends on the pixel size of the
map, the spectral energy distribution of the source, and the algorithm used
to make the image \citep{lutz15}. For these reasons, we attempted to
estimate the real PSF for our observations from the observations
themselves.

Since, not surprisingly, there is no very bright point source
on our images,
we used a statistical ``stacking analysis''
to estimate the PSF.
To do this, we started with the optical counterparts to the H-ATLAS sources
presented by \citetalias{bourne16}. We used only the 
optical counterparts with spectroscopic or photometric
redshifts $<1$, since it is unlikely a galaxy at higher redshift will
have any discernible emission in the PACS bands, and with 
optical sizes (SDSS parameter ISOA) less
than 5\,arcsec, in order to ensure that the counterpart is likely to be unresolved in
the PACS wavebands.
There are $\sim 10^4$ of these
counterparts that satisfy these criteria 
in each field. We then extracted a $80\times80$
pixel image centered on each
counterpart from the overall PACS image, and then
added these together,
which should provide a good estimate of the real PSF of our observations.


Our estimates of the beams show no obvious artefacts from our processing method, in particular
from the {\it Jscanamorphos} imaging algorithm. {\bf The empirical PSFs we derived are not recommended
to be used as a filter to smooth the map, because of the uncertainties in the peak. We 
therefore do not release them. For filtering the map, we recommend instead the use of our Gaussian fit.} 
We fitted a two-dimensional Gaussian to the empirical PSF, finding that
the full-width half-maximum (FWHM) is 
11.8$\times$11.0\,arcsec at 100\,$\mu$m and 14.6$\times$12.9\,arcsec
at 160\,$\mu$m. These are slightly larger than the measurements from the PACS team's
simulated PSF: 10.7$\times$9.7\,arcsec at 100\,$\mu$m and 13.7$\times$11.5\,arcsec
at 160\,$\mu$m (Table 6, \citealt{lutz15}). We also fitted an azimuthally-symmetric
Gaussian to our empirical PSFs, which is the one we recommend for smoothing the maps, 
obtaining a value for the FWHM of 11.4 and 13.7\,arcsec at 100 and 160\,$\mu$m, respectively. It is
these values we used when carrying out photometry on the PACS images
(\S~\ref{sect:pacsphot}).

\begin{figure}
\includegraphics[width=84mm,keepaspectratio,angle=0]{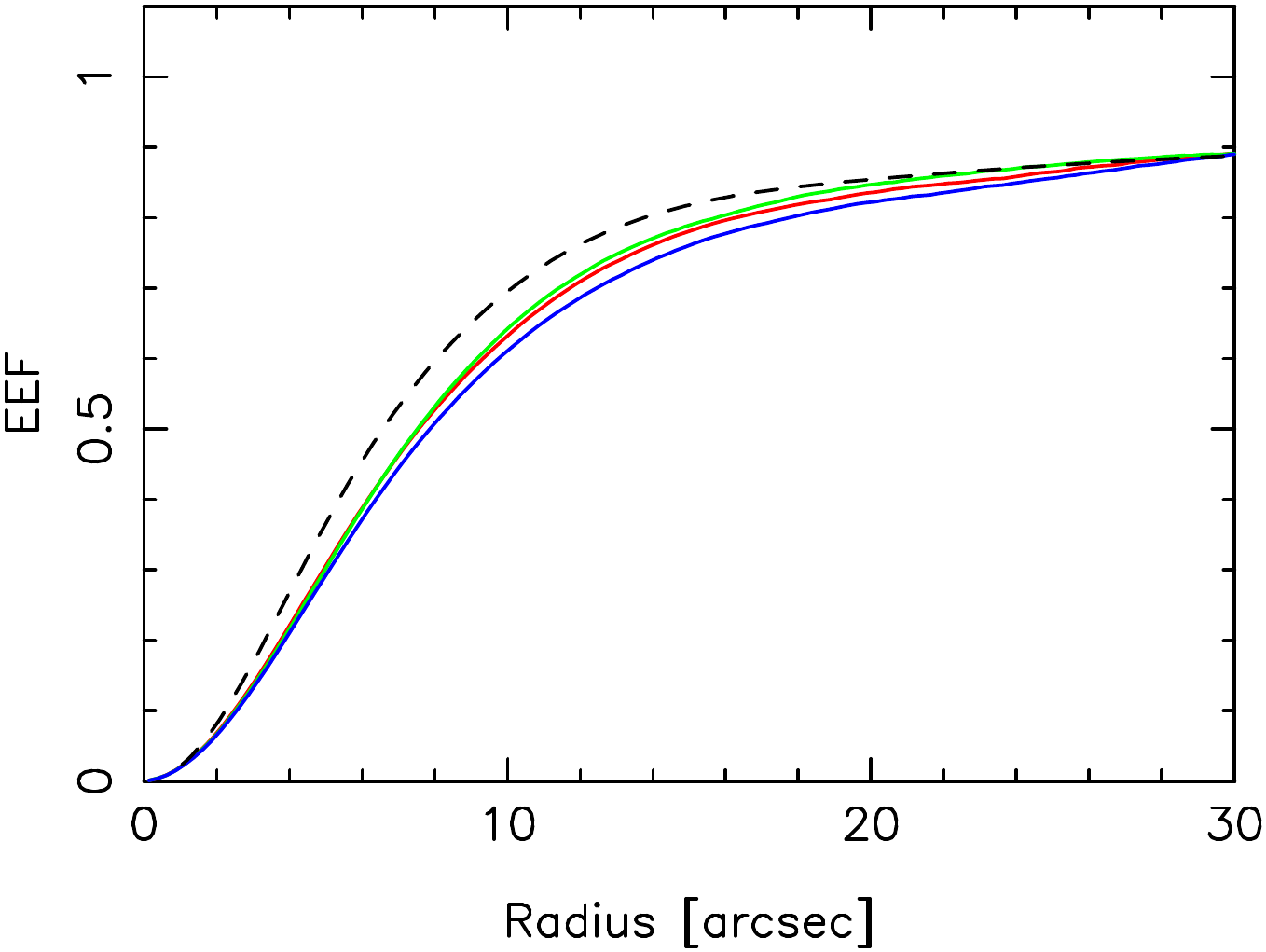}
\includegraphics[width=84mm,keepaspectratio,angle=0]{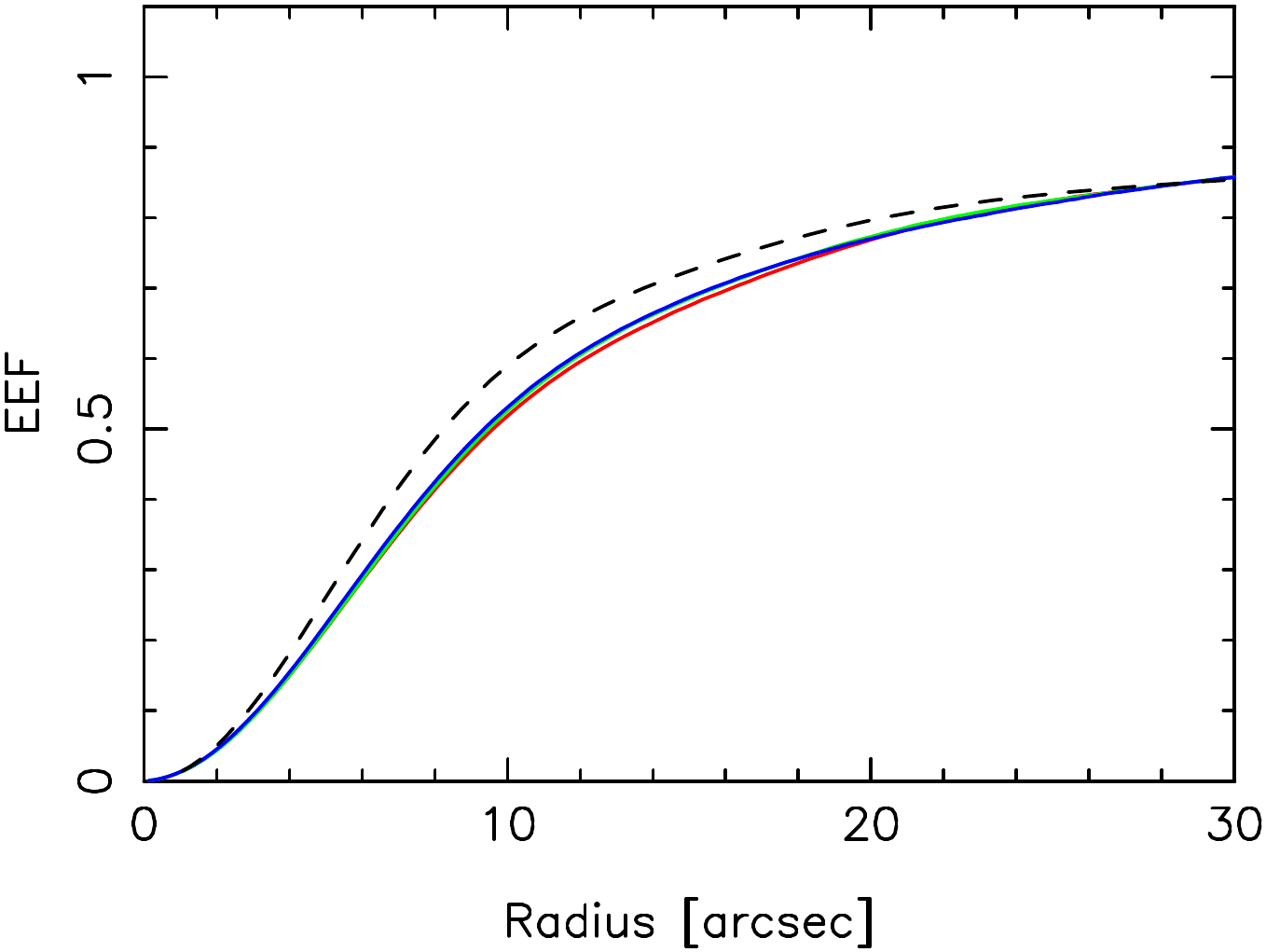}
\caption{The Encircled Energy Fraction (EEF) at 100\,$\mu$m ({\it top}) and 160\,$\mu$m
({\it bottom}). The dashed black line shows the PACS team's estimate of the EEF for our
observing mode \citep{lutz15}. The coloured lines
show our estimates for the three GAMA fields from the stacking procedure described
in the text (see \S~\ref{sect:pacspsf}). These curves have been normalised so that they have the same value
as the PACS team's EEF at a radius of 30\,arcsec. 
}
\label{fig:growth}
\end{figure}

Figure~\ref{fig:growth} shows the encircled energy fraction (EEF) plotted against radius
for the PSF derived from each of the GAMA fields. We have also plotted
in the figure the EEF derived from the PACS team's simulated PSF \citep{lutz15}. 
We have normalised all the curves
to the value of this EEF at a radius of 30\,arcsec.
The PACS team's EEF increases slightly faster
with radius than the estimates from the three GAMA fields, as expected given the
difference between the values of the FWHM.

The EEF is the crucial function for correcting aperture photometry (\S~\ref{sect:pacsphot})
to a standard radius. We can not derive an empirical EEF from our images much beyond
a radius of 30\,arcsec because the signal-to-noise produced by our
stacking method is too low.
We therefore constructed an EEF that can be used for aperture
photometry by using the PACS team's EEF for radii between 30\,arcsec
and 1000\,arcsec and the average value of our empirical EEFs for the
three GAMA fields
at smaller
radii, since these should best reflect our actual observations, normalising 
our empirical curve so that it has the same value as the PACS team's EEF
at a radius of 30\,arcsec. We supply these EEFs in the data release.

\section{Detection and Photometry of the Sources}\label{sect:phot}

\subsection{The Unresolved SPIRE Sources}\label{sect:madx}

Almost all the sources detected on the H-ATLAS images are unresolved by the telescope's PSF.
We have developed our own program to find unresolved sources on {\it Herschel} images: the Multi-band Algorithm
for source Detection and eXtraction (MADX)\footnote{The source code is available from the author Steve Maddox, maddox@roe.ac.uk}.
This will be described in detail elsewhere (Maddox, in preparation). Here we describe the basic algorithm.

The first step in the MADX source extraction is to use {\it Nebuliser} to remove the diffuse Galactic dust emission
from all the maps in the three bands (see \S~\ref{sect:dataproc}). 

In the second step, the images are convolved with the 
matched-filter (\S~\ref{sect:mapfiltering}, Figure~\ref{fig:profile}).
In carrying out the convolution, the 
contribution of each pixel of the
input image is weighted by the inverse of the square of the
instrumental noise in that pixel (\S~\ref{sect:noise_from_jn}). 
The matched filter is constructed
using the values of the instrumental and confusion noise given in the top one of the 
three panels in Table~\ref{tab:noisegauss}.

A map of the variance in this convolved map is created in the following
way. First, the map of the variance of the instrumental noise
(\S~\ref{sect:noise_from_jn}) 
is convolved with the matched-filter. Then the square of the confusion noise is added to the convolved map.
For this confusion term we used the mean values of the standard deviation from the
middle panel of Table~\ref{tab:noisegauss}, which are
2.0, 3.1 and 3.9\,mJy at 250, 350 and 500\,$\mu$m,
respectively. 
This variance map is used by MADX to determine the signal-to-noise
with which a source is detected.
The mean values of the square root of the variance given by this map
are 5.2, 5.6, 6.7\,mJy/beam.
{\bf Note the important point that these values are less than the more conservative values
we have adopted for the errors in the flux densities of the sources (see \S~\ref{sect:fluxunc}).}
We supply both the convolved images used by MADX and the noise maps of
these convolved images as part of this data release
(App.~\ref{app:spire}).

In the third step, the maps at 350 and 500\,\micron\ are 
interpolated onto images with the same pixel scale as the 250\,\micron\ map, 
and the three images and their associated variance maps
are then combined together to form a single signal-to-noise or ``detection''
image. In practice, we chose to give zero weighting to the
images at 350 and 500 $\mu$m, so our detection image was simply the 250\,$\mu$m
image. The reason for this choice was that
this image contains many more sources than the other SPIRE images, 
and the smaller size of the PSF leads to more accurate positions for the sources.
MADX produces a list of potential sources by finding all peaks in the detection
image with signal-to-noise $>2.5$.

The disadvantage of this approach is that we might miss sources that are very
faint at 250\,$\mu$m, but bright at the two longer 
wavelengths. By searching for $2.5\sigma$ detections at 250\,$\mu$m, but 
only releasing a catalogue of sources detected at a higher
significance in at 
least one of the three wavelengths, the 
catalogues at the two longer wavelengths should be fairly complete. 
\citet{rigby11} investigated the alternative 
``red prior'' method in which the three images are combined in a way that
makes the detection image sensitive to sources that are bright at the
two long wavelengths. By comparing catalogues
made in the two ways, they concluded that
in our 250\,$\mu$m-only approach, of the sources that should appear in a
$5\sigma$ catalogue, 
$\simeq 7\%$ of the sources at 350\,$\mu$m and $\simeq 12\%$ of the
sources
at 500\,$\mu$m will have been missed because they were
too 
faint to be detected at 250\,$\mu$m. We have already developed 
an extraction method that allows for the detection of 250\,$\mu$m drop-outs. 
The method is still under testing, but it has been used to select samples of ultra-red 
sources (Ivison et al., in prep).

After finding the peaks
detected at $>2.5\sigma$,
the next step in the algorithm is to sort the peaks in order of decreasing significance. 
The program fits a Gaussian to each peak of the detection map to provide 
an estimate of the source position accurate to a fraction of a pixel. 
The position of the source can be 
significantly affected by the presence of nearby sources \citep{rigby11},
but the negative sidelobes of the matched-filter 
(see Figure~\ref{fig:profile}) help to reduce this problem.

In the next step, MADX makes a first approximation of the flux density of 
sources in all three wavebands simply using the pixel value nearest to the source 
position in each of the filtered maps. Then for each band the sources are sorted in 
order of decreasing brightness. The flux density of the brightest source is estimated 
at the precise sub-pixel position determined from the detection image, using a bi-cubic 
interpolation between the flux densities in the surrounding ($3\times 3$) pixels. Using 
this measured flux density and the matched-filter-convolved point-spread function, this 
source is subtracted from the map. The program then moves to the next brightest  source 
and goes through the same set of steps. Note the sequence of image subtraction goes 
from brightest to faintest in each band, so the source subtraction can happen in a different 
order in each band. The subtraction of sources is important because this reduces the errors 
in flux densities of the faint sources caused by the wings of brighter sources.

The final step is to correct the flux densities for a small bias arising from the fact that sources 
are generally not found at the centre of a pixel, and the bicubic interpolation does not quite 
recover the true peak flux-density.
We multiplied the flux densities by 1.009, 1.013 and 1.010 at 250, 350 and 500\,$\mu$m, 
respectively, factors we estimated from the simulations
described in \S~\ref{sect:simulations}. For a similar reason, we need a further correction 
to the 350 and 500\,\micron\ flux densities: we noticed that, when sources are fainter than 40\,mJy at 250\,\micron,
their flux densities in the two longer waveband are slightly underestimated. We believe this is because the 
350 and 500\,\micron\ flux densities are measured at the 250\,\micron\ positions, and when the 250\,\micron\
flux densities are very low, the positions will be inaccurate. These corrections depend on
the 250\,\micron\ flux density only and are applied exclusively to sources fainter than 40\,mJy at 250\,\micron. 
We divided the flux densities at 350 and 500\,$\mu$m by $C_{350}$ and $C_{500}$, 
respectively, factors we estimated from the simulations described in \S~\ref{sect:simulations} and equals to:
\begin{align}\label{eq:faintcorr}
C_{350} = 0.877 + F250 \times 3.02 \nonumber\\
\\
C_{500} = 0.935 + F250 \times 1.53. \nonumber
\end{align}
These estimates of $C_{350}$ and $C_{500}$ are consistent with the analytical derivation of the same corrections
calculated assuming a Gaussian beams and the signal-to-noise ratio expected for sources fainter than 40\,mJy at 
250\,\micron.

The final flux densities are monochromatic flux densities
at 250, 350 and 500\,$\mu$m, calculated on the assumption that
$F_{\nu} \propto \nu^{-1}$, in which $\nu$ is frequency.

On top of the basic errors in the flux densities, which we discuss in \S~\ref{sect:fluxunc},
there is an additional error due to the uncertain photometric calibration
of {\it Herschel}. At the time of writing (the calibration of {\it Herschel} is
still improving), the error in the flux densities in the SPIRE
bands arising from the
absolute
uncertainty in the flux density of Neptune is 4\%
in all three bands and there is an additional error of 1.5\%
that is uncorrelated between the bands (SPIRE Handbook).
As recommended in the SPIRE Handbook, we add these errors
linearly and use a value of 5.5\% as the total calibration error.

A few of the sources detected by MADX are asteroids.
We removed these from the catalogues, although the positions on the images
are given in App.~\ref{app:asteroids}.
 
\subsection{The Extended SPIRE Sources}\label{sect:spireext}

The flux density measured 
by MADX will be an underestimate of the true flux density if the source is
resolved by the PSF of the telescope. It is often hard to be sure from the
SPIRE images alone whether a source is genuinely extended or whether it is
multiple unresolved sources confused together.
Apart from a handful of Galactic sources, a source is only 
likely to be extended
if it is associated
with a nearby galaxy. We have therefore 
used our catalogue of optical counterparts to the
Herschel sources, which is described in \citetalias{bourne16}, to decide whether a source
is likely to be extended.

\citetalias{bourne16} investigated whether a potential optical counterpart
on the SDSS $r$-band image is genuinely associated with a Herschel
source using a Likelihood method, based on the positional error
of the Herschel source, the angular distance between the potential counterpart
and the Herschel source and the $r$-band magnitude of the counterpart. 
Using this method, they estimated the probability (the reliability - R) 
that the potential counterpart is genuinely associated with the Herschel source.
They treated any object with $R \ge 0.8$ as being the likely counterpart to the {\it Herschel}
source\footnote{The only exception are gravitational lenses, which also
often have $R > 0.8$
\citep{negrello10, gonzalez12}, but the lenses are at high enough redshifts
that they have small optical sizes, and thus their existence does not distort
our analysis.}.

One of the quantities available for each galaxy from the SDSS catalogues is
the optical size (ISOA, in SDSS nomenclature)
which is defined as the length of the
semi-major axis
out to an isophote corresponding to 25 mag arcsec$^{-2}$.
We only used values of ISOA for
galaxies with SDSS {\it r}-band magnitude (the SDSS quantity {\it rmodelmag}) $<$ 19, because
values of ISOA for galaxies fainter than this limit are unreliable. 
In deciding whether a SPIRE source is likely to be extended, and thus whether aperture
photometry is necessary, our
basic assumption is that the radius of the submillimetre emission is proportional to ISOA.

We measured flux densities using
aperture photometry for all reliable optical counterparts 
($R > 0.8$)
of the sources detected
by our MADX source-finding algorithm (\S~\ref{sect:madx})
and with ${\rm ISOA}>10$\,arcsec, centering the aperture at
the position given in the SDSS for the galaxy.
We converted the images from Jy/beam to Jy/pixel by dividing the 
values in each pixel by a factor $C_{\rm conv}$, which is given
by the 
area of the beam divided by the area of a pixel. 
We used the current estimates of beam from the SPIRE handbook,
which are
469, 831 and 1804\,arcsec$^2$ at
250, 350 and 500\,$\mu$m, 
respectively. 

We measured the flux densities from the images from which the background
had been removed using {\it Nebulizer} (\S~\ref{sect:dataproc}) {\bf and after setting the mean of the maps to zero}. 
The alternative would have been to
measure the flux densities from the raw maps, using a annulus around each galaxy to
estimate the level of the background. However, in tests we found that
the two methods gave very similar results for even the biggest galaxies, whereas the
errors in the flux densities when starting from the raw maps were larger.

We used a circular aperture with a radius, $r_{\rm ap}$, given by

\begin{equation}\label{eq:aperture_spire}
r_{\rm ap} = \sqrt{{\rm FWHM}^2 + {\rm ISOA}^2}. 
\end{equation}

\noindent 
In this equation ISOA is in units of arcsec (we converted the values in the
SDSS database, which are in units of SDSS pixels, using the SDSS pixel size of 0.3958\,arsec).
To increase photometric accuracy when the aperture size is small relative to
the pixel size, we divided each pixel into 16 sub-pixels, assigning one sixteenth
of the flux density in the real pixel to each sub-pixel.
We corrected all the aperture flux densities for the fraction of the emission
that falls outside the aperture because of the extended profile of the PSF \citep{griffin13}.
We used a version of the PSF appropriate for a source with a spectral energy distribution
with monochromatic flux density, $F_{\nu} \propto \nu^{-1}$, which is
the spectral energy distribution assumed in the pipeline (SPIRE Handbook). 
We provide the table with the SPIRE aperture corrections as part of this data release. 
In App.~\ref{app:cat}
we describe how corrections should be made to the flux densities 
in the catalogues to obtain flux densities for more
realistic spectral energy distributions.

We estimated the uncertainties by placing at random positions around each sources 2000 apertures with the same size as the one
used to measure the flux of the source and at distance between 2 and 15 times the aperture radius. 

In general, the flux error scales as $r_{\rm ap}^{3/2}$ for small apertures and is flatter for aperture radi larger than $\sim 45\,{\rm arcsec}$. 
This relationship is different from the one expected for a Gaussian noise ($\propto r_{\rm ap}$) because of the confusion component. 
The flattening at larger scales is also expected, because {\it Nebuliser} removes the large-scale power (\S~\ref{sect:dataproc}). No flattening is indeed 
visible if the same process is applied to the raw maps.

This Monte-Carlo technique takes also into account fluctuations in the background and variation of the confusion noise within the map,
because it is applied to each source independently.

\begin{figure}
\center
\includegraphics[width=80mm,keepaspectratio,angle=0]{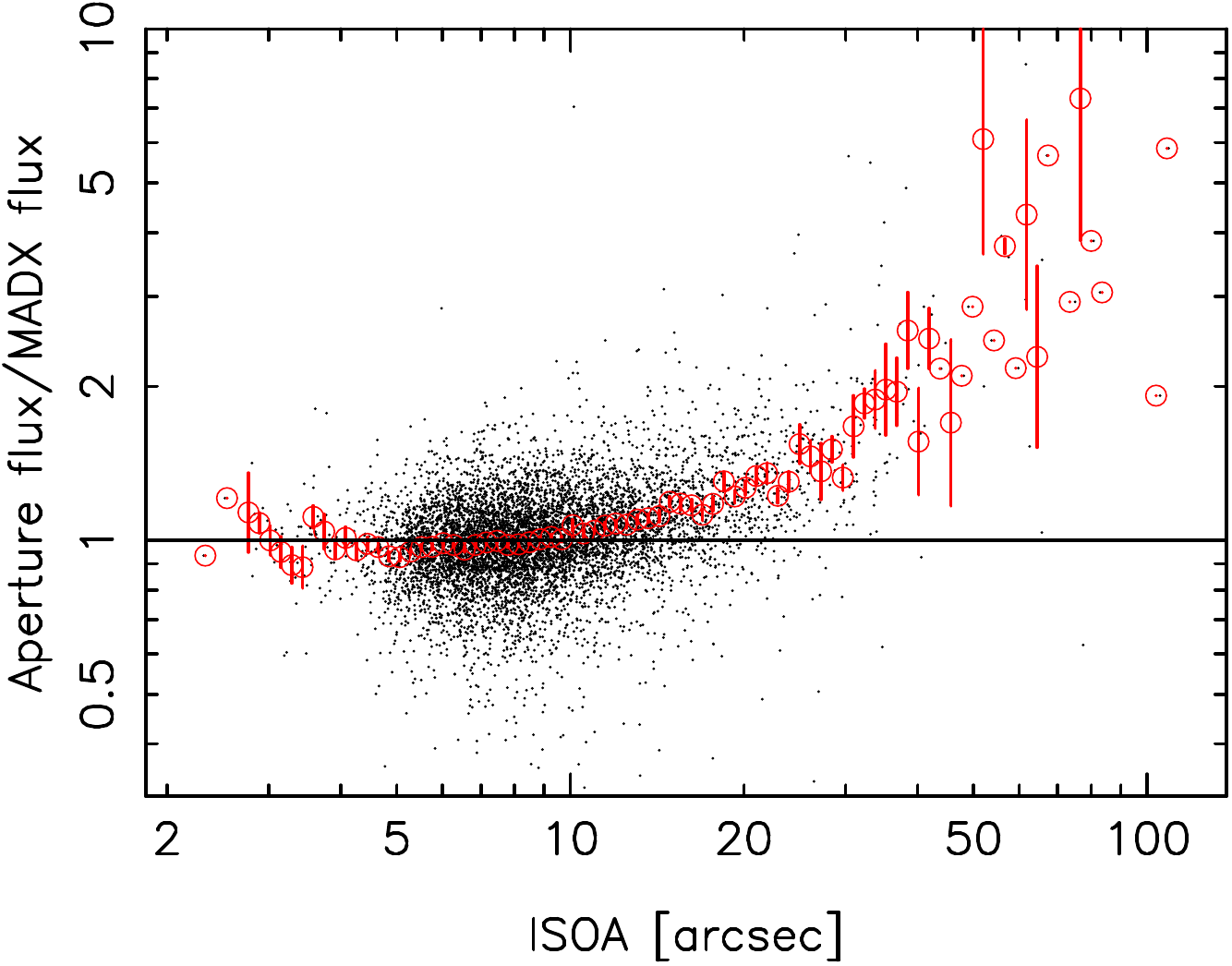}
\caption{Ratio of flux density measured through the aperture defined by Eq.~\ref{eq:aperture_spire}
to the flux density measured by the MADX source-detection algorithm plotted against
the value of the ISOA parameter, which is a measurement of the semi-major axis
of the optical counterpart. The points represent the galaxies in the three GAMA fields
with values of ${\rm ISOA}>1$\, arcsec. The red points are the mean values of the
ratio of the flux densities for 100 bins of ISOA, with the error bars
showing the errors on the mean. The horizontal line shows where the MADX and aperture
flux densities are equal.
}
\label{fig:apphot}
\end{figure}

\begin{figure}
\center
\includegraphics[width=80mm,keepaspectratio,angle=0]{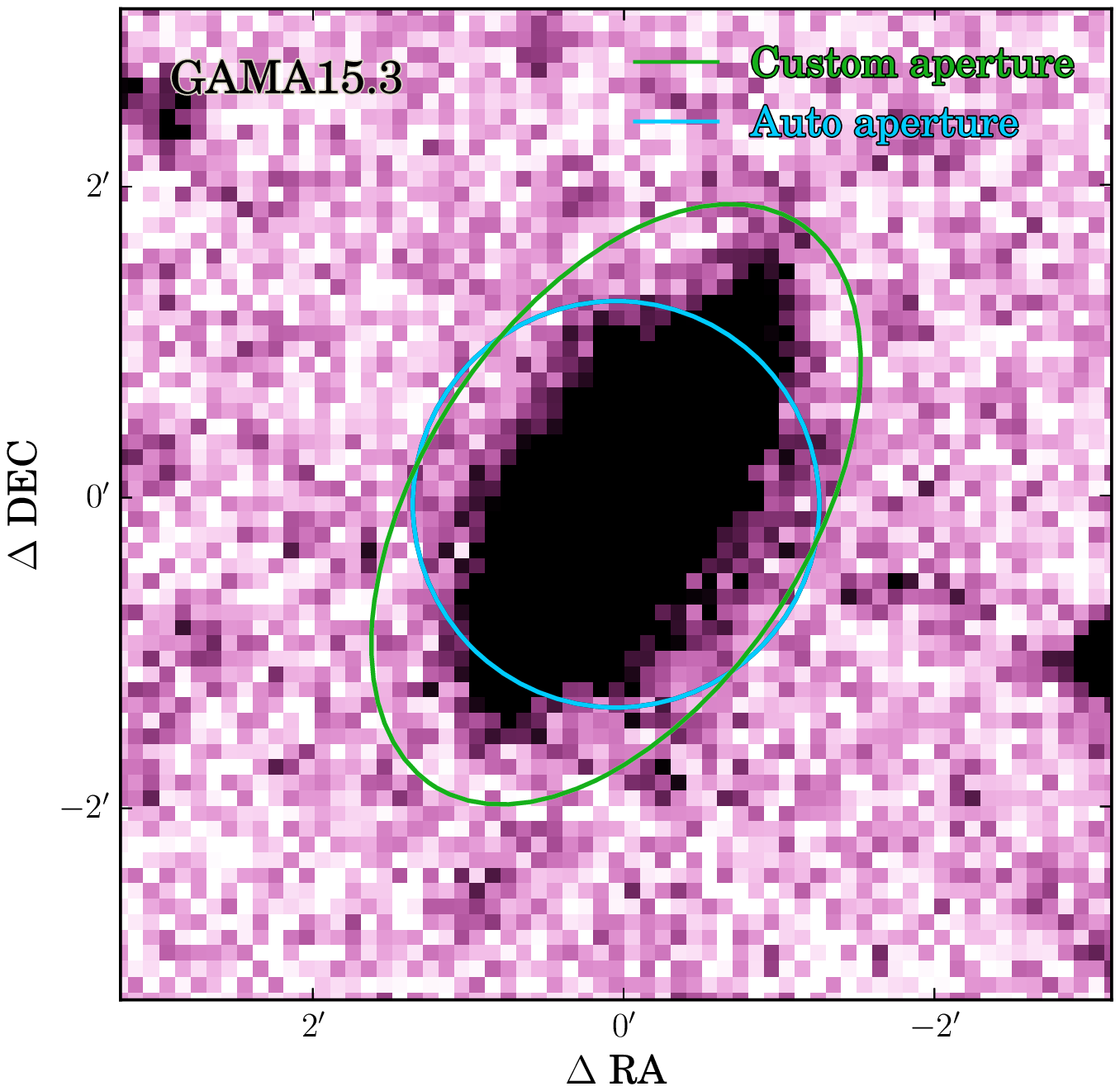}
\includegraphics[width=80mm,keepaspectratio,angle=0]{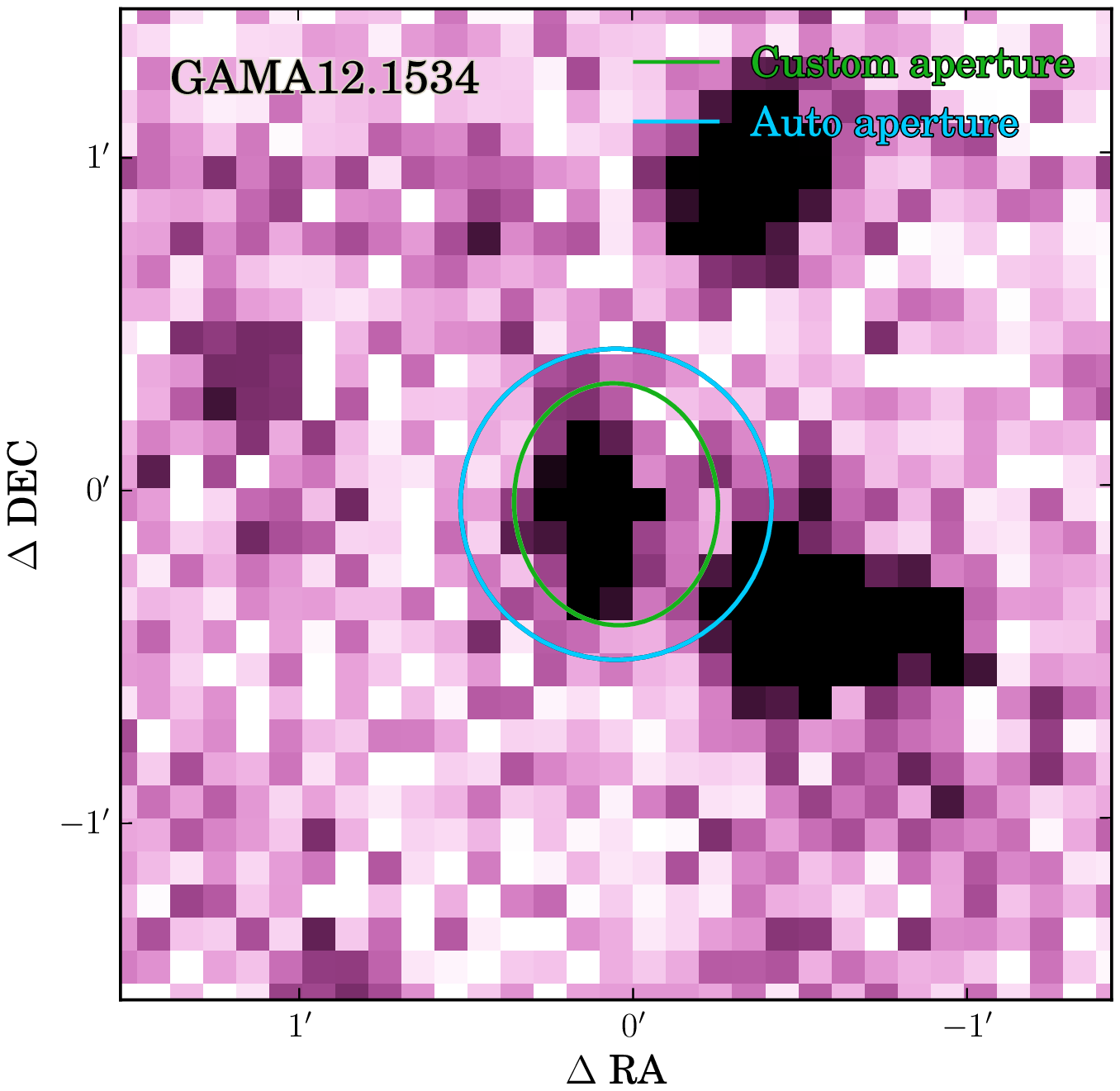}
\caption{Two examples of sources that, after visual inspection, required a customised aperture 
defined manually ({\it green line}), because the automatic aperture defined by Eq.~\ref{eq:aperture_spire} ({\it blue line}) was
inappropriate. {\it Top}: automatic aperture having a wrong shape. {\it Bottom}: automatic aperture including
flux from neighbouring sources. Images realised using {\sc cubehelix} \citep{green11}.
}
\label{fig:custom}
\end{figure}

Figure~\ref{fig:apphot} shows the ratio of the flux density measured from aperture photometry to
the flux density measured by MADX (\S~\ref{sect:madx}), in which it
is assumed that the sources are unresolved, plotted against ISOA. The red points
show the mean ratio of flux densities in bins of ISOA, with the error
bars showing the errors on the means. At ${\rm ISOA}>10$ arcsec,
flux densities are systematically greater than the flux densities measured
by the source-detection algorithm. As the result of this analysis, in first approximation
we would expect the aperture flux density to be the best estimate of the flux 
for galaxies with ${\rm ISOA} >10$\,arcsec. However, we prefer this estimate of the flux
instead of the the flux densities measured by the
source-detection algorithm, only when the result of the aperture photometry is significantly 
larger than the flux measured by MADX. This is true when
\begin{equation} 
(F_{\rm ap}-F_{\rm ps})>\sqrt{\sigma_{\rm ap}^2-\sigma_{\rm tot\_Var}^2} ,
\end{equation}
where $F_{\rm ap}$ and $\sigma_{\rm ap}$ are respectively the flux density estimated by the aperture
photometry and the uncertainty from the Monte-Carlo process and $F_{\rm ps}$ and $\sigma_{\rm tot\_Var}$
are the flux density measured by the source-detection algorithm and the uncertainty estimated for unresolved
sources (see \S~\ref{sect:madx} and Eq.~\ref{eq:sigma_tot_var2}). When this condition is not satisfied, the fluxes estimated by MADX are preferable, 
since the algorithm is optimised for unresolved sources and the flux-density errors will thus be smaller
than in the aperture photometry.

For galaxies with large optical sizes, we visually compared the aperture given by
Eq.~\ref{eq:aperture_spire} with the 250\,$\mu$m emission of the galaxy. In $\sim 100$ cases,
the aperture was not well matched to the 250\,$\mu$m emission, either being too small, too
large, with the wrong shape or including the flux from a neighbouring galaxy (see Fig.~\ref{fig:custom} for 
examples). In these cases, we chose a more appropriate aperture for the galaxy. The details about
the customised apertures of these galaxies are given as part of the data release.

\subsection{PACS Photometry}\label{sect:pacsphot}

\begin{figure*}
\includegraphics[width=84mm,keepaspectratio,angle=0]{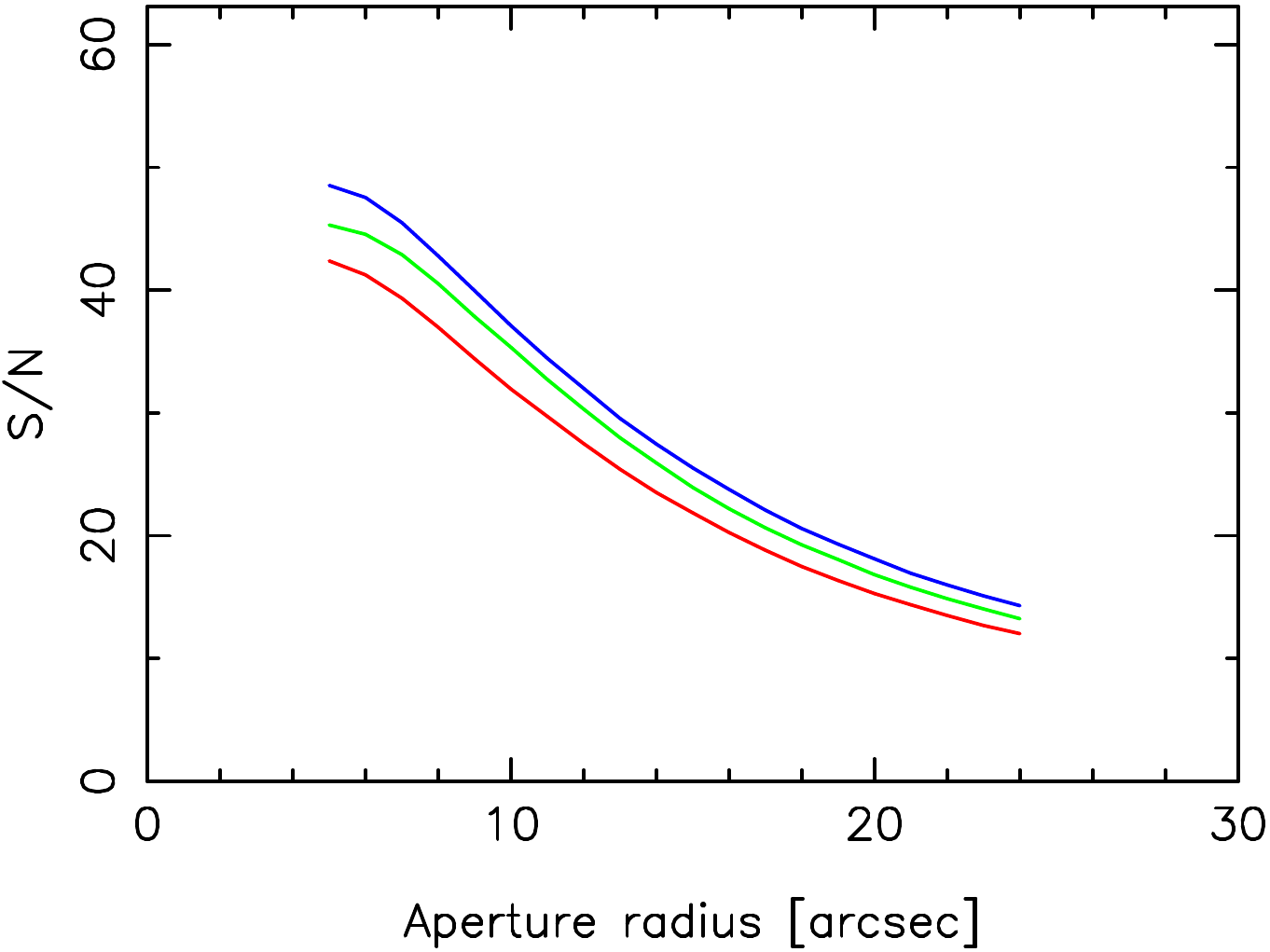}
\includegraphics[width=84mm,keepaspectratio,angle=0]{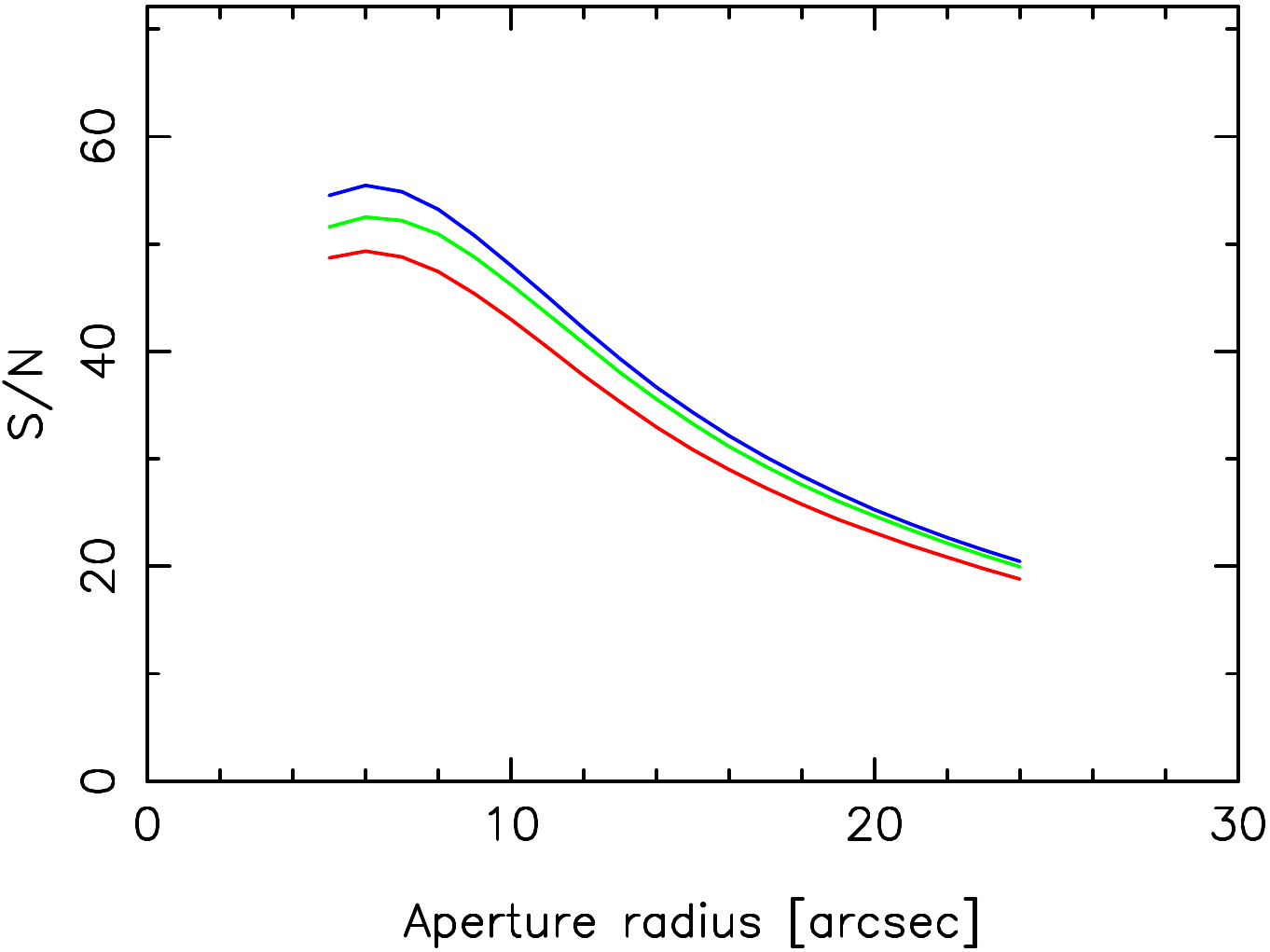}
\caption{Results of a simulation of how the average signal-to-noise for an unresolved source
depends on the radius of the aperture used to carry out the photometry (see \S~\ref{sect:pacsphot} for
details). The two panels show the results for 100\,$\mu$m ({\it left}) and 160\,$\mu$m ({\it right}).
Each colour shows the results for one of the GAMA fields.
}
\label{fig:ston}
\end{figure*}

As a result of the lower sensitivity of the PACS images,
all the sources detected on the PACS images are detected
on the SPIRE 250\,$\mu$m image. Therefore, 
we have measured flux densities in the two PACS bands
for the sources detected with MADX on the
SPIRE 250\,$\mu$m image. We have measured these flux densities
using aperture photometry,
centering the aperture either on the position of the
SPIRE source
or, when possible, on the position of the optical counterpart to
that source.

To determine the aperture size that would
maximise the signal-to-noise
of the photometry, we followed a similar ``stacking procedure'' to that described
in \S~\ref{sect:pacspsf}. We started with all the optical counterparts to the SPIRE
sources which have $z<1$ (spectroscopic or photometric) and
optical size (SDSS parameter ISOA) $<5$\,arcsec. We measured
the flux density for each galaxy from the PACS images through 20 apertures
with radii equally spaced from 5 to 25\,arcsec (we did not need to measure a sky value because
of the application of {\it Nebuliser} to the images\footnote{We checked whether the application of {\it Nebuliser} led to any loss of flux for big galaxies by also carrying out aperture photometry on images to which Nebuliser had not been applied. In this photometry, we estimated the background from an annulus centred on the galaxy with radii equal to 1.5 and 2.5 times the radius of the aperture. We found that the application of Nebuliser leads to a significant loss of flux for galaxies if their diameter is larger than about 1/3 of the filter scale, confirming the results of our Monte-Carlo simulation (see \S~\ref{sect:pacsmaps}).}).   
In our signal-to-noise calculation, the signal for an aperture
of a given radius is the sum of the measurements for all the
galaxies.

To estimate the noise, we used a Monte-Carlo simulation in which, for
each aperture size, we placed 3000 apertures randomly on the images,
calculating the standard deviation of the flux densities measured in all the apertures. 
In the Monte-Carlo simulation, we used aperture sizes radius of 5,10,15...\,arcsec, 
up to a maximum radius of 100\,arcsec.
From this simulation, the error for aperture photometry, $\sigma_{\rm ap}$, 
on the PACS images
is given by the following relationship:

\begin{equation}\label{eq:pacs_noise}
\sigma_{\rm ap} = C (r/10\,{\rm arcsec})^{\beta} \sqrt{2/N_{\rm scan}}
\end{equation}

\noindent in which $r$ is the radius
of the aperture, the constant $C$ is 0.023 Jy and 0.020 Jy at 100 and 160\,$\mu$m, 
respectively; $\beta$ is 1.47 and 1.46 at 100 and 160\,$\mu$m, respectively; 
and $N_{\rm scan}$ is the number of PACS observations
contributing to the pixels in the aperture (two in most cases, but four
in the regions in which the quadrants overlap).
If the noise was completely uncorrelated between pixels, the value of
$\beta$ should be one. The fact that $\beta>1$ shows the limitation on the
accuracy of photometry on large scales produced by the noise characteristics of
PACS (\S~\ref{sect:pacsmaps}).

Figure~\ref{fig:ston} shows signal-to-noise
plotted against aperture radius for the three GAMA fields. 
At both wavelengths, the signal-to-noise
is at a maximum at very small apertures
($r< 8$\,arcsec). Thus the way to produce PACS photometry with the best possible
signal-to-noise would be to carry out photometry using this aperture, and then use the
Encircled Energy Function (EEF, \S~\ref{sect:pacspsf}) to correct the photometry to our
standard reference radius of 1000\,arcsec.
This argument only holds if one knows precisely the position of the
object for which one wants PACS photometry; if there is 
significant uncertainty in the position, using a very small aperture may lead
to some of the flux being missed, which is the position we face when using 
the positions of SPIRE sources.

In practice, because of the uncertainties in the positions of the
SPIRE sources, we did not use such a small aperture. Instead,
we use an aperture with a radius given by the following 
relationship:

\begin{equation}\label{eq:aperture_pacs}
r_{\rm ap} = \sqrt{{\rm FWHM}^2 +  {\rm ISOA}^2}. 
\end{equation}

\noindent We used the value for the FWHM obtained from our empirical
PSF (see \S~\ref{sect:pacspsf}): 11.4 and 13.7\,arcsec at 100 and 160\,$\mu$m, respectively.
We deliberately made the apertures smaller for the PACS observations than for
the SPIRE observations because of the
poorer sensitivity of the PACS images.
To improve the accuracy when the aperture sizes were small relative to the
pixel sizes, we divided each pixel into four, assigning one fourth of the
flux density of the real pixel to each sub-pixel.

If a source had a reliable optical counterpart ($R>0.8$, \S~\ref{sect:spireext}), we centered
the aperture on the optical position and used the value of ISOA for the
counterpart. If the source did not have an optical counterpart, we
centered the aperture on the SPIRE position and set the value of ISOA
to zero. In the latter case, the uncertainty in the SPIRE positions
leads to a loss of flux density.
Using a 
positional error for the SPIRE sources of 3.4\,arcsec \citepalias{bourne16}, we estimated 
that to correct for this loss of flux we needed to increase the measured flux
densities by 1.10 and 1.05 at 100 $\mu$m and 160 $\mu$m, respectively,
and therefore if the source did not have an optical counterpart, we multiplied
the measured PACS flux densities by these factors.
We corrected all flux densities to our reference radius of 1000\,arcsec using
our empirical PACS EEF (\S~\ref{sect:pacspsf}).
We used Eq.~\ref{eq:pacs_noise} to estimate the error for each flux density measurement, scaling
the error to the reference radius using the EEF.

The absolute accuracy of the PACS flux scale is 5\% (PACS Observer's Manual). The reproducibility
of measurements of the flux density of individual point sources is better than 2\%. To be conservative,
we combine these errors linearly, giving a calibration uncertainty on our flux densities
in the two PACS bands of 7\%. As for SPIRE (see \S~\ref{sect:spireext}), all our measurements of flux density
are based on the assumption that the flux density, $F_{\nu}$, of a source has the spectral
dependence $F_{\nu} \propto \nu^{-1}$. Most sources have very different spectral energy distributions
and in App.~\ref{app:cat} we describe how corrections should be made to flux densities
in the catalogues to obtain flux densities for more realistic spectral energy 
distributions.

\section{The In-Out Simulations} \label{sect:simulations}

\subsection{Overview}

We need simulations to determine the completeness of the catalogues and the real errors in the
flux densities, the random errors but also the systematic errors (flux
bias, \S~\ref{sect:intro}). 
Although flux bias and Eddington bias (see \S~\ref{sect:intro}), the systematic errors on the number counts, affect 
all astronomical surveys \citep{hogg98}, they have always been a 
particular problem in submillimetre catalogues because the submillimetre source counts are much steeper 
than the number counts in other wavebands \citep{clements10,oliver10}. Quantitative 
estimates of flux bias in ground-based surveys at 850\,$\mu$m have a range 
of 20-40\% of the measured flux densities \citep{eales00,scott02,coppin05}. 

The approach we followed for the catalogue we released at the end of the {\it Herschel} SDP \citep{rigby11} 
was to use a theoretical model of the galaxy population to generate artificial
submillimetre catalogues, which we then compared with the real catalogues. 
The drawback of this is that the method depends on the
model being a good representation of the galaxy population. In practice, the model reproduced
well the slope of the submillimetre number counts but not the overall normalisation of the counts.
A second disadvantage of this kind of method is that its complexity makes it very difficult
for anyone else to reproduce.

Our approach in this paper is simpler and more empirical. We have broken down the problem into two parts. In this
section, we describe simple simulations in which we place artificial sources on the real
images and then try to find them with our source-extraction algorithm (\S~\ref{sect:madx}). 
These simulations allow us to determine, for a known true flux ($F_{\rm t}$), the conditional probability 
distribution of 
the measured flux ($F_{\rm m}$): $P(F_{\rm m}|F_{\rm t})$. 
From the simulations, we derive
analytic 
distributions of $P(F_{\rm m}|F_{\rm t})$ for a large range
of $F_{\rm t}$ in the three SPIRE bands. The simulations also allow us to
estimate the completeness of the 250\,$\mu$m catalogue 
as a function of true flux density ($F_{\rm t}$)
by measuring
$N_{\rm d}/N_{\rm i}$, the ratio of the number of detected sources to the number of injected sources. 

This simple approach does not, however, allow us to estimate the more 
observationally useful function which is the 
completeness of the survey as a function
of {\it measured} flux density. It also does not allow us to estimate the flux bias in the
three wavebands, the completeness in the two longer wavebands, and the effect of Eddington bias.
All these questions form the second part of the problem, which we address in \S~\ref{sect:counts250} 
and \S~\ref{sect:corrections}.

\begin{table*}
\centering
\caption{Statistics of the In-Out Simulations. For each 250\,\micron\ flux bin, the table reports the ratios of the number of recovered sources
to the number of injected sources, $N_{\rm d}/N_{\rm i}$, averaged over the whole survey and for the deepest areas only, and the 
parameters of the shape of the conditional probability distribution, $P(F_{\rm m}|F_{\rm t})$, described as two Gaussians with different 
standard deviations (see \S~\ref{sect:simresults}, Eq.~\ref{eq:2Gauss}).}
\begin{minipage}{100mm}
\centering
\label{tab:stats}
\begin{tabular}{@{}ccccccc@{}}
\hline
      &   &    & \multicolumn{4}{c}{Double-Gaussian Fits} \\
$F_{\rm t}$ & \multicolumn{2}{c}{$N_{\rm d}/N_{\rm i}$} &  $\hat{F}$ & $\sigma_1$ & $\sigma_2$ & $N_0$  \\
    {[}mJy]       &  whole survey         &  4 scans regions  only   &             [mJy]        &    [mJy]         &        [mJy]      &    [mJy]        \\
\hline
   6.0    &    0.093 &   0.131 &        14.7    &   2.1    &   1.9    &   244     \\
   7.4    &    0.136 &   0.144 &        15.7    &   2.5    &   2.1    &   592	\\
   9.1    &    0.190 &   0.205 &        15.4    &   2.3    &   2.9    &   661 	\\
 11.1    &   0.277  &   0.373 &       15.1    &   2.1    &   3.6    &   719 	\\
 13.7    &   0.417  &   0.527 &        15.2    &   2.0    &   4.8    &   726 	\\
 16.8    &   0.598  &   0.699 &        16.3    &   2.4    &   6.2    &   1668 	\\
 20.6    &   0.761  &   0.788 &        18.3    &   2.8    &   8.3    &   1890 	\\
 25.4    &   0.871  &   0.874 &        22.8    &   3.7    &   9.4    &   2434 	\\
 31.2    &   0.915  &   0.903 &        29.0    &   4.4    &   9.0    &   2646 	\\
 38.3    &   0.961  &   0.960 &        35.7    &   4.4    &   9.4    &   2824 	\\
 47.0    &   0.980  &   0.971 &        44.6    &   4.3    &   9.2    &   2989        \\
 57.8    &   0.991  &   0.992 &        54.8    &   4.3    &   9.2    &   3383	\\
 71.0    &   0.992  &   0.989 &        67.5    &   4.0    &   9.6    &   3936 	\\
 87.2    &   0.994  &   0.991 &        84.2    &   4.5    &   9.3    &   4425 	\\
107.2   &   0.997  &   0.996 &     104.6    &   4.6    &   8.5    &   3515 	\\
131.7   &   0.997  &   0.995 &      129.0    &   4.7    &   8.8    &   5613 	\\
161.8   &   0.996  &   0.998 &     159.1    &   4.5    &   9.0    &   7647 	\\
198.7   &   0.997  &    1.00  &   195.7    &   4.8    &   9.4    &   7178 	\\
244.2   &   0.999  &    1.00  &   241.2    &   5.0    &   9.3    &   14303 	\\
300.0   &   0.998  &    0.998&    297.2    &   5.0    &   9.0    &   3842 	\\
\hline
\end{tabular}
\end{minipage}
\end{table*}

\subsection{The Method}\label{sect:method}

We started our simulations 
with an empty image (no noise or sources) with 1-arcsec pixels.
We then added sources to this image at positions that lay on a grid. The grid spacing was
sufficiently large that any overlap between injected sources was avoided and that
when these artificial sources were added on to the real images (see below)
they did not affect the
statistics of the images, especially the confusion noise. We added a small random scatter to the
injected positions, so that when they were added onto the real images
they did not fall at the same position within a real pixel.

We then created three images 
from the high resolution maps by convolving them with the PSF
for the three SPIRE bands and then 
re-binnning these images to the actual pixel sizes of 6, 8 and 12\,arcsec at the three SPIRE wavelengths.
As a ``sanity test'' on our software, we first injected 100 1\,Jy sources and ran MADX (see \S~\ref{sect:madx}) to find these sources on these noiseless maps, using the same settings
we used for the real images. 
The mean of the 100 flux densities we measured was 0.991, 0.987 and 0.990\,Jy at 250, 350 and 500\,$\mu$m, 
respectively, with a scatter of less than 1\%. This error is caused by the fact that sources are not generally found at the centre of a pixel. The 
bi-cubic interpolation in MADX partially corrects for this effect but does not do it perfectly. We used these factors to correct the flux densities
of the real sources measured by MADX (\S~\ref{sect:madx}).

We then ran simulations by adding artificial sources onto the real images.
We created noiseless images with point sources covering a wide range of flux densities (see Table~\ref{tab:stats}).
The flux densities were chosen in order to be equally spaced on a logarithmic scale and cover a range of 
$\sim 1-50\,\sigma_{\rm tot\_Gauss}$.
About 3000 sources for each flux density were injected.
We then added the noiseless images to the real maps.

We ran MADX on these images, the real maps plus the artificial sources, using exactly the same procedure we used for the real data.
We then compared the new MADX catalogue with the catalogue obtained before the artificial sources 
were added, in order to find which artificial sources were detected. We used a radius of 12\,arcsec to look for matches 
between the sources in the new catalogue and those in the old catalogue and between those in the new catalogue and the positions of the injected sources. If there was a match 
between a source in the new catalogue and the position of an injected source but 
not with
a source in the old catalogue, it was clear that we had simply found the injected source, 
allowing us to compare the measured flux density and 
position with the true (injected) flux density and position. An ambiguity occurred if there was a 
source in the new catalogue that matched both the position of  
an injected source and a source in the old catalogue. In this case, 
we made a decision by comparing the injected flux density with the flux density
of the matched source
in the old catalogue. If the injected flux density was larger than the flux density of the source 
in the old catalogue, we concluded that the source in the new catalogue
was a match for the injected source. Otherwise, we concluded that we had not detected the injected source. 
\citet{hogg01} used the same criterion in interpreting his simulations of the submillimetre sky.

\subsection{The Conditional Probability Distributions}\label{sect:simresults}

\begin{figure*}
\center
\includegraphics[trim=1cm 5mm 9mm 3mm,clip,width=65mm,angle=0]{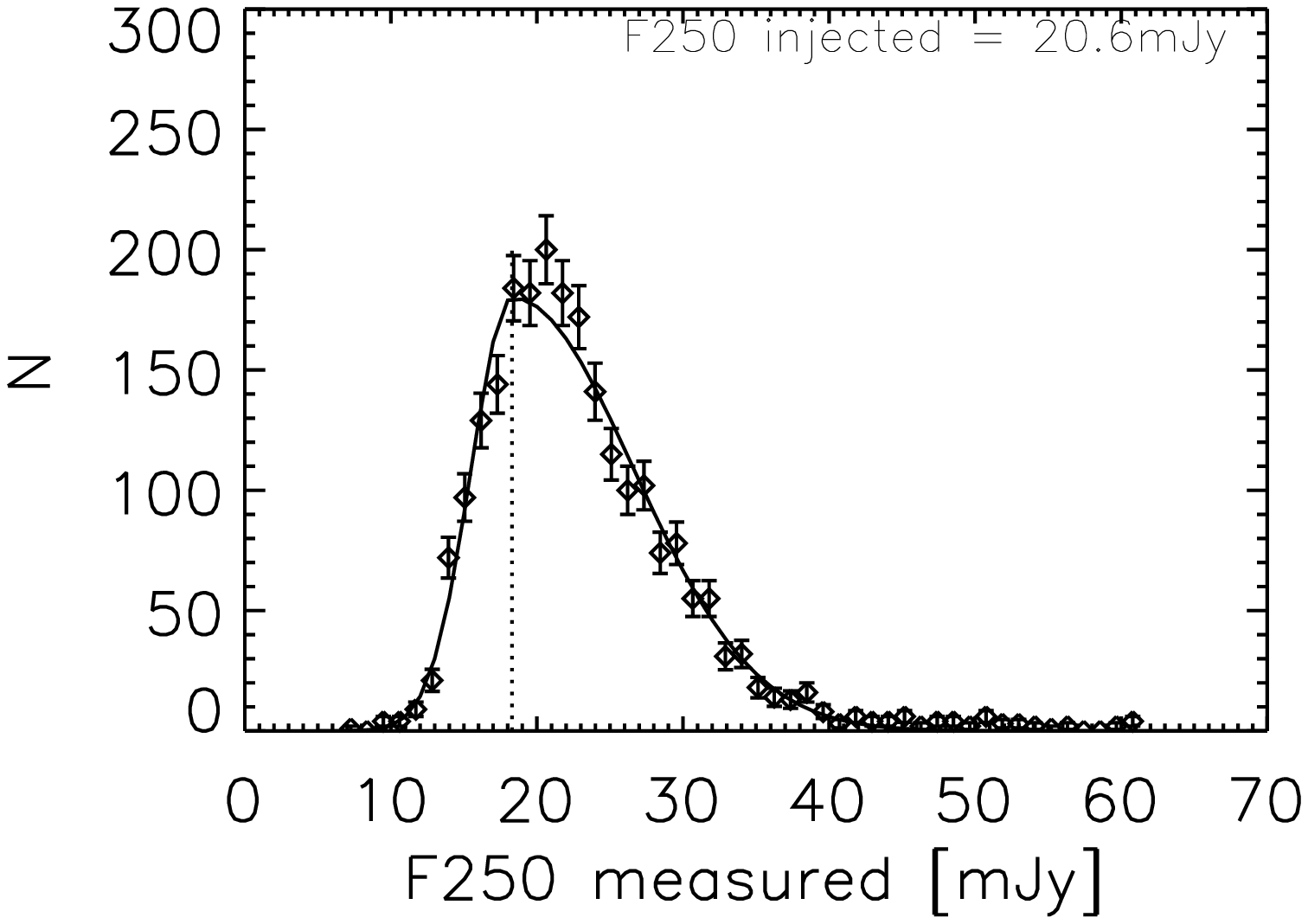}
\includegraphics[trim=42mm 5mm 7mm 3mm,clip,width=52.7mm,angle=0]{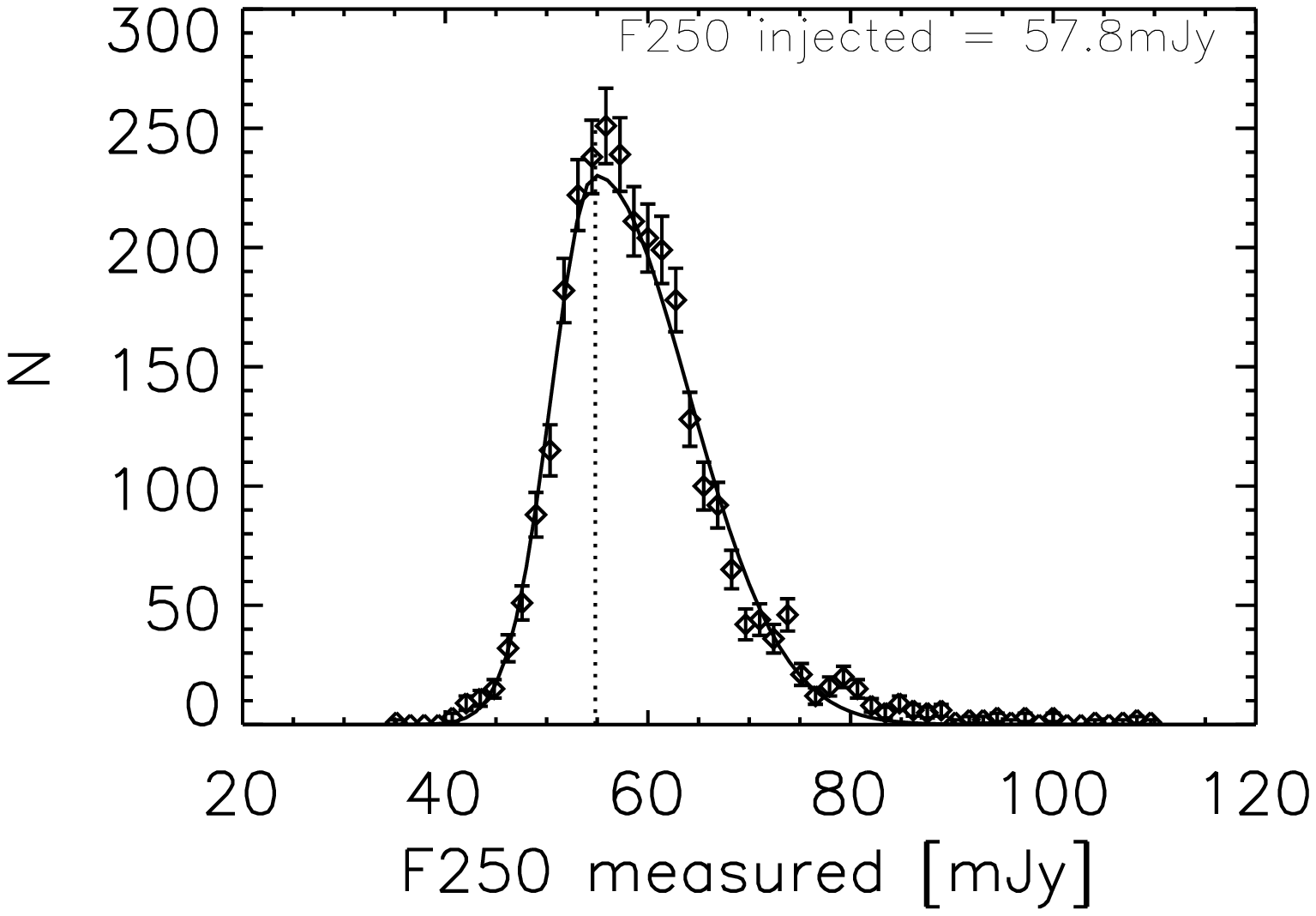}
\includegraphics[trim=42mm 5mm 0mm 3mm,clip,width=55.6mm,angle=0]{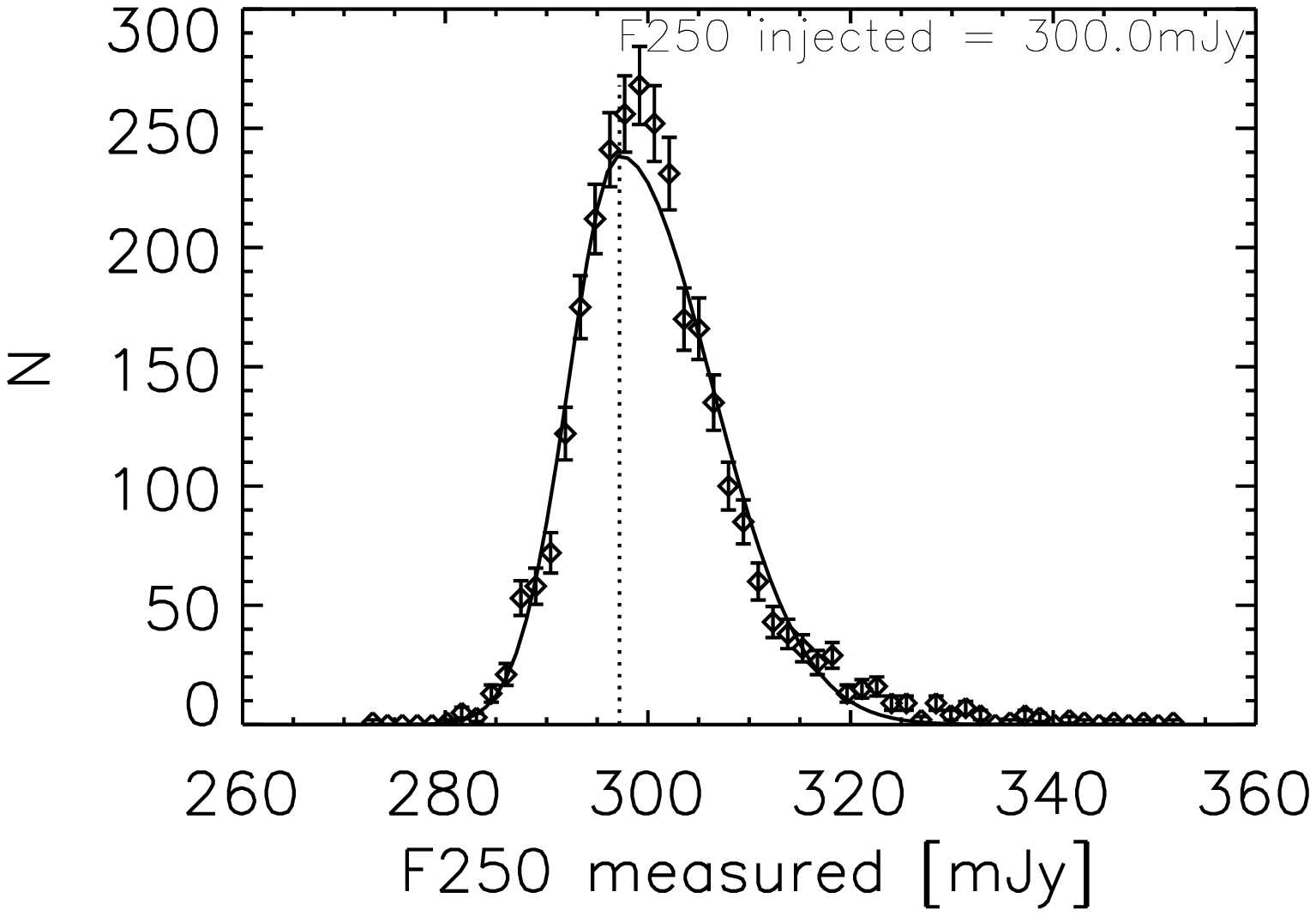}
\caption{Histograms of the measured 250\,$\mu$m flux density for 
the In-Out simulations in which we add artificial sources onto the real images. Simulations are
made for 20 different bins of flux density; here only 3 representative bins are presented.
Each panel shows the distribution of measured fluxes for ~3000 artificial sources injected with 
the flux density shown in the panel. The distributions follow the pixel distribution of the map, with a suppression
for faint fluxes due to the $2.5\sigma$ limit in the source detection. The continuous line is the best-fitting double-Gaussian 
distribution and the vertical dotted line is the fitted value of $\hat{F}$, described in \S~\ref{sect:simresults}. 
The basic statistics of the histograms and the parameters of 
the double-Gaussian are given in Table~\ref{tab:stats}.}
\label{fig:scatterhist}
\end{figure*}

Figure~\ref{fig:scatterhist} shows histograms for the recovered
injected sources of the measured flux density, $F_{\rm m}$, for three injected 
flux densities, $F_{\rm t}$, at 250\,$\mu$m. About three 
thousand sources were injected for each true flux density,
although the number of sources in each histogram is less than 3000, because not all 
sources were recovered.

Each histogram provides an estimate of the conditional probability function, 
$P(F_{\rm m}|F_{\rm t})$, at a particular
value of $F_{\rm t}$.
However, the histograms are quite noisy and to
provide a less noisy estimate of this function, we
have fitted an analytic function 
to the histograms.
The distributions are asymmetric and they follow the shape of the pixels 
distribution of the map with a suppression
for faint fluxes due to the $2.5\sigma$ limit in the source detection: we found that a good fit is obtained 
by fitting the following function, which represents the two sides of the histogram by 
Gaussians with different standard deviations, $\sigma_1$ and $\sigma_2$, 
centred in $\hat{F}$ and normalised to $N_0$:

\begin{align}\label{eq:2Gauss}
N = \frac{N_0}{2\sqrt{2\pi}} \left( \frac{1}{\sigma_1} + \frac{1}{\sigma_2} \right) e^{-\frac{(F_{\rm m}-\hat{F})^2}{2\sigma_1^2}}\ \ F_{\rm m}<\hat{F} \nonumber\\
\\
N =  \frac{N_0}{2\sqrt{2\pi}} \left( \frac{1}{\sigma_1} + \frac{1}{\sigma_2} \right) e^{-\frac{(F_{\rm m}-\hat{F})^2}{2\sigma_2^2}}\ \ F_{\rm m}>\hat{F}.\nonumber
\end{align}

\noindent Table~\ref{tab:stats} lists the parameters of the distribution 
for each value of $F_{\rm t}$.
It also lists
$N_{\rm d}/N_{\rm i}$, the ratio of
the number of recovered sources to the number of injected sources.

\subsection{The Errors in the Flux Densities}\label{sect:fluxunc}

In \S~\ref{sect:conf_noise} we concluded that a good estimate of the error in the flux density for a source
is more likely to come from the variance of the image than from fitting the negative part
of the histogram of pixel values. Following Eq.~\ref{eq:sigma_tot_var}, we now write the variance 
for a source of flux density, $F_{\rm s}$, as

\begin{equation}\label{eq:sigma_tot_var2}
\sigma_{\rm tot\_Var} = \sqrt{\frac{1}{N_{\rm pix}} \sum_i^{N_{\rm pix}} (F_i - <F_{\rm map}>) ^2}\ \ \ F_i < F_{\rm s}.
\end{equation}

\noindent This means that each source will have a different confusion noise, depending on its flux $F_{\rm s}$. The justification for the use of a different 
noise for each source
is that pixels brighter than the flux density of the source cannot contribute to
the calculation of the confusion for that source.
If a brighter source was also in the pixel, we would have assigned all of the flux in the pixel to that source;  
the fainter source would contribute to the confusion noise of the brighter source, 
but it would not have been detected as a source in our catalogue.
This approach has been suggested
before \citep{crawford10} and also takes into account, indirectly, the variation of confusion noise within the map: in a more confused region, the 
flux of the detected sources will probably have a larger contribution from undetected sources, thus it will be larger and, according with 
Eq.~\ref{eq:sigma_tot_var2}, also their confusion noise will be larger. On different premises, also 
\citet{leiton15} have recently
calculated a ``customised confusion error'' for the sources
in the GOODS-\textit{Herschel} survey, using source density arguments instead of 
the flux density argument used here.

We now test if Eq.~\ref{eq:sigma_tot_var2} is a good estimate of the uncertainty in the flux density using the results of the In-Out simulations.
These simulations are as close to
reality as we can make them,
since we are injecting artificial individual sources on to the
real images, running our detection software on the images, and comparing the measured flux densities
with the injected flux densities.
As our estimate of the error on the flux density from the simulations, we combine the parameters
from the double-Gaussian fits to give

\begin{equation}\label{eq:noisesim}
\sigma_{\rm sim}=\sqrt{\frac{\sigma_1^2+\sigma_2^2}{2}}.
\end{equation}

\begin{figure}
\center
\includegraphics[trim=2.5cm 5mm 5mm 0cm,clip,width=84mm]{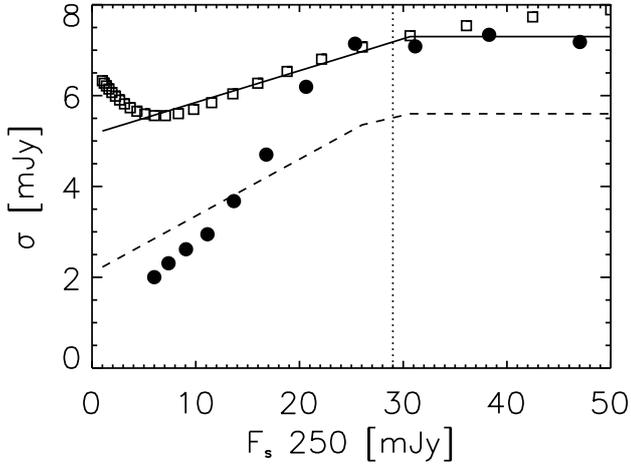}
\caption{Plot of the relationship between noise at 250\,$\mu$m, $\sigma$, and the 250\,$\mu$m
flux density. 
The full circles show $\sigma_{\rm sim}$ the noise value calculated from
the simulations using Eq.~\ref{eq:noisesim}. The squares show $\sigma_{\rm tot\_Var}$
measured from the images using Eq.~\ref{eq:sigma_tot_var2}. The upturn at low flux densities
is due to the fact that $<F_{\rm map}>$ used in Eq.~\ref{eq:sigma_tot_var2} is calculated on the whole
map so, for faint $F_{s}$, it is larger than the flux limit used to calculate $\sigma_{\rm tot\_Var}$.
The continuous line shows the assumption we have made
about the relationship between 
flux density error and flux density to derive the flux density errors in the catalogue. The dashed
line (Eq.~\ref{eq:confnoise_cat})
shows the confusion noise that is predicted from this relationship, after 
using Eq.~\ref{eq:confnoise} to
remove the contribution of instrumental noise.
The vertical dotted line shows the $4\sigma$ catalogue limit at 250\,\micron.}  
\label{fig:noiseforcat}
\end{figure}

The circle points in Figure~\ref{fig:noiseforcat} show this quantity plotted against flux density. The variation
of $\sigma_{\rm sim}$ 
with flux density does in general 
support the idea that the error in the flux density of a source does depend on the
flux density of the source. The squares show $\sigma_{\rm tot\_Var}$
calculated using Eq.~\ref{eq:sigma_tot_var2}. 
The results from the In-Out simulations and from the variance measurements agree well at a flux density
of $\sim 30$\,mJy, but $\sigma_{\rm sim}$ does not increase with flux density
at brighter flux densities in the way that 
$\sigma_{\rm tot\_Var}$ does, and it falls off faster with 
decreasing flux density than $\sigma_{\rm tot\_Var}$.

The flattening of $\sigma_{\rm sim}$ at bright fluxes is probably the result of small-number 
statistics together with the very non-gaussian pixel distribution of the maps. We would expect to see an 
increase in the average $\sigma_{\rm sim}$ with flux density, because the noise measured for an injected 
source comes from all pixels fainter than itself, but not from brighter pixels (\S~\ref{sect:method}); as the upper 
threshold of pixels that contribute to the noise is increased, the non-Gaussian positive tail (Figure~\ref{fig:noisegauss}) 
increases the variance. In the limit of infinite number of injected sources we would recover the error distribution 
expected from the pixel distribution of the map, but since there are only a very small number of bright pixels in the map, 
it is very likely that our simulations do not have any sources on bright pixels, and so the estimated variance is low. 
These rare cases of very high errors lead to a high average error which is not appropriate for a ``typical'' source, so we 
chose to use the errors from the simulations, as these are more representative of the ``typical'' error for a source.

The second difference, is that the apparent noise in the simulated source fluxes falls rapidly for fluxes fainter than 
20\,mJy, and this is due to the fact that $\sigma_{\rm sim}$ is calculated from only those injected sources that are detected. 
A measured source has to be brighter than $2.5\sigma$ (\S~\ref{sect:madx}), so we recover only a narrow tail of a broad 
error distribution. The simulations show that the scatter in the flux density of a source that is just detected above the $2.5\sigma$ 
threshold is extremely low, but this does not imply that the uncertainty is small. When we modelled this effect by measuring the 
variance on an image with an upper and lower flux-density limit, we found values that followed the results of the In-Out simulations 
much more closely than when the lower limit is not used. A better estimate of the uncertainty for a faint source is simply the total 
noise as estimated from the Gaussian fit of the pixel histogram (Table~\ref{tab:noisegauss}).

Based on the comparison of the variance measurements and the results
of the In-Out simulations, we adopted the following 
approach for
estimating the error in a source's flux density.
We assume that the variance is constant at flux densities $\gtrsim 30\,$mJy and
linear below this flux density.
This relationship is shown by the solid line
in Figure~\ref{fig:noiseforcat}. The justification for the constancy above 30\,mJy is
that this is what the simulations tell us the error is. At lower flux densities,
the simulations will drastically underestimate the true flux density errors because they
only give us information about the dispersion in the flux densities
of the {\it detected} sources.
The relationship we assume below 30\,mJy
is a conservative one, since while it follows the behaviour of the simulations  at brighter flux
densities, it is systematically higher than it is at lower
flux densities, where it follows the variance measurements. 
We then use our estimate of the instrumental noise (see \S~\ref{sect:noise_from_jn}) and 
Eq.~\ref{eq:confnoise} to obtain an estimate
of the confusion noise for a source of flux density $F_{\rm s}$, which is
given by:

\begin{equation}\label{eq:confnoise_cat}
\sigma^2_{\rm conf250\_Cat} = \min(0.0049,F_{\rm s}/5.6)^2+0.00253^2
\end{equation}

\noindent in which the flux density of the source is given in Jy.

The actual uncertainty in the flux density of an individual source depends
on the source's position because the instrumental noise is
not uniform. We assigned an error to the flux density of each source
in the catalogue (see \S~\ref{sect:catalogue}) by using Eq.\ref{eq:confnoise_cat} to estimate the
confusion noise and then adding this 
in quadrature
to the instrumental noise given by the
map of the instrumental noise 
(\S~\ref{sect:noise_from_jn}). 

At 350 and 500\,$\mu$m, 
the values for the errors in the flux density estimated from
the In-Out simulations 
are fairly constant (8.0 and 8.5\,mJy at 350 and 500\,$\mu$m, respectively) for input
flux densities above 15\,mJy.
At the two longer wavelengths we are simply measuring the brightness of a source
that was detected at 250\,$\mu$m, and so
we expect that
the relationship between flux-density error and flux density should be much
weaker than at 250\,$\mu$m. For this reason we have assumed that the
contribution to the error in flux density from confusion is a constant.
Subtracting the average instrumental noise in quadrature from
the errors above (see Tables~\ref{tab:noisevar} and \ref{tab:noisegauss}), 
we obtain a confusion noise of 6.59 and 6.62\,mJy
at 350 and 500\,$\mu$m, respectively.
As at 250\,$\mu$m, we estimate the total uncertainty on the flux density of each source
in the catalogue (see \S~\ref{sect:catalogue}) by adding the confusion noise in quadrature
to the instrumental noise given by the map of the instrumental noise
(\S~\ref{sect:noise_from_jn}).

\subsection{The positional uncertainties}\label{sect:posunc}

\begin{figure*}
\center
\includegraphics[trim=1cm 5mm 9mm 3mm,clip,width=65mm,angle=0]{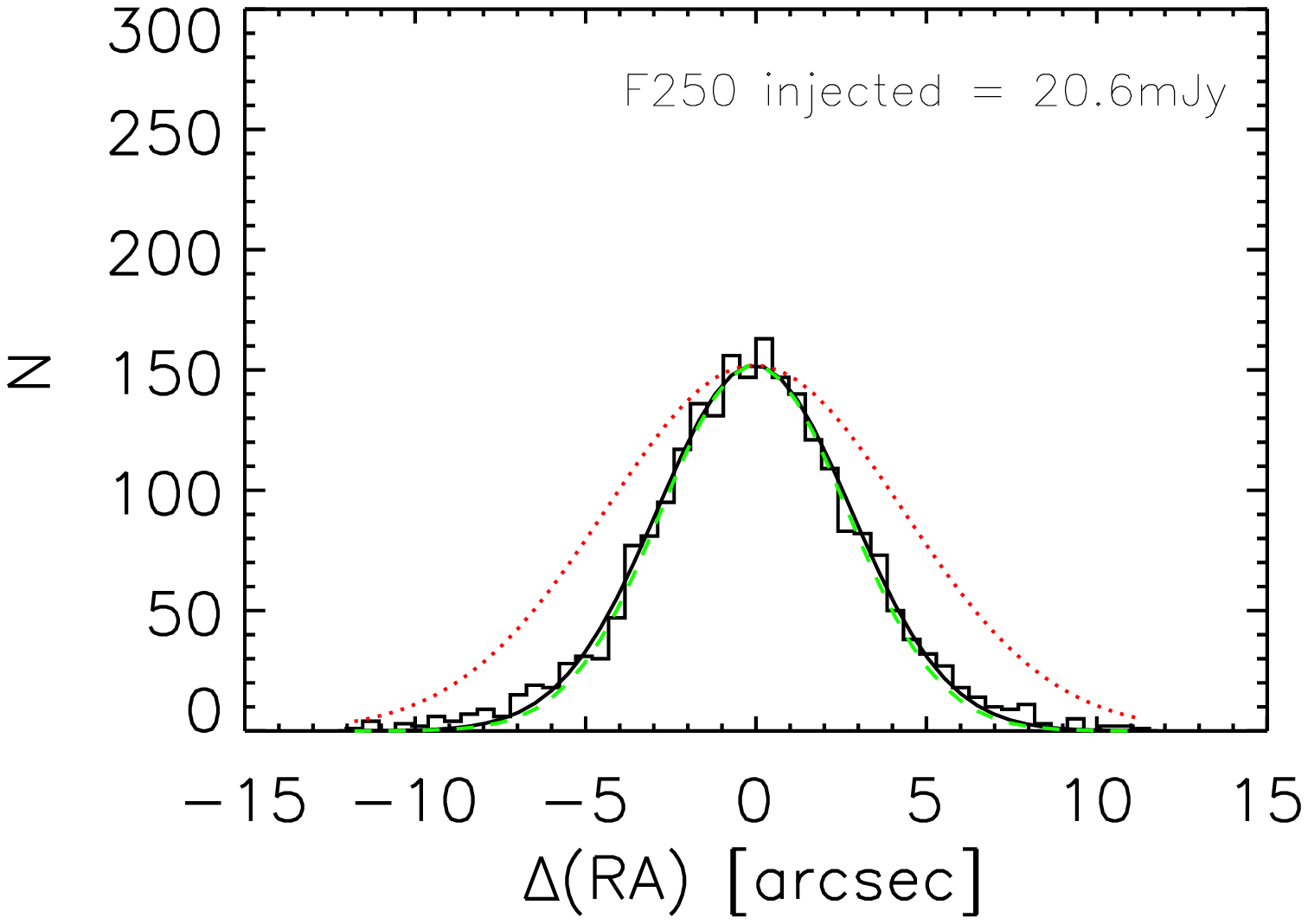}
\includegraphics[trim=42mm 5mm 7mm 3mm,clip,width=52.7mm,angle=0]{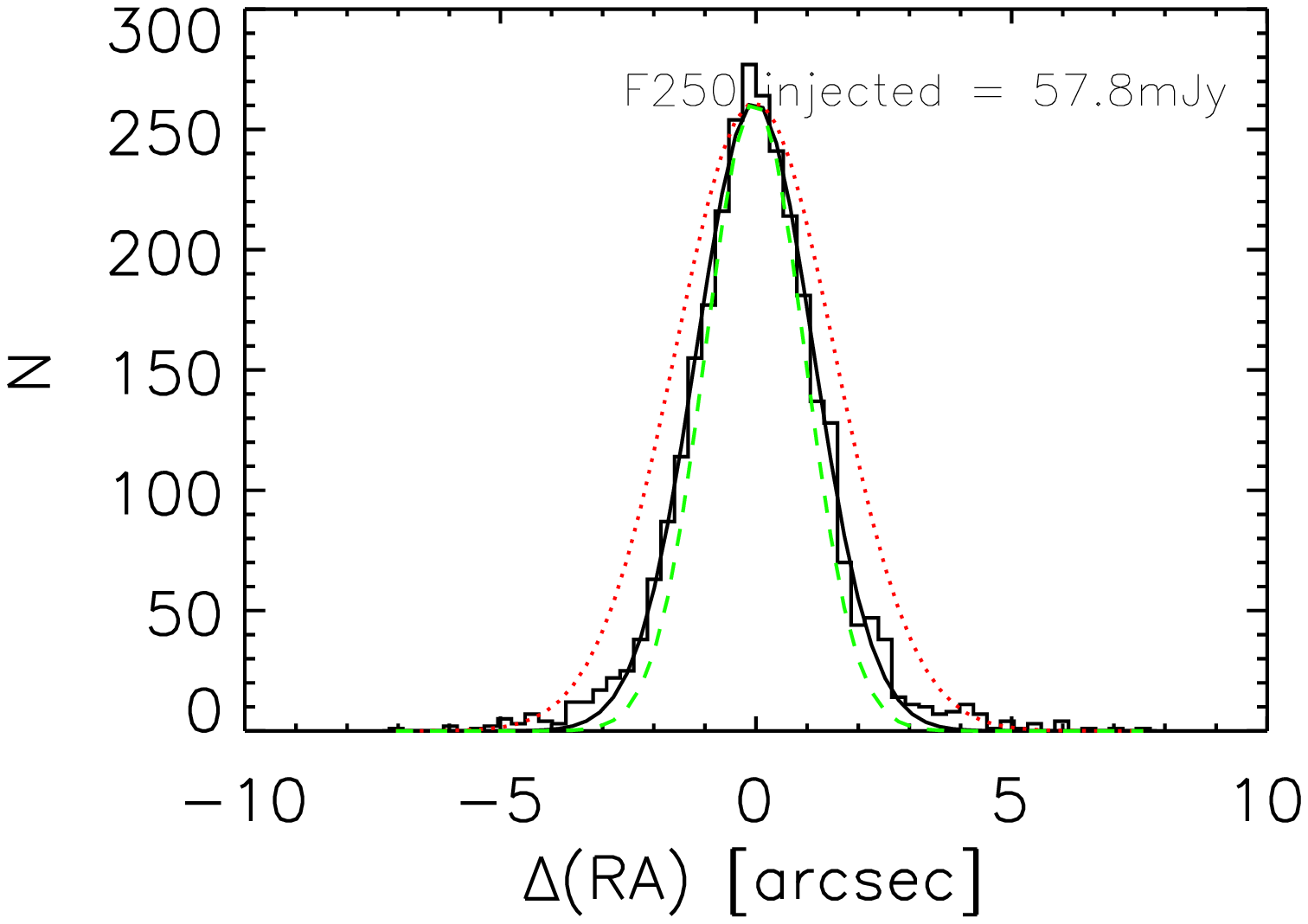}
\includegraphics[trim=42mm 5mm 0mm 3mm,clip,width=55.6mm,angle=0]{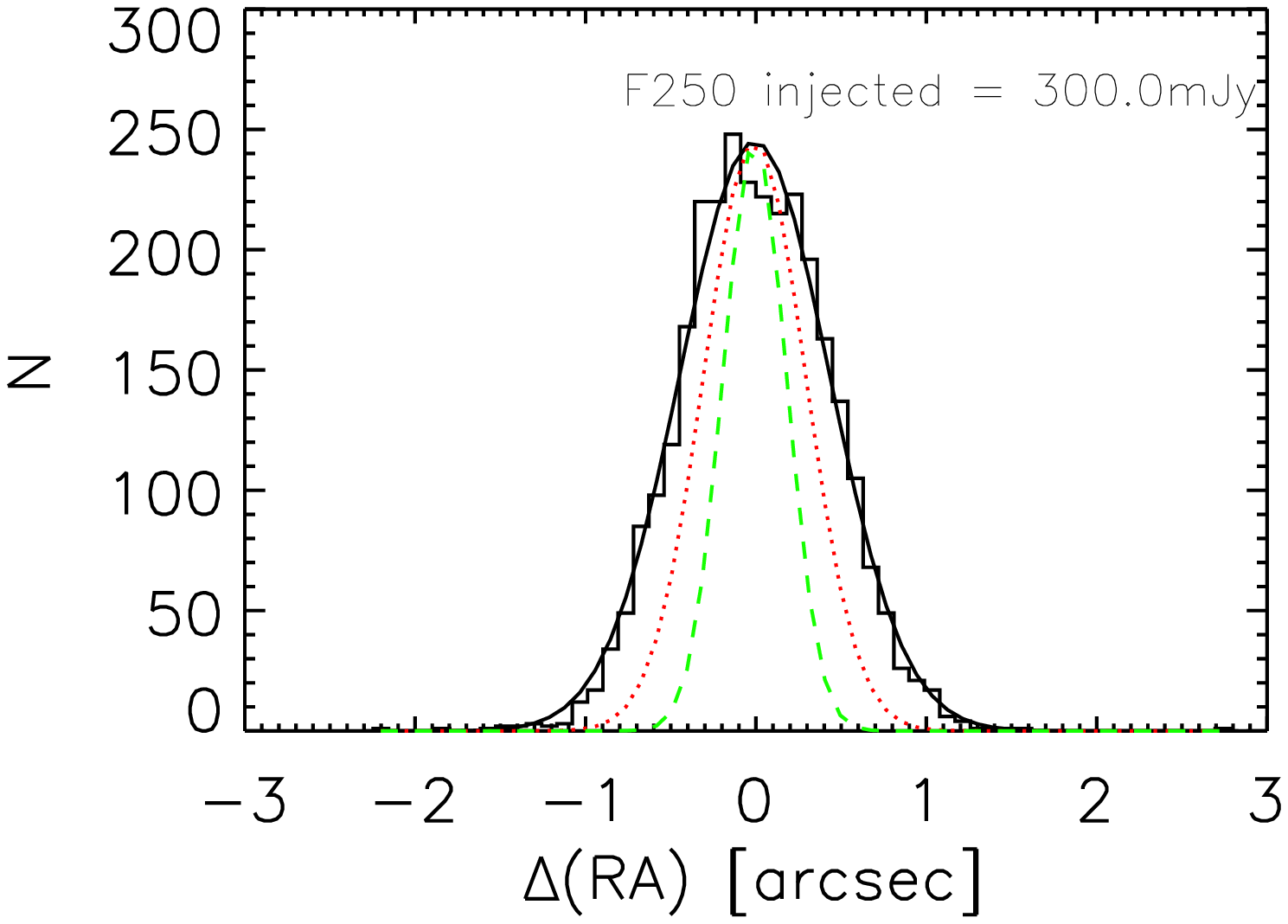}
\caption{The difference between the measured positions of the injected sources and their true positions.
The black histogram shows this distribution for RA (the distribution for Dec is very similar).
The black continuous line shows the Gaussian fit to this histogram. The
red dotted line shows the predicted distribution on the assumption that the
uncertainty in the flux densities is given by
$\sigma_{\rm tot\_Var}$ and the green dashed line shows the
predicted distribution on the assumption that the uncertainty in the flux
density is given by $\sigma_{\rm tot\_Gauss}$.}
\label{fig:posoffset1}
\end{figure*}

\begin{figure}
\center
\includegraphics[trim=2cm 5mm 1cm 3mm,clip,width=84mm,angle=0]{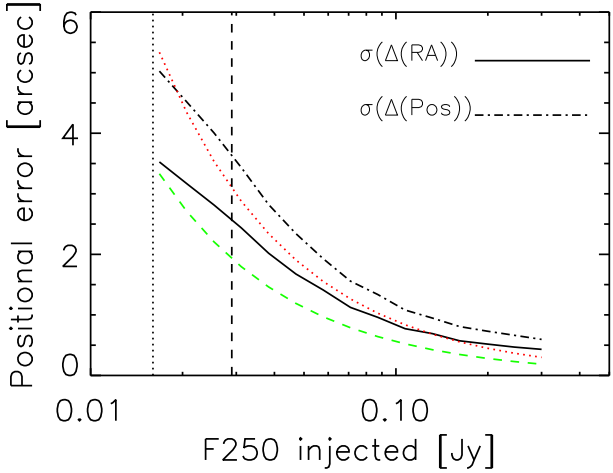}
\caption{Relationship between positional error and 250\,$\mu$m flux density derived
from the In-Out simulations. The continuous black line shows the standard deviation
in the RA positional error derived from a Gaussian fit to the histograms shown in 
Figure~\ref{fig:posoffset1}. The green dashed line show the prediction from the relationship
given in text if we assume that the error on the flux density of the source
is given by $\sigma_{\rm tot\_Gauss}$, the red dotted line the predicted relationship
if we assume that the flux-density error is given by $\sigma_{\rm tot\_Var}$.
The vertical
black dotted and black dashed lines show respectively the $2.5\sigma$ detection and $4\sigma$ catalogue limits.} 
\label{fig:posoffset2}
\end{figure}

The In-Out simulations can also be used to
investigate the accuracy of the source
positions. 
Figure~\ref{fig:posoffset1} shows histograms of the differences
between the true positions of the injected sources and the positions
measured with MADX for three different injected flux densities.
The histograms are well fit by a Gaussian.
Figure~\ref{fig:posoffset2} shows the standard deviation of the Gaussian
as a function of injected flux density. As expected, the accuracy of the
measured positions decreases with decreasing flux density.

Theoretically, we expect the differences to follow a Gaussian distribution with
the standard deviation of the Gaussian given by (0.6$\times$FWHM)/SNR, where FWHM
is
the full-width half-maximum of the PSF and SNR is the signal-to-noise
ratio \citep{ivison07}. The FWHM at 250\,$\mu$m, the wavelength
at which the positions are measured in MADX, is 18\,arcsec.
The green lines in Figures~\ref{fig:posoffset1} and \ref{fig:posoffset2} show the theoretical prediction
when $\sigma_{\rm tot\_Gauss}$, 
the noise calculated from the Gaussian fitting technique (\S~\ref{sect:conf_noise}), is used
in the model. The red lines shows the theoretical prediction
when $\sigma_{\rm tot\_Var}$, the variance in the image, is used. The actual distributions
fall between the two predictions.

In the accompanying data-release paper \citepalias{bourne16}, 
we estimate the accuracy of the positions by comparing the positions of
the H-ATLAS sources with the positions of galaxies on the SDSS.
Since this technique is likely to be even closer to ground truth than the
In-Out simulations, we prefer the estimates of positional accuracy given
in that paper, although the estimates of the
positional accuracy from the In-Out simulations are actually very similar.

\subsection{Completeness at 250\,\micron}\label{sect:compl250}

\begin{figure}
\includegraphics[trim=1.8cm 5mm 1cm 3mm,clip,width=84mm,angle=0]{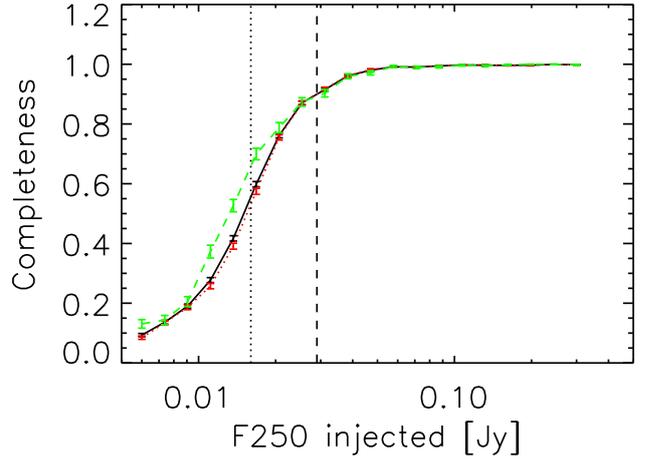}
\caption{Completeness of the survey
derived from the In-Out simulations
plotted
against {\it true} 250\,$\mu$m flux density.
The completeness values are given by $N_{\rm d}$/$N_{\rm i}$, in which $N_{\rm d}$ and $N_{\rm i}$ are the
number of detected and injected sources 
at each flux density in the In-Out simulations.
The errors
are given by
$\sqrt{N_{\rm d}\times (N_{\rm i}-N_{\rm d})/N_{\rm i}^3}$. 
The black points show the results for the survey as
a whole, the red points show the results for
the ``two-scan'' regions and the green points for the deeper ``four-scan''
regions. The values for the survey as a whole and for the ``four-scan''
regions are listed in Table~\ref{tab:stats}.
The vertical black dotted 
and black dashed lines show respectively the $2.5\sigma$ detection and $4\sigma$ catalogue limits.}
\label{fig:compl250}
\end{figure}

Figure~\ref{fig:compl250}
shows 
the ratio of the number of recovered sources to the number
of injected sources 
($N_{\rm d}/N_{\rm i}$ listed in Table~\ref{tab:stats}) plotted against
injected flux density.
This figure represents the completeness of the survey but as a function
of {\it true} flux density rather than {\it measured} flux
density.

We also carried out versions of the In-Out simulations restricted
to the small deeper regions of the maps in which the quadrants
overlap (see Figure~\ref{fig:coverage} to visualise this) and also to the part
of the map outside these regions.
Figure~\ref{fig:compl250} shows the completeness verses flux
density for these regions, made from four nine-hour datasets and
two nine-hour datasets, respectively (see \S~\ref{sect:observations}).
As one would expect, the results for the ``two-scan'' regions are almost
exactly the same as for the survey as a whole, because most of
the survey consists of these regions. The completeness for the
deeper ``four-scan'' regions is slightly higher than for the
survey as a whole, again as one would expect.
Table~\ref{tab:stats} also lists the values of the 
$N_{\rm d}/N_{\rm i}$ for these deeper regions.
Although we do not use these values in the
rest of the paper, they give enough information for anyone in the future
who is interested specifically in these regions to carry out the analysis
of flux bias and completeness that is described in \S~\ref{sect:corrections}.

The measured flux densities are likely to be systematically different from the
true flux densities as the result of flux bias. Therefore, to
derive the completeness
of the survey as a function of measured flux density, which is sometimes
what one needs from a practical point of view,
we need also to be able to estimate the flux bias of the
survey. This not possible from the In-Out simulations by themselves.
Although we can not derive the flux bias and completeness of the surveys directly
from the In-Out simulations, they do provide an important stepping stone.
The flux bias can be represented by the conditional probability
of the true flux density given the measured flux density, $P(F_{\rm t}|F_{\rm m})$.
The In-Out simulations, however, provide an estimate of the conditional probability
of the measured flux density given the true flux density, $P(F_{\rm m}|F_{\rm t})$.
To go from one to the other requires knowledge of the number counts of sources
down to a flux density well below the measured flux densities in
the H-ATLAS catalogue. Fortunately, the deeper surveys with
{\it Herschel} have provided this information. 
In \S~\ref{sect:counts250} we use the conditional probability distributions provided by
the In-Out simulations to show that the number counts of the sources
from our survey are consistent, over the range of flux density covered
by our catalogue, with the results of the deeper surveys in other parts
of the sky.
In \S~\ref{sect:boost250} we
use the results from the deeper surveys to estimate the flux bias.

\section{The Source Counts}\label{sect:counts250}

We can combine the  conditional probability distributions from the In-Out simulations
as a matrix, 
$\mathbf P$, each element of which, $P_{ij}$, is the probability that a galaxy with true 
flux $F_i$ is detected with a 
measured flux $F_j$. This matrix contains the same information as the
conditional probability distribution, and it also
contains information about the incompleteness of the survey, which is given by the recovered fraction
of sources in the In-Out simulations; the sums of the columns of the matrix 
are the recovered fractions listed in Table~\ref{tab:stats}.
 
Let us now represent the number of measured sources in each bin of flux density as a 
vector, $\bf \hat{N}_m$, and the mean number of sources predicted, from the true
source counts,
to fall in these bins as another vector, 
$\bf \hat{N}_t$. We calculate the number of sources in each bin, given that the flux densities we used in the simulations 
(see Table~\ref{tab:stats}) are the centre of the bins and that the bin limits are half way between two contiguous fluxes. 
We can then relate the two vectors by the matrix equation:

\begin{equation}\label{eq:matrixeq}
\bf \hat{N}_m = P \cdot \hat{N}_t.
\end{equation}

\noindent This equation expresses in matrix form the combined effects of 
incompleteness and Eddington Bias on the intrinsic number counts. The matrix 
$\mathbf{P}$ contains all the information we need about the effects of the noise properties of our images and of the 
algorithm we have used to find the sources.

We want to know the true source counts $\bf \hat{N}_t$, so in principle
we 
should simply invert Eq.~\ref{eq:matrixeq}. Unfortunately, 
the elements of the matrix are not perfectly known, and simply inverting the
matrix amplifies the errors in the matrix elements, producing a very uncertain and unstable
solution (this is analogous to a deconvolution, in which errors in the deconvolution
kernel lead to an increase of noise in the solution).

To solve the problem, we need to use some extra prior information about the source counts. It is not necessary to assume any strong hypothesis, we can
simply introduce what is called a {\it stabilising functional} or {\it regularising operator}. For example, one assumption that is reasonable to make is that we expect the 
source counts to have a smooth form.

A detailed account of the theory of linear regularisation and its implementation is given
in Numerical Recipes (\citealt{press82}, \S~18.5). Using the 
method of Lagrange multipliers, we determine our vector ${\bf \hat{u}} = {\bf \hat{N}_t}$ by 
minimising the quantity:

\begin{equation}\label{eq:functionals}
\mathcal{A}+\lambda \mathcal{B}.
\end{equation}

\noindent In this equation
$\mathcal{A}$ and $\mathcal{B}$ are two positive functionals 
of ${\bf \hat{u}}$ and $\lambda$ is a Lagrange multiplier.
The first functional comes from minimizing the Chi-squared difference between the
observed source counts and the model, and is given by:

\begin{equation}
\mathcal{A} = | \mathbf{A\cdot  \hat{u}-b|^2}.
\end{equation}

\noindent The elements of the matrix $\bf A$ are 
$A_{ij} = P_{ij}/\sigma_i$, with $\sigma_i$ being the Poisson error on the number of
sources with measured flux densities $F_i$.
The vector $\bf b$ contains the observational data and its elements are
$b_i = N_{m,i}/\sigma_i$, in which
$N_{m,i}$ is the number of sources with measured flux densities $F_{i}$.

The second functional in Eq.~\ref{eq:functionals}
 represents our prior assumptions
about the source counts
and is given by 

\begin{equation}
\mathcal{B} = {\bf \hat{u} \cdot H \cdot \hat{u} }.
\end{equation}

\noindent The matrix $\bf H$ is based on our assumptions about the source
counts.
In our analysis we assume that the logarithm of the source counts are approximately linear,
and we therefore minimise the second derivatives of the counts.
With this assumption, ${\bf H = B^T \cdot B}$, in which $\bf B$ is given by
(see section 18.5 of Numerical Recipes):

\begin{equation}\label{eq:matrixb}
\mathbf{B} =
 \begin{pmatrix}
  -1         & 2  & -1 &  0        &\cdots & 0 \\
   0          & -1&  2 & -1        & \cdots & 0 \\
   \vdots &     &      & \ddots &            & \vdots  \\
   0          & \cdots & 0 & -1  & 2          & -1
 \end{pmatrix}.
 \end{equation}

The value of $\lambda$ will give different 
weights to the two parts of the minimisation. When 
$\lambda=\mathbf{{\mathrm Tr}(A^T\cdot A)/{\mathrm Tr}(H)}$, the two functionals have comparable weights. 

The minimum value of $\mathcal{A}+\lambda \mathcal{B}$ occurs when

 \begin{equation}
 \mathbf{\hat{u}}=\mathbf{(A^T\cdot A+\lambda H)^{-1}(A^T\cdot b)}.
 \end{equation}
 
The standard uncertainties on the elements of $\mathbf{\hat{u}}$ are given by the diagonal terms of the covariance matrix:

 \begin{equation}\label{eq:cov}
 \mathbf{C=(A^T\cdot A+\lambda H)^{-1}}.
 \end{equation}

\noindent The covariance between the elements of the solution strongly depends on the choice of $\lambda$: the larger the weight given to the prior, the larger will be the
covariance.

We do not want our results being too dependant on the choice of the prior and we also 
want to reduce the covariance between the elements of $\mathbf{\hat{u}}$. For these 
reasons, we have chosen the minimum value of $\lambda$ which gives a solution in which all the
elements of $\mathbf{\hat{u}}$ are positive, and thus a solution that is
physically reasonable. 
We never need the weight of the prior to be larger than 1/100 of the 
weight of the data to find a solution that satisfies this condition.

The matrix $\bf B$ in Eq.~\ref{eq:matrixb} is only one
of the possible priors that can be used. We also tried a different prior (minimising the first 
derivative of the counts) and the results did not change significantly, due to the small weight we give to the prior.

This approach to retrieve the number counts shows some similarities with Reduction C used for the $850$\,\micron\ survey 
SHADES (see \S~5.2 from \citealt{coppin06}), which in turn was built on the methods used by \citet{borys03} and \citet{laurent05} 
for the analyses of SCUBA and Bolocam data, respectively. The new element we have introduced in this study is the
regularising operator. The use of a regularising operator is quite common in imagine reconstruction, for example in the 
study of lensed objects (see for example \citealt{warren03}), but, to our knowledge, this is the first time it has been used to 
derive the source number counts of a galaxy population. 
We postpone a more exhaustive analysis of this technique to future papers.

\begin{figure*}
\includegraphics[width=150mm,keepaspectratio]{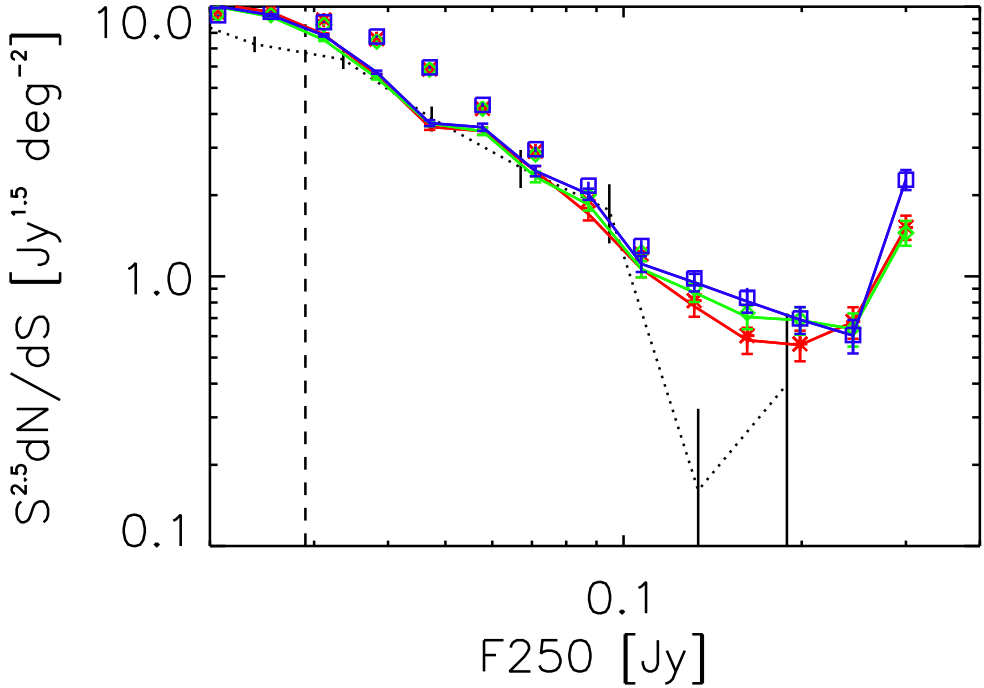}
\caption{The number counts of sources at 250\,$\mu$m. The red, green and blue colours represent the GAMA9, GAMA12 and GAMA15 fields, respectively. 
The symbols show the measured counts, $\hat{\rm N}_{\rm m}$, while the continuous 
lines show the true numbers counts, $\hat{\rm N}_{\rm t}$, estimated from
our inversion technique. Uncertainties do not include the correlation 
between flux bins. 
The dotted lines show the source counts from the deeper HerMES survey \citep{bethermin12}. 
They underestimate the counts at the bright end because the 
HerMES area is smaller than the H-ATLAS one, so they do not have adequate statistics above $\sim 100\,$mJy. 
The vertical dashed line shows the 4$\sigma$ limit of the H-ATLAS catalogue (see \S~\ref{sect:catalogue}).
}
\label{fig:counts250}
\end{figure*}

The source number counts obtained from the matrix inversion are shown in 
Figure~\ref{fig:counts250} and Table~\ref{tab:counts250}.
The coloured symbols in the plots show the observed number counts, $\bf \hat{N}_m$, 
in each of the three GAMA fields, while the coloured lines show the true number counts
for each field,
$\bf \hat{N}_t$, estimated using 
our inversion technique. Points are slightly correlated
because of the covariance between flux bins.

The uncertainies are derived by Monte Carlo simulations
of synthetic data sets: we generate 10000 times a vector of ``measured'' number counts, $\bf \hat{N}_m$,
assuming a Poisson uncertainty on the real $\bf \hat{N}_m$, and we apply the inversion procedure each time,
obtaining 10000 $\bf \hat{N}_t$. The scatters in the results are the uncertainties shown in Figure~\ref{fig:counts250} 
and listed Table~\ref{tab:counts250} and may be up to a factor 2 larger than the nominal uncertainties calculated
from the covariance matrix (see Eq.~\ref{eq:cov}). The quoted uncertainties do not include the correlation between
flux bins. Because our counts are derived from simulations only and do not assume any prior on the existing galaxy population,
we can only trust them for fluxes brighter than $\sim 20$\,mJy. At fainter fluxes, the conditional probability
distribution $P(F_{\rm m}|F_{\rm t})$ is too uncertain.

The number counts for the three GAMA fields
are quite similar except at the brightest flux densities, where the difference
is due to cosmic variance.
The observed counts differ quite markedly from the source counts measured by the HerMES team
from their much deeper observations, both the source counts inferred
from a $P(D)$ analysis \citep{glenn10} and the source counts
measured by a stacking analysis \citep{bethermin12}.
However, after we have applied our
inversion technique, the corrected source counts show consistency 
with the HerMES measurements, except at the faintest flux densities, well below the flux limit
of our catalogue. 

In \S~\ref{sect:compl2bands}, we show that it also possible to apply this method, with some
modifications, to the source counts at 350 and 500\,$\mu$m.

\begin{table*}
\centering
\caption{Number counts at 250\,\micron, $\hat{\rm N}_{\rm t}$, estimated from
the inversion technique discussed in \S~\ref{sect:counts250}. Uncertainties do not include the correlation between flux bins.
}
\begin{minipage}{140mm}
\centering
\label{tab:counts250}
\begin{tabular}{@{}rrcrrcrrcr@{}}
\hline
F250        & \multicolumn{9}{c}{Number counts (${\rm d}N/{\rm d}S\times S^{2.5}$)}  \\
 {[}mJy]     & \multicolumn{9}{c}{[Jy$^{1.5}$deg$^{-2}$]}    \\
 		& \multicolumn{3}{c}{GAMA9} &  \multicolumn{3}{c}{GAMA12} &  \multicolumn{3}{c}{GAMA15}                         \\
\hline
25.4&9.557&$\pm$&0.084&9.181&$\pm$&0.083&9.305&$\pm$&0.086\\
31.2&7.891&$\pm$&0.073&7.531&$\pm$&0.072&7.824&$\pm$&0.073\\
38.3&5.542&$\pm$&0.075&5.443&$\pm$&0.075&5.716&$\pm$&0.076\\
47.0&3.582&$\pm$&0.089&3.692&$\pm$&0.091&3.701&$\pm$&0.094\\
57.8&3.457&$\pm$&0.107&3.450&$\pm$&0.109&3.573&$\pm$&0.110\\
71.0&2.457&$\pm$&0.106&2.337&$\pm$&0.105&2.458&$\pm$&0.108\\
87.2&1.703&$\pm$&0.091&1.845&$\pm$&0.093&2.018&$\pm$&0.095\\
107.2&1.067&$\pm$&0.074&1.065&$\pm$&0.074&1.112&$\pm$&0.076\\
131.7&0.774&$\pm$&0.066&0.872&$\pm$&0.069&0.950&$\pm$&0.072\\
161.8&0.580&$\pm$&0.064&0.707&$\pm$&0.070&0.807&$\pm$&0.074\\
198.7&0.557&$\pm$&0.071&0.687&$\pm$&0.079&0.691&$\pm$&0.078\\
244.2&0.677&$\pm$&0.091&0.639&$\pm$&0.088&0.603&$\pm$&0.084\\
300.0&1.522&$\pm$&0.157&1.452&$\pm$&0.156&2.284&$\pm$&0.190\\
\hline
\end{tabular}
\end{minipage}
\end{table*}

\section{Corrections for Flux Bias and Completeness}\label{sect:corrections}

By injecting artificial sources
onto the images with a low enough surface density that they do not affect the
statistical properties of the images and by using the real
data-reduction pipeline, the In-Out simulations (\S~\ref{sect:simulations}) come as close as possible to
the ground truth of how real sources on the sky are turned into sources in the H-ATLAS
catalogue. In \S~\ref{sect:counts250} we showed that this information can be used directly
to correct the 250\,$\mu$m source counts
for Eddington bias. However, we also need to know the
completeness of the
survey and the flux bias correction for individual sources at all three wavelengths.
The flux bias problem is an
inverse problem, requiring
approximations and 
prior assumptions about either
the submillimetre sky (e.g. \citealt{rigby11}) or about the submillimetre source
counts \citep{crawford10}. 
This section describes our best current attempt to estimate
the flux bias, but we
recognise that it may be possible to improve
our analysis in the future, if only because
of improved knowledge of the submillimetre sky. We recommend that anyone
interested in improving our estimates
should start from
the ground truth provided by the In-Out simulations.

\subsection{Flux Bias and Completeness at 250\,$\mu$m}\label{sect:boost250}

There are analytic methods to correct for flux bias
(e.g. \citealt{hogg98,crawford10}) and also ones based
on simulations (e.g. \citealt{coppin05}). 
In either case, it is necessary to make assumptions about the underlying source counts down
to flux densities well below the detection limit of the catalogue
because it is these faint sources which produce the majority of the
confusion noise. In the previous section, we showed that the 250\,$\mu$m source
counts in the H-ATLAS fields, after a correction for Eddington bias, are in
good agreement above the H-ATLAS detection limit with the much deeper source
counts of \citet{glenn10} and \citet{bethermin12}.
In this section, we estimate the flux bias at 250\,$\mu$m, using these deeper
source counts to provide a prior probability distribution for the flux density of a source.
We emphasise that we only use these deeper source counts as a {\it prior} probability distribution,
and the method does not require the assumption that the source counts
in the H-ATLAS fields below the catalogue limit are exactly the same as
those measured in the deeper surveys.

We have used the method of  \citet{crawford10}.
In this method, the 
measured flux density in a pixel is $F_{\rm m}$ and the true
flux density of the brightest individual source in that pixel
is $F_{\rm t}$. Given a measured flux density in a pixel, the probability
of the true flux density of the brightest source in that pixel is then:

\begin{equation}
P(F_{\rm t}| F_{\rm m}) \propto P(F_{\rm m}| F_{\rm t})P(F_{\rm t})
\end{equation}  

\noindent in which $P(F_{\rm m}| F_{\rm t})$ is the likelihood of 
measuring a flux density equal to $F_{\rm m}$ in a pixel given that the brightest source 
in that pixel has flux density  $F_{\rm t}$, and $P(F_{\rm t})$ is the 
prior probability that the brightest source in that pixel has flux density $F_{\rm t}$.

\begin{figure}
\centering
\includegraphics[trim=1.5cm 0cm 1cm 0cm,clip,width=85mm]{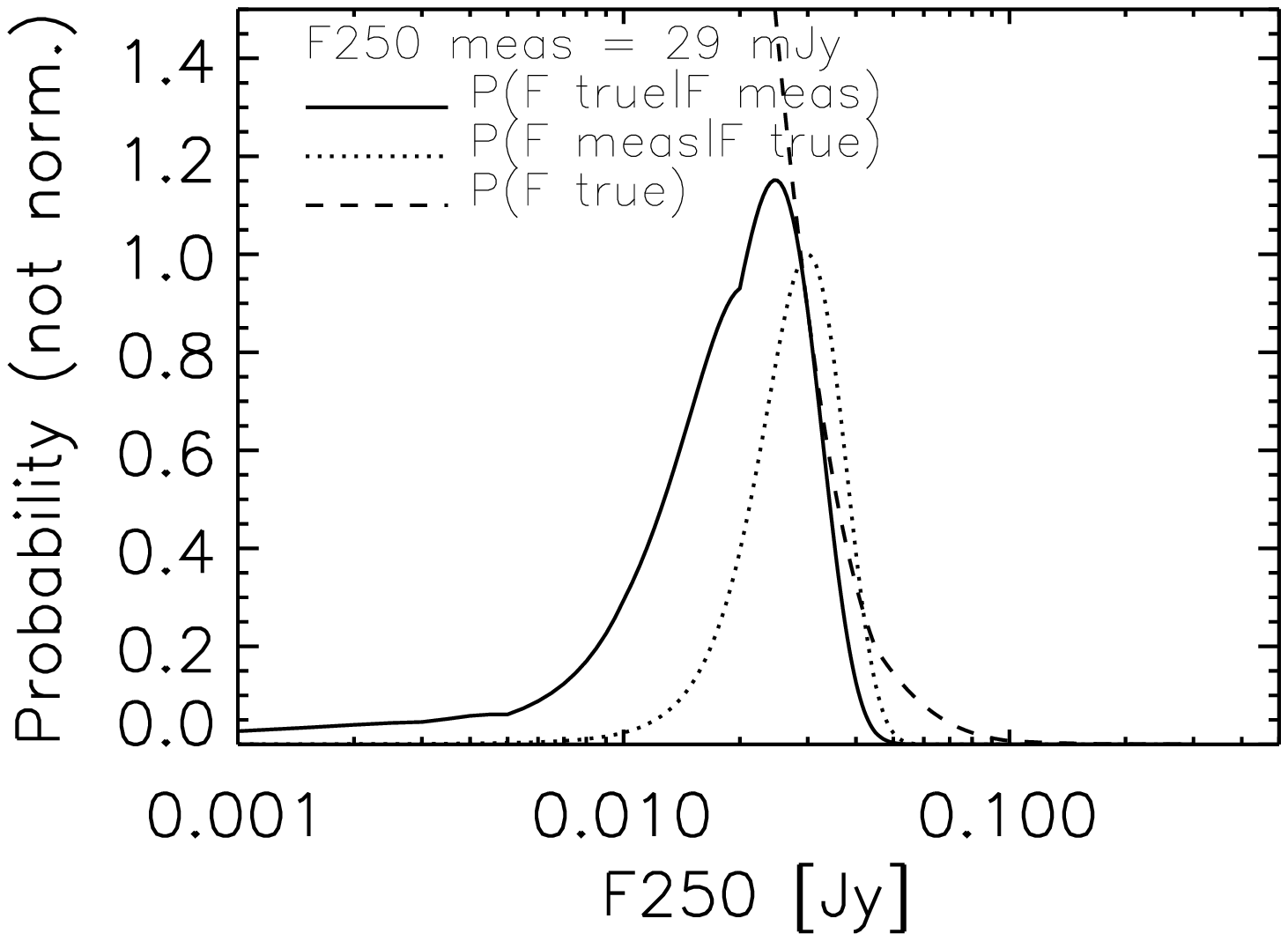}
\includegraphics[trim=1.5cm 0cm 1cm 0cm,clip,width=85mm]{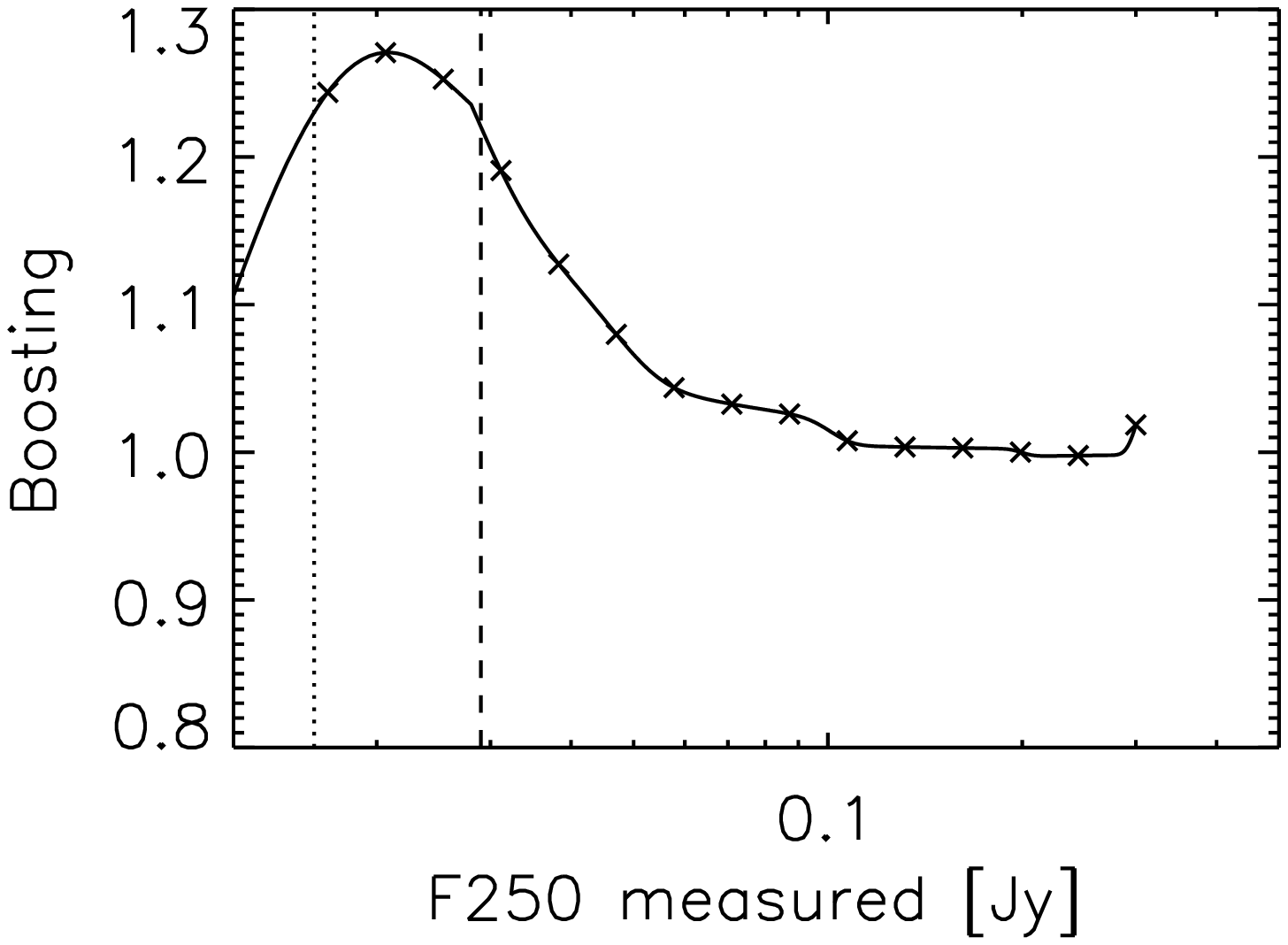}
\caption{The results of our analysis, using 
the method of  \citet{crawford10}, of the
flux bias at 250\,$\mu$m. In this analysis, 
the true flux density of the brightest source in a pixel is
$F_{\rm t}$ and the measured flux density in the pixel is $F_{\rm m}$.
\textit{Top}: probability distributions for a pixel with
$F_{\rm m} = 29$\,mJy: $P(F_{\rm t} | F_{\rm m})$,  
$P(F_{\rm m} | F_{\rm t})$ and $P(F_{\rm t})$ (see text for
more details).
Note that the probabilities in this figure have not been normalised and have been scaled
so that the shapes of the three distributions can be easily compared.
\textit{Bottom}: 
Flux bias plotted against
measured 250\,$\mu$m flux density. 
We have calculated the flux bias in each bin of measured flux density by calculating the
mean value of 
$F_{\rm m}/F_{\rm t}$ for
the sources in this bin. This relationship is
listed in Table~\ref{tab:boost}.
The 
vertical black dotted and black dashed lines show respectively the 2.5$\sigma$ detection limit (\S~\ref{sect:madx})
and the
$4\sigma$ catalogue limit (see \S~\ref{sect:catalogue}).}
\label{fig:boost250}
\end{figure}

\begin{table}
\centering
\caption{Flux Biases. These numbers are the ratio between the measured and the true flux density of a source ($F_{\rm m}/F_{\rm t}$).
In order to derive the corrected flux densities, the flux values reported in the released catalogues need to be divided by the numbers in this table.}
\begin{minipage}{80mm}
\centering
\label{tab:boost}
\begin{tabular}{@{}cccc@{}}
\hline
$F_{\rm m}$ & 250\,$\mu$m & 350\,$\mu$m & 500\,$\mu$m  \\
    {[}mJy]       &    & &                          \\
\hline
   6.0    &    -    &  1.11 & 1.12       \\
   7.4    &    -    &  1.10 & 1.10  	\\
   9.1    &    -     & 1.09 & 1.09 	\\
 11.1    &    -    & 1.09 & 1.09  	\\
 13.7    &    -     & 1.09 & 1.09  	\\
 16.8    &1.24  & 1.09 & 1.08 	\\
 20.6    &1.27  & 1.09 & 1.08  	\\
 25.4    &1.25  & 1.08 & 1.07  	\\
 31.2    &1.19  & 1.07 & 1.06  	\\
 38.3    &1.13  & 1.05 & 1.04  	\\
 47.0    &1.08  & 1.04 & 1.03        \\
 57.8    &1.04  & 1.02 & 1.02  	\\
 71.0    &1.03  & 1.01 & 1.01  	\\
 87.2    &1.03  & 1.01 & 1.01  	\\
107.2   &1.01  & 1.00  & 1.01	\\
131.7   &1.00  & 1.01  & 1.00  	\\
161.8   &1.00  & 1.01  & 1.00 	\\
198.7   &1.00  & 1.01  & 1.00 	\\
244.2   &1.00  & 1.00 & 1.00  	\\
300.0   &1.02  & 1.00 & 1.00 	\\
\hline
\end{tabular}
\end{minipage}
\end{table}

The prior probability, $P(F_{\rm t})$, is the
differential source counts with an exponential suppression at low flux density
(see \S~2.1 of  \citet{crawford10} for more details). 
In calculating the prior probability distribution, we use the differential
250\,$\mu$m source counts
given in \citet{bethermin12}.
The likelihood, $P(F_{\rm m}| F_{\rm t})$, can be written using a 
Gaussian likelihood approximation  \citep{crawford10}:

\begin{equation}
P(F_{\rm m}| F_{\rm t}) = \frac{1}{\sqrt{2\pi 
\sigma^2}}e^{-(F_{\rm m}-F_{\rm t}-\overline{F})^2/2\sigma^2}
\end{equation} 

\noindent in which $\overline{F}$ is the mean of the image, which 
in general is not precisely zero (see \S~\ref{sect:dataproc}).
The value of
$\sigma$ we use for each source is the uncertainty in the flux density of
the source, derived individually for each source using the
method of
(\S~\ref{sect:fluxunc}). In this way,
we produce an individual estimate
of the flux bias for each source.

Figure~\ref{fig:boost250} (\textit{top}) shows the probability 
distributions $P(F_{\rm t}|F_{\rm m})$,  $P(F_{\rm m}|F_{\rm t})$ and $P(F_{\rm t})$ for a 
pixel with a measured flux density equal to the catalogue limit
of 29\,mJy.
We estimate the true flux density
of the brightest source in the pixel from the maximum in the posteriori probability
distribution $P(F_{\rm t} | F_{\rm m})$. The bottom panel in Figure~\ref{fig:boost250}
show the mean flux bias factor ($F_{\rm m} / F_{\rm t}$)  plotted against
the measured flux density $F_{\rm m}$.
We have calculated the mean flux bias for each bin
of $F_{\rm m}$ by estimating $F_{\rm t}$ for each pixel that falls in this
bin and then calculating the mean value of $F_{\rm m} / F_{\rm t}$ for all the pixels
in the bin.
This relationship is listed in Table~\ref{tab:boost}.

\begin{figure}
\centering
\includegraphics[trim=1cm 0cm 1cm 1cm,clip,width=84mm,angle=0]{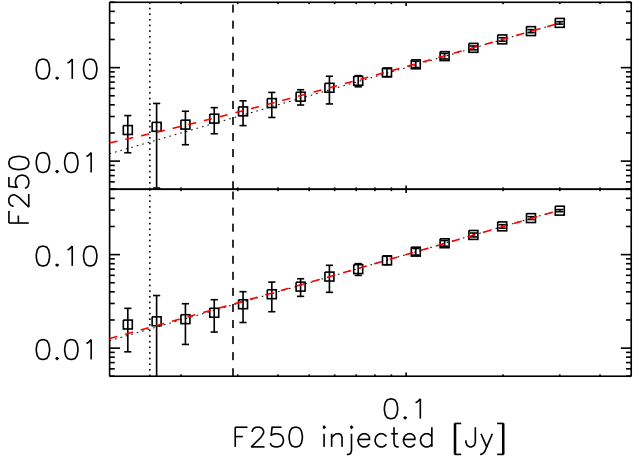}
\caption{\textit{Top}: The mean value and standard deviation of the measured flux density plotted
against the
injected flux density for
the artificial sources in the In-Out simulations.
The dotted line shows where the two flux densities are equal
and the
red line
is a linear relationship fitted
to the points.
The vertical dotted and dashed lines show, respectively, the
2.5$\sigma$ detection and 4$\sigma$ catalogue limits.
\textit{Bottom}: The same as in the top panel except that the measured flux density of each
source has been corrected for flux bias using the relationship shown in Figure~\ref{fig:boost250}
({\it bottom panel}) and listed in Table~\ref{tab:boost}.
}
\label{fig:boost250simtest}
\end{figure}

We have tested the results of our flux-bias analysis using the 
In-Out simulations
(see \S~\ref{sect:simulations})\footnote{We made one technical adjustment to the
simulations to make the results suitable for testing the results of the
flux-bias analysis. In situations where the recovered source was within
12\,arcsec of both the injected source and a source in the real catalogue,
we made no attempt to distinguish whether the recovered source was the real
source or the injected source, but simply compared the flux density of the
recovered source with that of the injected source.}.
In the top panel of Figure~\ref{fig:boost250simtest} 
we have plotted the mean value of the measured flux density agains the injected
flux density of the artificial sources
in the simulations. The red lines show a linear
fit to these mean flux-density values.
As expected, the measured flux densities are 
systematically higher than the injected flux densities, particularly at fainter
flux densities. The bottom panel shows same plot after we have used the
flux-bias corrections, shown in Figure~\ref{fig:boost250} and listed in Table~\ref{tab:boost},
to correct the measured flux densities.
The measured flux densities are now much closer to the
injected flux densities.

Table~\ref{tab:boost} shows that the flux bias at the flux-density limit of the
catalogue is approximately 20\%.
This is larger than the value of 6\%, which we estimated for our SDP
catalogue \citep{rigby11}. In our SDP analysis we were forced to
use a theoretical model of the submillimetre sky to estimate the
distribution of true flux densities, whereas now we have had the major
advantage of being able to use the deeper {\it Herschel} surveys to
produce a prior probability distribution for the true flux densities, and so
our new estimates supersede the former ones.

The completeness of the survey shown in Figure~\ref{fig:compl250} was derived from the
In-Out simulations and is the
completeness of the survey as a
function of true flux density. Our derivation of the flux bias now allows us
to calculate the completeness of the
survey as a function of measured flux density.
First, we use the
flux-bias relationship shown in the bottom panel of Figure~\ref{fig:boost250}
(listed in Table~\ref{tab:boost}) to estimate the true flux density of each source
in the catalogue.
We then use this estimate of the true flux density 
and the completeness estimates 
shown in Figure~\ref{fig:compl250} (listed
in Table~\ref{tab:stats})
to estimate the probability that this source would have been detected
in the survey. 
We then derive the completeness for each bin of measured flux density
by calculating the mean probability value 
for all the sources in the catalogue
in this bin. 
These completeness estimates are given in 
Table~\ref{tab:compl} and shown in Figure~\ref{fig:compl2bands}.

\begin{figure}
\centering
\includegraphics[trim=0.5cm 0cm 0.5cm 0cm,clip,width=85mm]{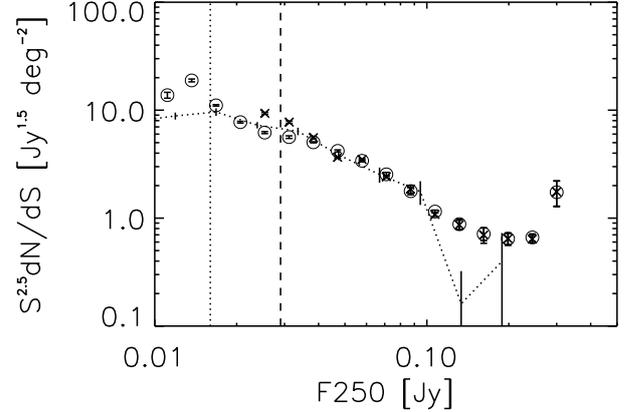}
\caption{The number counts at 
$250\,\micron$ plotted against flux density. 
We first corrected the measured flux density of each source for flux bias
using the relationship in Figure~\ref{fig:boost250} ({\it bottom panel}) and listed in Table~\ref{tab:boost}. We then
used this estimate of the true flux density 
and the completeness estimates from the In-Out simulations
(Table~\ref{tab:stats}) to calculate the number counts.
The circles show the average source counts for the three GAMA fields after
these corrections for flux bias and incompleteness have been made.
The error bars take into account uncertainties in 
the corrections and the scatter between the fields. For comparison, the dotted line
shows the
numbers counts for the HerMES survey from \citet{bethermin12} and the crosses are the counts derived in \S~\ref{sect:counts250}
and reported in Table~\ref{tab:counts250} averaged among the three fields. The vertical black dotted and black 
dashed lines show respectively the 2.5$\sigma$ detection (\S~\ref{sect:madx})  and $4\sigma$ catalogue
limits (\S~\ref{sect:catalogue}).
}
\label{fig:counts250fromcat} 
\end{figure}

In \S~\ref{sect:counts250}
we corrected the observed H-ATLAS counts for the effect of Eddington bias using a method that
did not require us to make corrections to the flux densities of individual sources.
After making this correction we found good agreement between the H-ATLAS
source counts and the HerMES source counts \citep{bethermin12}
at flux densities $\gtrsim 30$\,mJy. We can now derive the source
counts in a different way, using the results of our flux-bias analysis and
the estimates of completeness from the In-Out simulations. First, we use the
flux-bias relationship
(listed in Table~\ref{tab:boost}) to estimate the true flux density of each source
in the catalogue, which allows us to calculate the source
counts as a function of {\it true} flux density.
We then correct these source counts for incompleteness, this time
using the completeness values in Table~\ref{tab:stats}, which give the completeness as a function
of true flux density.
Figure~\ref{fig:counts250fromcat} shows the corrected source counts compared with the ones
from \citep{bethermin12}. Once again there is good agreement between the HerMES and
H-ATLAS counts.

The two methods to derive the number counts described in \S~\ref{sect:counts250} and  \S~\ref{sect:corrections} 
give quite similar results, but they have different positive and negative aspects. The matrix inversion method requires no prior
and take into account all the effects included in the real maps that can influence the measure of the flux 
density: residual cirrus emission, large extended sources, clustering (even if this information is partially lost during the simulation). 
Unfortunately, running the in-out simulation is quite computationally expensive and the accuracy of the conditional probability distribution
is not good enough at the faintest flux densities to provide reliable number counts. On the other side,
correcting the flux density of the single sources it is possible to derive reliable number counts down to the
detection limit of the survey, but requires the assumption of a prior. We let the reader to decide which method
to use, according with the scientific goals pursued.

\subsection{Completeness at 350 and 500\,$\mu$m}\label{sect:compl2bands}

\begin{figure}
\center
\includegraphics[trim=1cm 0cm .5cm 1cm,clip,width=84mm]{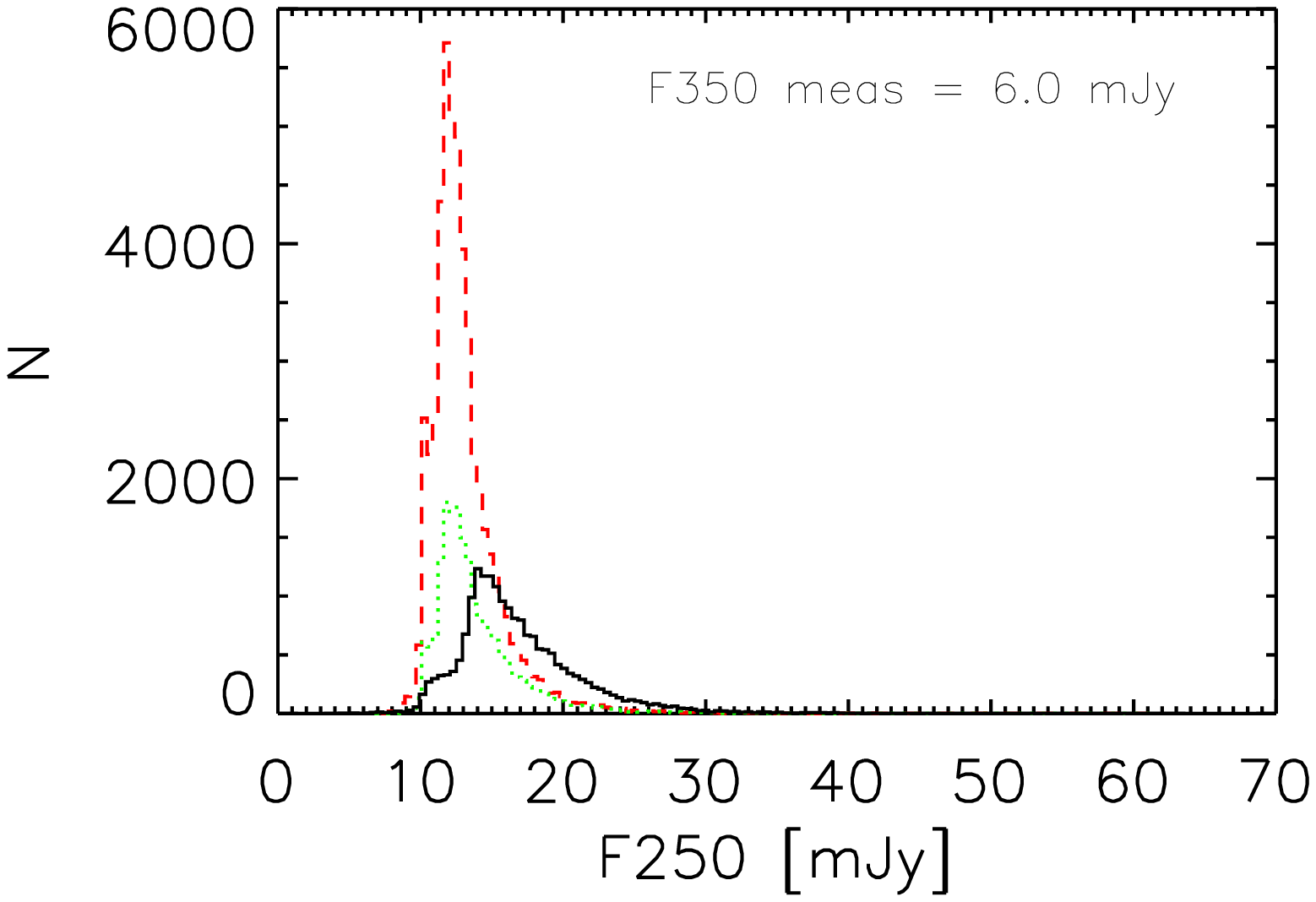}
\includegraphics[trim=1cm 0cm .5cm 1cm,clip,width=84mm]{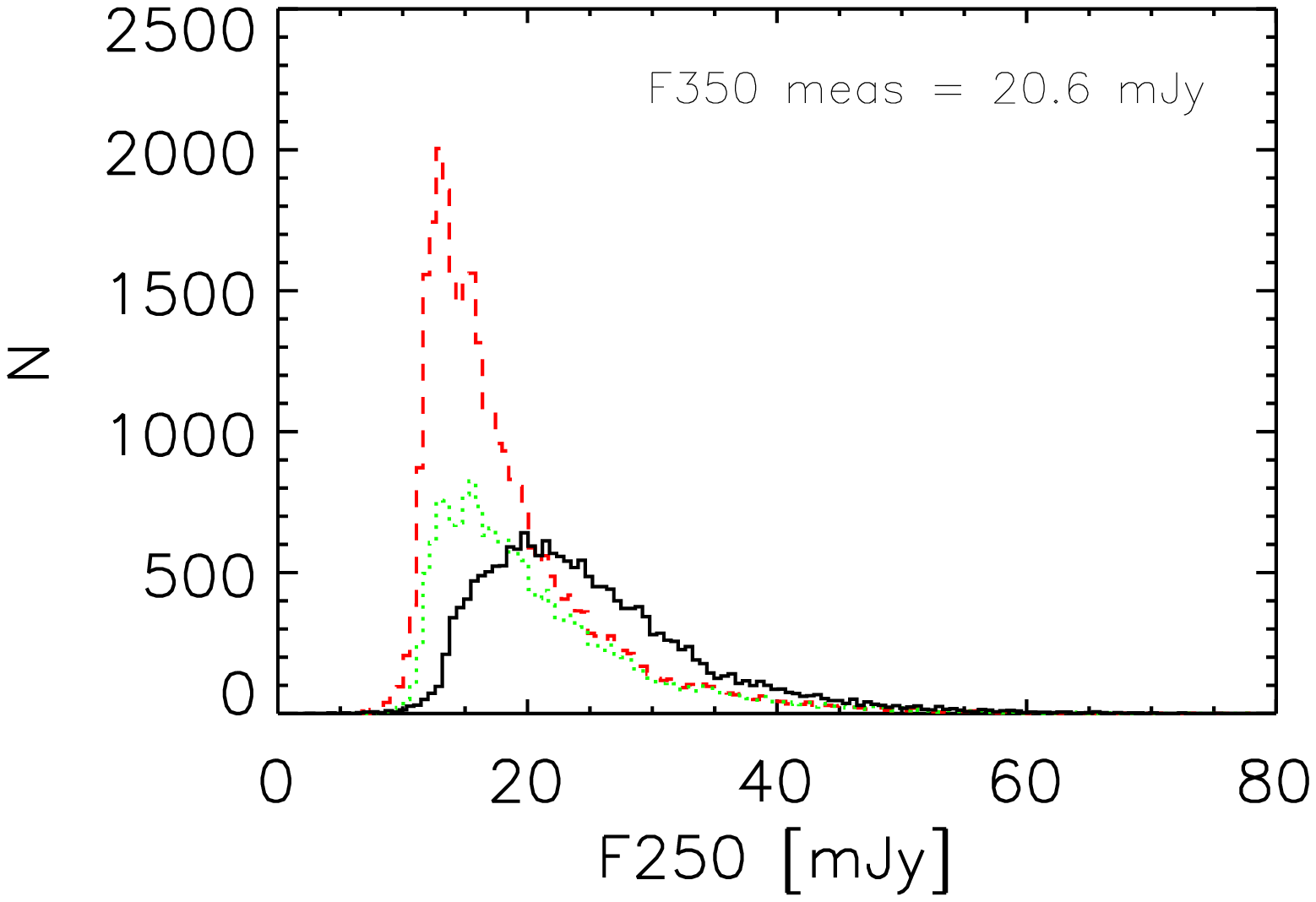}
\includegraphics[trim=1cm 0cm .5cm 1cm,clip,width=84mm]{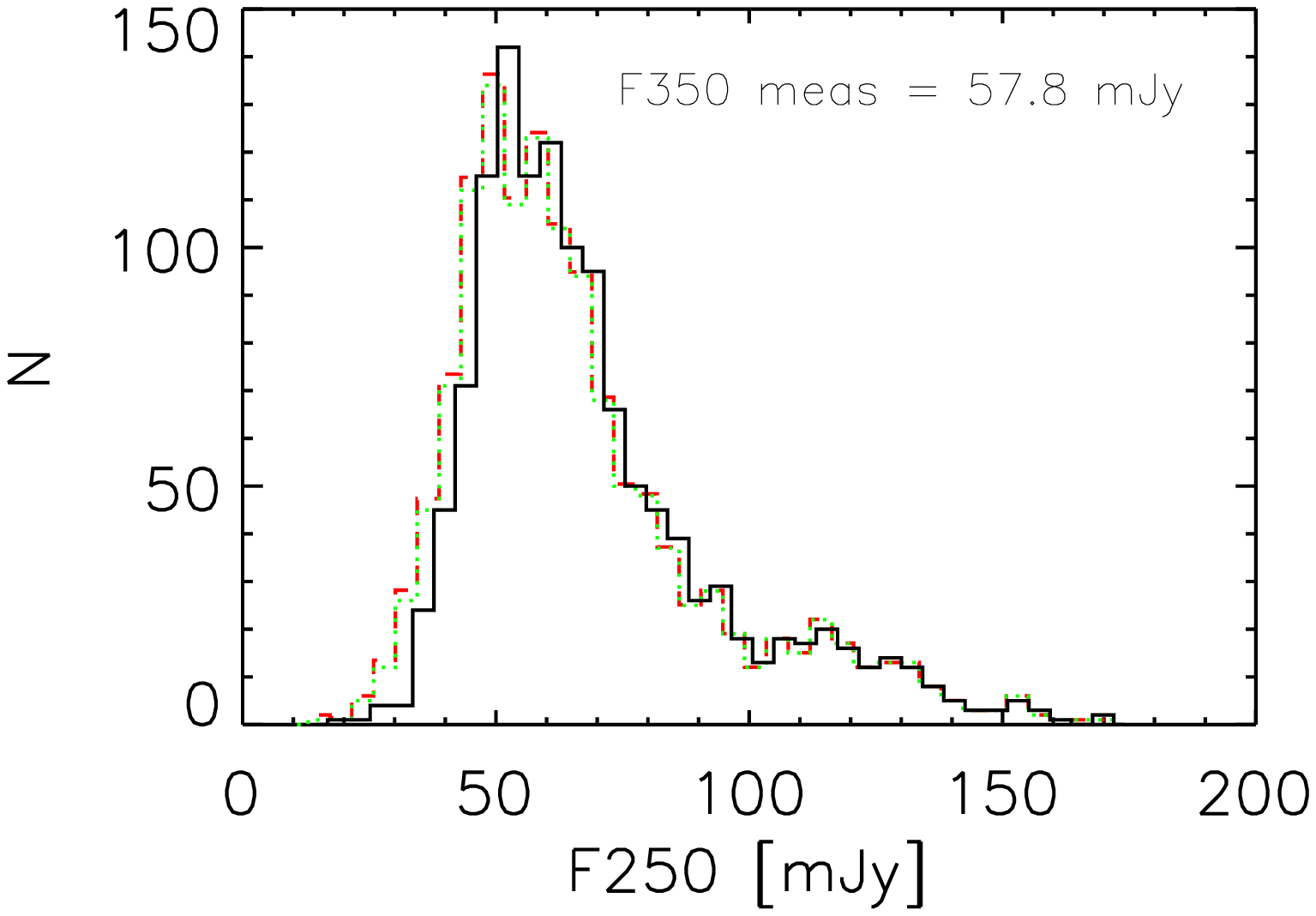}
\caption{Examples of the application of our empirical technique for estimating the
completeness of the survey at 350 and 500\,$\mu$m.
The panels show histograms of 250\,$\mu$m flux density
for three representative bins of 350\,$\mu$m flux density.
The measured distribution of the $250\,\micron$ flux density
 (\textit{black histogram}) is first corrected for flux bias, giving the 
dotted green histogram, and then for completeness at 250\,$\mu$m, producing the 
dashed red histogram. The completeness estimate
for each bin of 350\,$\mu$m flux density is given by the
ratio of the area under the black histogram to the area under the red histogram.} 
\label{fig:compl350hist} 
\end{figure}

The completeness of the catalogue at 350 and 500\,$\mu$m must be derived with a different method 
because the 350\,$\mu$m and 500\,$\mu$m sources were actually detected 
on the 250\,$\mu$m images.  We have used the following empirical technique to estimate the completeness of the catalogues 
at these wavelengths.

Whether a source is detected should only depend on its true 
250\,\micron\, flux density and not on its 350\,$\mu$m flux density.
We take the measured 250\,$\mu$m flux density of each source in the catalogue, correct 
it for flux bias (Table~\ref{tab:boost}) and use the completeness estimated from the
In-Out simulations for the true 250\,$\mu$m flux densities
(Table~\ref{tab:stats}) to calculate
the probability this source would have been detected 
by the survey.
In each bin of 350\,$\mu$m flux density,
we then take the average value of these
probabilities as our estimate
of the completeness in this bin. We follow the same
procedure at 500\,$\mu$m.

We used the results of the In-Out simulations (\S~\ref{sect:simulations}) to perform a
simple sanity test of this technique.
Although the distribution of 
$350/250\,\mu$m flux-density ratios in the simulations was
not the same as in the real sky\footnote{In the In-Out simulations, 
equal numbers of sources are injected
at each of
the flux densities listed in Table~\ref{tab:stats} in each waveband, which means
that at each
injected 350\,$\mu$m flux density, there are equal numbers of sources
with each possible
250\,$\mu$m flux density.}, 
our correction technique should work equally
well. 
In the simulations, the distribution of this ratio for the detected
sources was quite different from the distribution for the
injected sources, because sources with high values for this
flux ratio, and thus low values
of 250\,$\mu$m flux density, will tend not to be detected.
However, after we applied the correction for flux bias at 250\,$\mu$m and then
the corrections for incompleteness from Table~\ref{tab:stats}, we recovered a distribution
of $350/250\,\mu$m flux density ratios that was similar
to the distribution for the injected sources.

Figure~\ref{fig:compl350hist} shows how this technique is applied to the real data.
The figure shows histograms 
of observed 250\,$\mu$m flux densities for several bins of 350\,$\mu$m flux density.
The black lines in the figure show the observed distributions of 
250\,$\mu$m flux density. The green line in each figure shows the
effect of the first correction, after we have corrected each 250\,$\mu$m flux density
for flux bias (Table~\ref{tab:boost}). The red dashed line shows the effect of the
second correction, after we have corrected for the incompleteness at 250\,$\mu$m
(Table~\ref{tab:stats}). The completeness in each 350\,$\mu$m flux density
bin is the ratio of the number of sources under the black line to the the number
of sources under the red line. 

\begin{figure}
\center
\includegraphics[trim=1cm 0cm 1cm 1cm,clip,width=84mm]{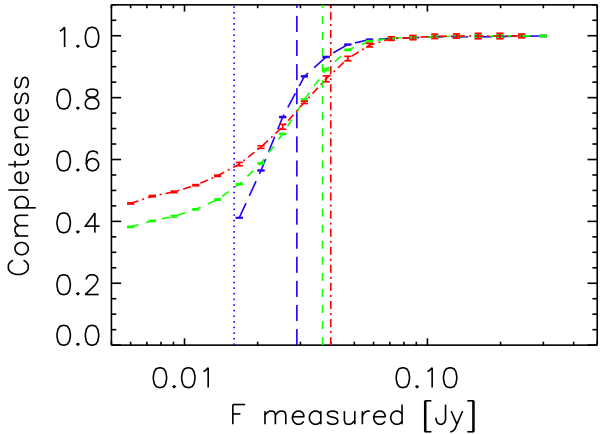}
\caption{Completeness at 250 ({\it blue long-dashed}), 350 ({\it green dashed}) and 500\,\micron\ 
({\it red dot-dashed}), as listed in Table~\ref{tab:compl},
plotted against {\it measured} flux density. The vertical lines
show the $4\sigma$ catalogue limit at each wavelength, 29, 37 and 40\,mJy at 250, 350 and 500\,\micron\
respectively. The vertical blue dotted line shows the $2.5\sigma$ detection limit at 250\,\micron\  (\S~\ref{sect:catalogue}).} 
\label{fig:compl2bands}
\end{figure}

Our completeness estimates are
listed in Table~\ref{tab:compl} and shown visually in Figure~\ref{fig:compl2bands}.
The reason why completeness does not decline so rapidly with decreasing
flux density as at 250\,$\mu$m is that a source may be very faint at 350 or 500\,$\mu$m but
still be bright enough at 250\,$\mu$m to be easily detected.

\begin{table}
\centering
\caption{Completenesses at 250, 350 and 500\,\micron. Note that
flux density in this table is
{\it measured} flux density. Column 2 in Table~\ref{tab:stats} shows how
completeness
at 250 $\mu$m depends on {\it true} flux density.}
\begin{minipage}{80mm}
\centering
\label{tab:compl}
\begin{tabular}{@{}cccc@{}}
\hline
$F$ & 250\,\micron & 350\,\micron & 500\,\micron  \\
    {[}mJy]       &                                         &             \\
\hline
  6.0     &    -     		&   0.382     &     0.458         \\
  7.4     &    -     		&   0.402     &     0.482    	\\
  9.1     &    -    	 	&   0.416     &     0.495     	\\
 11.1     &    -     		&   0.439     &     0.517     	\\
 13.7     &    -     		&   0.471     &     0.547     	\\
 16.8     &    0.412     &   0.520     &     0.585    	\\
 20.6     &    0.565     &   0.588     &     0.640    	\\
 25.4     &    0.737     &   0.683     &     0.706    	\\
 31.2     &    0.869     &   0.794     &     0.785    	\\
 38.3     &    0.932     &   0.891     &     0.861    	\\
 47.0     &    0.972     &   0.954     &     0.926          \\
 57.8     &    0.988     &   0.982     &     0.970    	\\
 71.0     &    0.991     &   0.992     &     0.991    	\\
 87.2     &    0.994     &   0.995     &     0.994    	\\
107.2     &    0.997     &   0.997     &     0.998   	\\
131.7     &    0.997     &   0.998     &     0.999   	\\
161.8     &    0.996     &   0.998     &     0.999   	\\
198.7     &    0.997     &   0.999     &     0.999   	\\
244.2     &    0.999     &   1.000     &     1.000   	\\
300.0     &    0.999     &   1.000     &       \\
\hline
\end{tabular}
\end{minipage}
\end{table}

\begin{table*}
\centering
\caption{Number counts at 350\,\micron, $\hat{\rm N}_{\rm t}$, estimated from
the inversion technique discussed in \S~\ref{sect:counts250} and \S~\ref{sect:compl2bands}. Uncertainties do not include the correlation between flux bins.
}
\begin{minipage}{140mm}
\centering
\label{tab:counts350}
\begin{tabular}{@{}rrcrrcrrcr@{}}
\hline
F350        & \multicolumn{9}{c}{Number counts (${\rm d}N/{\rm d}S\times S^{2.5}$)}  \\
 {[}mJy]     & \multicolumn{9}{c}{[Jy$^{1.5}$deg$^{-2}$]}    \\
 		& \multicolumn{3}{c}{GAMA9} &  \multicolumn{3}{c}{GAMA12} &  \multicolumn{3}{c}{GAMA15}                         \\
\hline
25.4&4.438&$\pm$&0.045&4.187&$\pm$&0.045&4.076&$\pm$&0.047\\
31.2&4.495&$\pm$&0.070&4.333&$\pm$&0.062&4.534&$\pm$&0.056\\
38.3&4.209&$\pm$&0.092&4.179&$\pm$&0.072&4.463&$\pm$&0.066\\
47.0&2.722&$\pm$&0.084&2.738&$\pm$&0.067&2.757&$\pm$&0.070\\
57.8&1.277&$\pm$&0.063&1.193&$\pm$&0.061&1.366&$\pm$&0.072\\
71.0&0.797&$\pm$&0.065&0.817&$\pm$&0.065&0.798&$\pm$&0.072\\
87.2&0.471&$\pm$&0.054&0.513&$\pm$&0.056&0.542&$\pm$&0.056\\
107.2&0.355&$\pm$&0.043&0.286&$\pm$&0.041&0.350&$\pm$&0.044\\
131.7&0.175&$\pm$&0.033&0.175&$\pm$&0.032&0.242&$\pm$&0.038\\
161.8&0.190&$\pm$&0.037&0.116&$\pm$&0.029&0.254&$\pm$&0.042\\
198.7&0.085&$\pm$&0.028&0.149&$\pm$&0.037&0.222&$\pm$&0.045\\
244.2&0.230&$\pm$&0.053&0.157&$\pm$&0.044&0.189&$\pm$&0.047\\
300.0&0.380&$\pm$&0.080&0.461&$\pm$&0.087&0.614&$\pm$&0.100\\
\hline
\end{tabular}
\end{minipage}
\end{table*}

\begin{table*}
\centering
\caption{Number counts at 500\,\micron, $\hat{\rm N}_{\rm t}$, estimated from
the inversion technique discussed in \S~\ref{sect:counts250} and \S~\ref{sect:compl2bands}.}
\begin{minipage}{140mm}
\centering
\label{tab:counts500}
\begin{tabular}{@{}rrcrrcrrcr@{}}
\hline
F500        & \multicolumn{9}{c}{Number counts (${\rm d}N/{\rm d}S\times S^{2.5}$)}  \\
 {[}mJy]     & \multicolumn{9}{c}{[Jy$^{1.5}$deg$^{-2}$]}    \\
 		& \multicolumn{3}{c}{GAMA9} &  \multicolumn{3}{c}{GAMA12} &  \multicolumn{3}{c}{GAMA15}                         \\
\hline
25.4&1.611&$\pm$&0.043&1.599&$\pm$&0.042&1.665&$\pm$&0.038\\
31.2&1.412&$\pm$&0.046&1.506&$\pm$&0.063&1.737&$\pm$&0.088\\
38.3&1.142&$\pm$&0.063&1.253&$\pm$&0.080&1.459&$\pm$&0.077\\
47.0&0.626&$\pm$&0.028&0.616&$\pm$&0.029&0.640&$\pm$&0.046\\
57.8&0.160&$\pm$&0.035&0.075&$\pm$&0.042&0.132&$\pm$&0.039\\
71.0&0.117&$\pm$&0.027&0.112&$\pm$&0.029&0.145&$\pm$&0.034\\
87.2&0.142&$\pm$&0.025&0.110&$\pm$&0.024&0.131&$\pm$&0.028\\
107.2&0.009&$\pm$&0.014&0.071&$\pm$&0.021&0.057&$\pm$&0.020\\
131.7&0.046&$\pm$&0.016&0.029&$\pm$&0.014&0.040&$\pm$&0.016\\
161.8&0.037&$\pm$&0.017&0.039&$\pm$&0.017&0.070&$\pm$&0.022\\
198.7&0.026&$\pm$&0.016&0.026&$\pm$&0.015&0.043&$\pm$&0.020\\
244.2&0.060&$\pm$&0.027&0.024&$\pm$&0.017&0.059&$\pm$&0.027\\
300.0&0.016&$\pm$&0.016&0.132&$\pm$&0.046&0.162&$\pm$&0.051\\
\hline
\end{tabular}
\end{minipage}
\end{table*}

In \S~\ref{sect:counts250}
we described an inversion technique for correcting the source counts
at 250 $\mu$m for Eddington bias.
We can now use the completeness estimates at 350 and 500\,$\mu$m
to extend this method to the two longer wavelengths.
As before, we represent the
the results of the In-Out simulations as a matrix, $\bf \widetilde{P}$,
in which each element, $\widetilde{P}_{ij}$, is the probability that a 
galaxy with intrinsic flux $F_i$ is detected with a measured flux $F_j$. 
At 250\,$\mu$m, the matrix contains all necessary information
about the effect of noise on the flux density of a source and
on the incompleteness of the survey, with the sum of the elements
in each column of 
the matrix giving the incompleteness of the survey for each
injected flux density.

At the two longer wavelengths, however, the sum of the columns of each matrix
does not give the true completeness of the survey because
the detection of the sources was performed at 250\,$\mu$m\footnote{The sum
of the columns does not give the true incompleteness because the
distribution of the 250/350\,$\mu$m flux-density ratios used in
the In-Out simulations was not the same as in the real sky. We could
have used a more realistic distribution in the simulations but chose
not to because we could not have been sure how close our
assumed distribution was to the real distribution, since
the distribution we observe in the survey is strongly distorted
by the effects of completeness and flux bias.}. Nevertheless, we can
still apply the inversion method using a simple modification.

First, we normalise each column of the matrix, so that the sum of
the elements in the column is unity, effectively removing the
incorrect completeness information. 
As before (see \S~\ref{sect:counts250}), the observed source counts are represented
by a vector, $\bf \hat{N}_m$. 
We now correct this vector using the completeness estimates from Table~\ref{tab:compl}. 
We now have an equation similar to Eq.~\ref{eq:matrixeq}, linking the observed
counts and the true counts $\bf \hat{N}_t$:

\begin{equation}\label{eq:matrixeq2}
\bf \hat{N}_{m\_corr} = \widetilde{P} \cdot \hat{N}_t,
\end{equation}

\noindent in which $\bf \widetilde{P}$ is the same as $\mathbf P$, but with sum of the columns
normalised to unity, and $\bf \hat {N}_{m\_corr}$ is $\bf \hat{N}_m$ divided
by the completeness at 350 or 500\,\micron.

\begin{figure*}
\includegraphics[width=150mm,keepaspectratio]{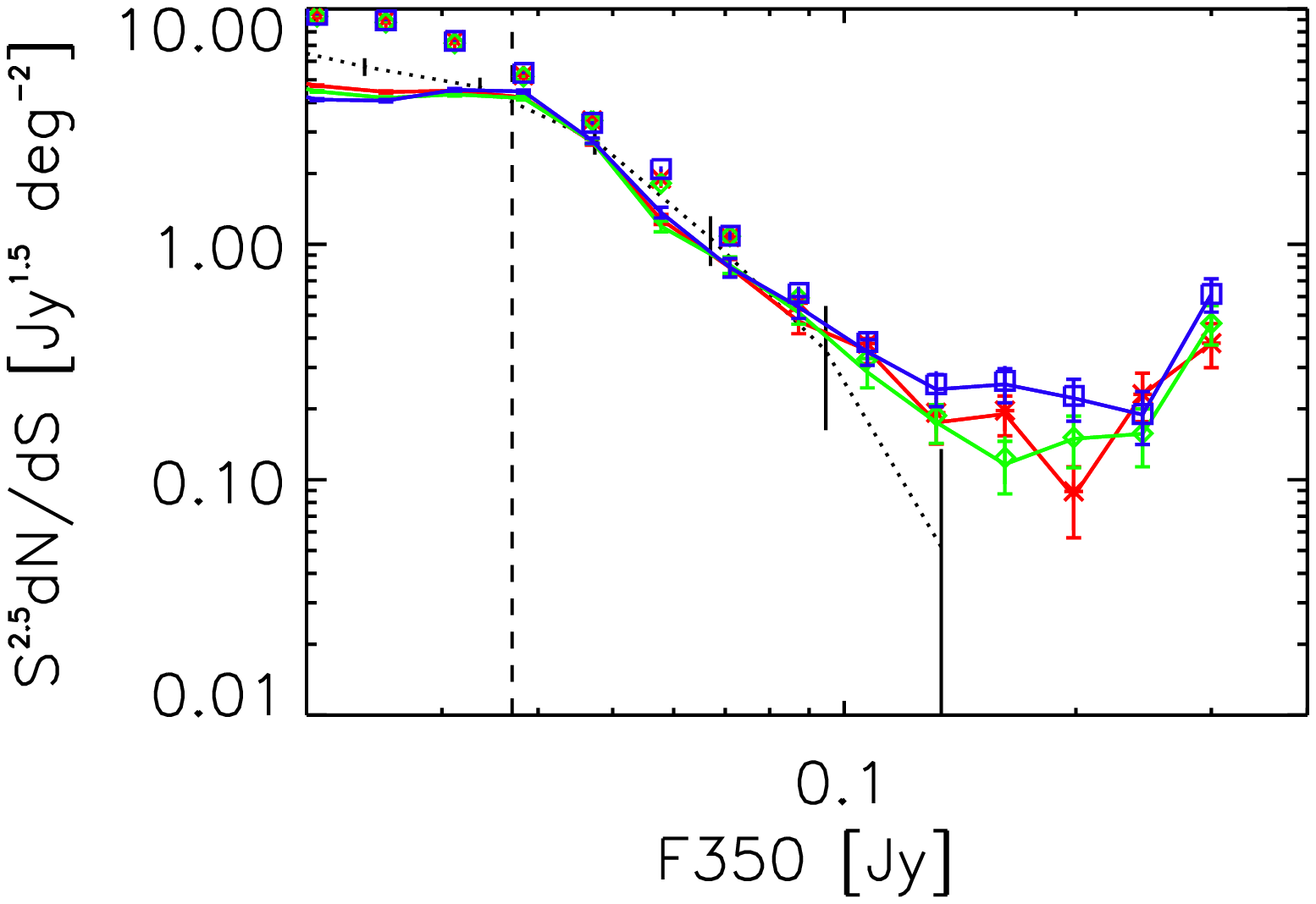}
\includegraphics[width=150mm,keepaspectratio]{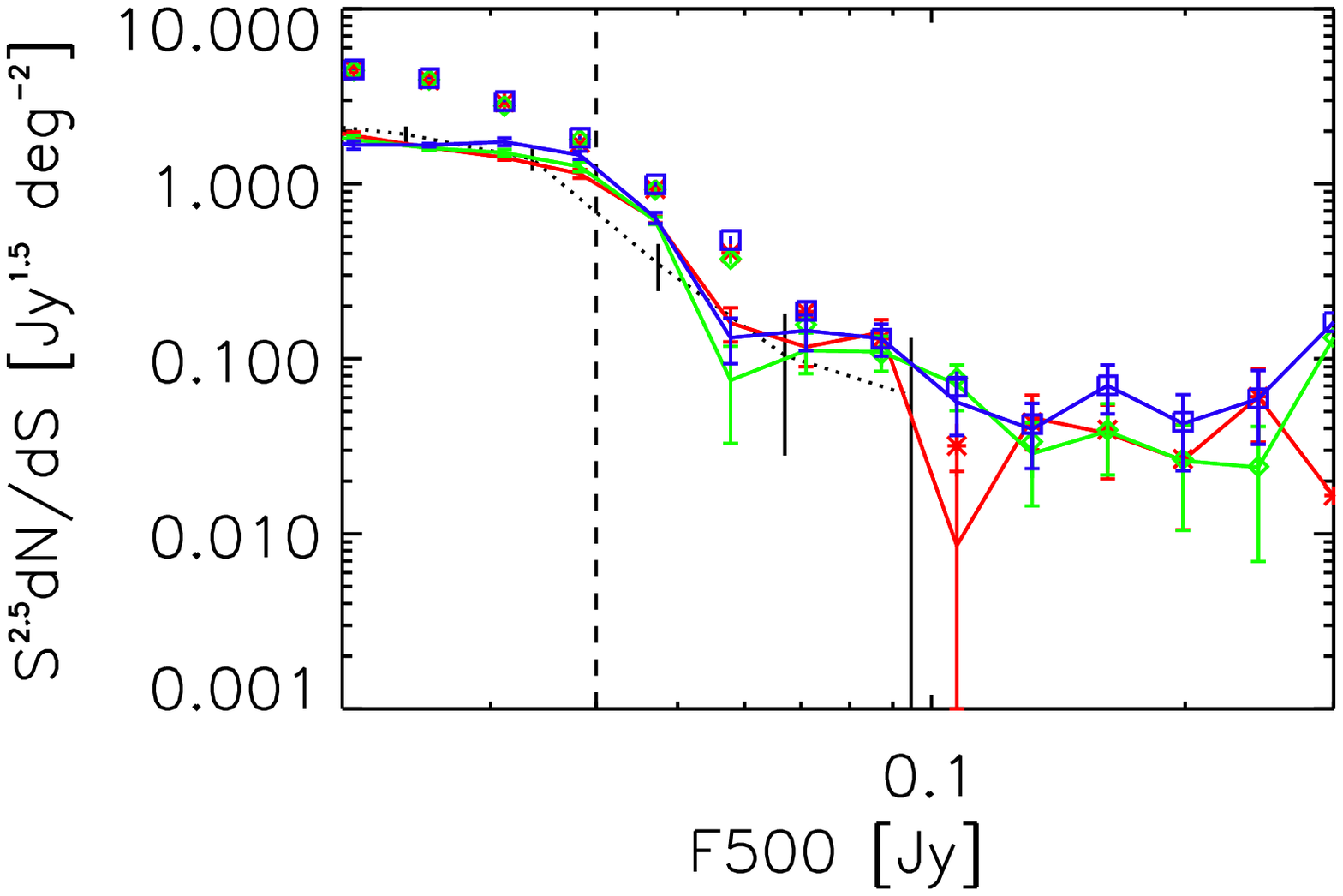}
\caption{The number counts of sources from the inversion technique. The red, green and blue colours represent the GAMA9, GAMA12 and GAMA15 fields, respectively. 
The symbols show the measured counts, $\hat{\rm N}_{\rm m}$, while the continuous 
lines show the true numbers counts, $\hat{\rm N}_{\rm t}$, estimated from our inversion technique. 
Uncertainties do not include the correlation between flux bins. 
The dotted lines are the source counts from the \citet{bethermin12}. 
They underestimate the counts at the bright end because the HerMES area is 
smaller than the H-ATLAS one, so they do not have adequate statistics above $\sim 100\,$mJy. 
The vertical dashed line shows the 4$\sigma$ limit in that band. {\it Top}: counts at 350\,\micron. {\it Bottom}: counts at 500\,\micron.}
\label{fig:counts2bands}
\end{figure*}

The results at 350 and 500\,$\mu$m 
are shown in Figure~\ref{fig:counts2bands} and listed in Tables~\ref{tab:counts350} and
\ref{tab:counts500}. Uncertainties are estimated by Monte Carlo simulations, in the same way
as explained in \S~\ref{sect:counts250} and do not include the correlation between fluxe bins. 
Because our counts are derived from simulations only and do not assume any prior on the existing galaxy population,
we can only trust them for fluxes brighter than $\sim 20$\,mJy. At fainter fluxes, the conditional probability
distribution $P(F_{\rm m}|F_{\rm t})$ is too uncertain. 
As at 250\,\micron\ (see Figure~\ref{fig:counts250}),
the corrected H-ATLAS counts at 350\,$\mu$m are in good agreement with 
the HerMES source counts
above the flux-density limit of our catalogue.
At 500\,$\mu$m, the corrected H-ATLAS source counts are a little higher than
the HerMES source counts but consistent within the errors.

\subsection{Corrections for Flux Bias at 350 and 500\,$\mu$m}\label{sect:boost2bands}

\begin{figure}
\includegraphics[trim=1cm 0cm 1cm 1cm,clip,width=84mm]{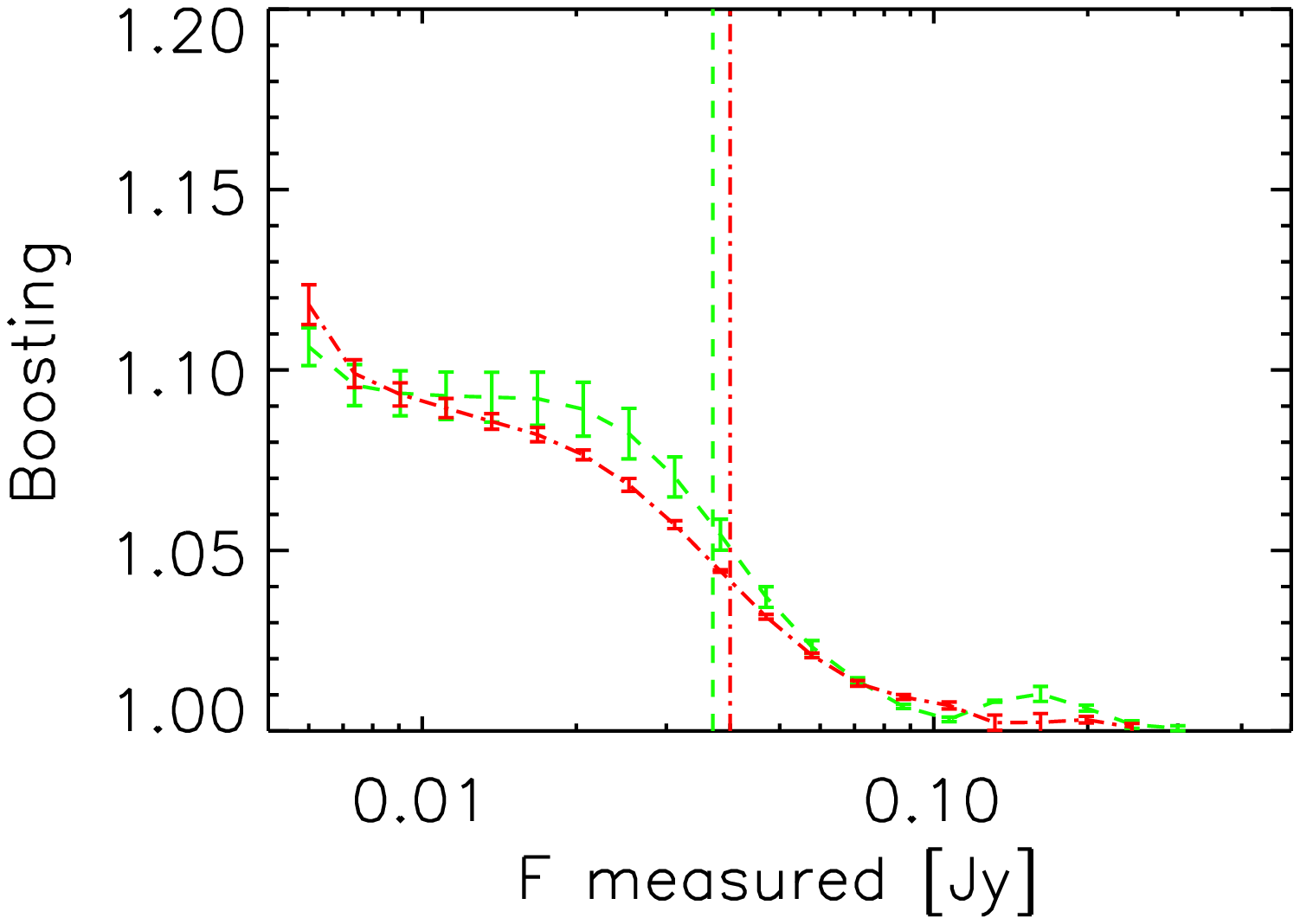}
\caption{Flux bias verses {\it measured} flux density at 350\,$\mu$m ({\it green dashed}) and 500\,\micron\ 
({\it red dot-dashed}), as listed in Table~\ref{tab:boost}.
Error bars take into account differences in covariance between the fields.
The vertical lines show the $4\sigma$ catalogue limit at each wavelength, 37 and 40\,mJy at 350 and 500\,\micron\
respectively (\S~\ref{sect:catalogue}).}
\label{fig:boost2bands}
\end{figure}

\begin{figure*}
\includegraphics[trim=5mm 5mm 1cm 1cm,clip,width=84mm]{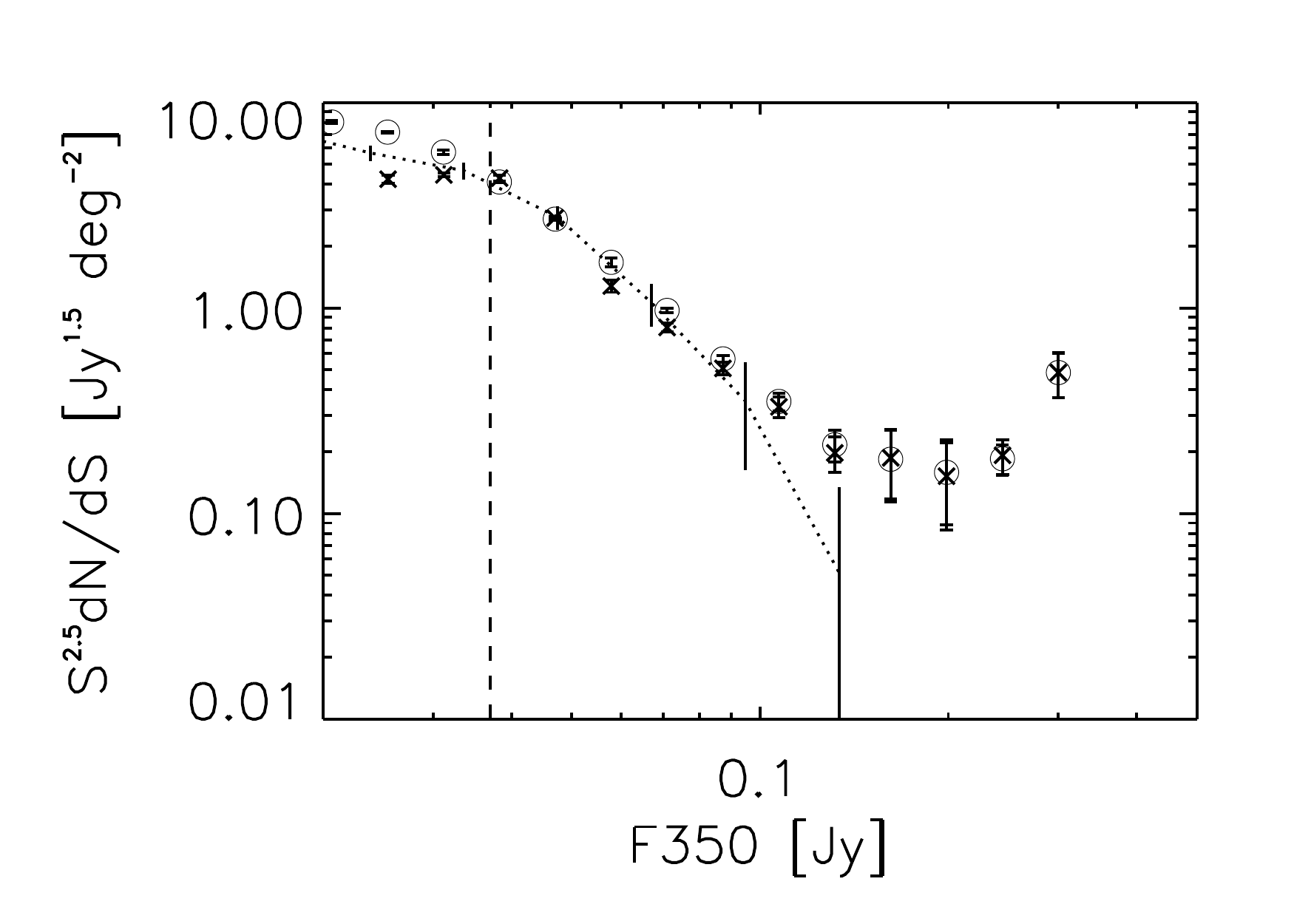}
\includegraphics[trim=5mm 5mm 1cm 1cm,clip,width=84mm]{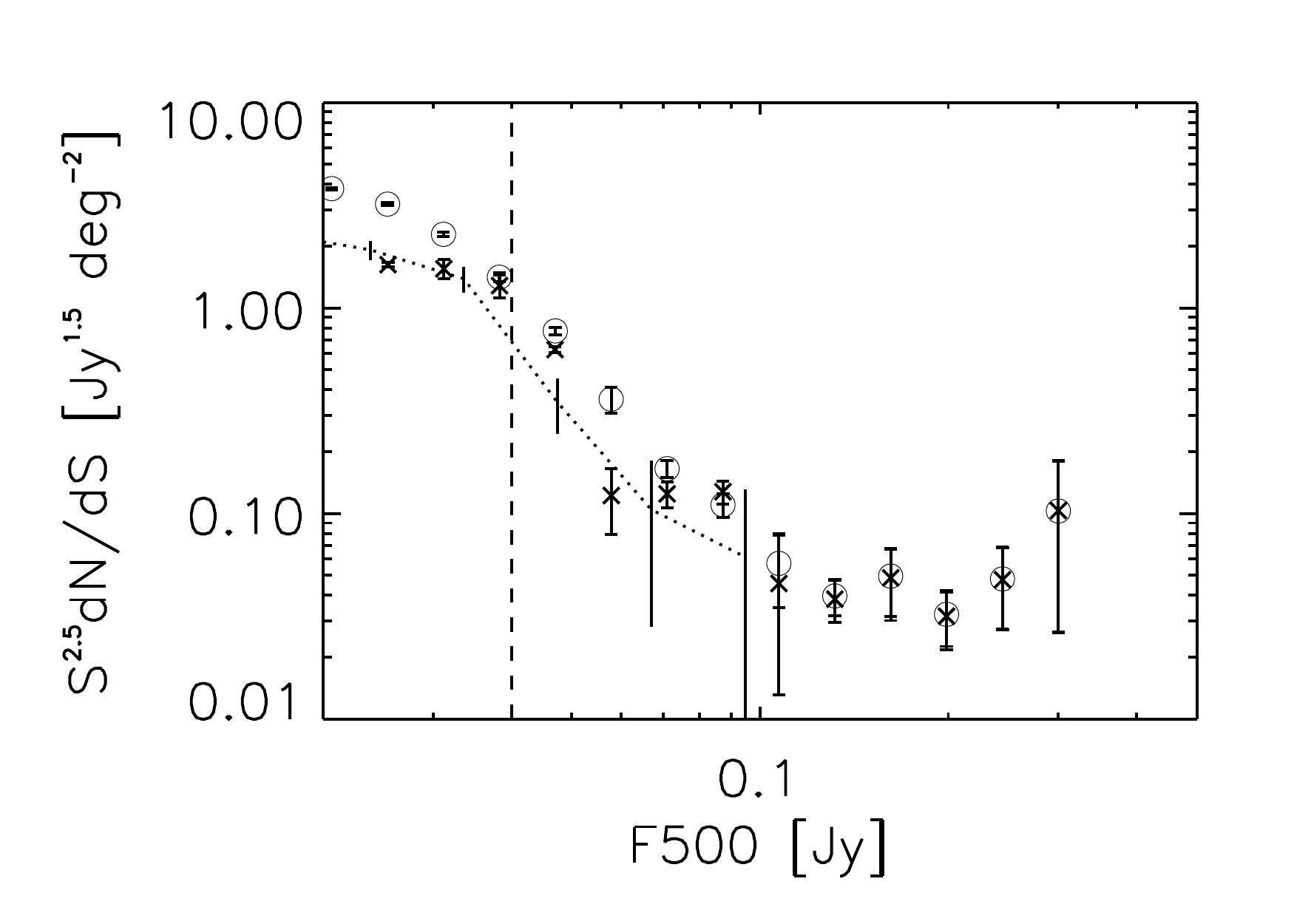}
\caption{Differential source counts at 350\,$\mu$m ({\it left}) and 500\,$\mu$m ({\it right}), normalised
to the value expected in a Euclidean universe. The dotted lines in both diagrams show the source
counts measured from deeper {\it Herschel} surveys \citep{bethermin12}. The circles show
our source counts after correcting for source incompleteness and flux bias (see \S~\ref{sect:compl2bands} 
and \ref{sect:boost2bands} for details). The crosses are the counts derived in \S~\ref{sect:compl2bands}
and reported in Tables~\ref{tab:counts350} and \ref{tab:counts500} averaged among the three fields. 
}
\label{fig:counts2bandsfromcat}
\end{figure*}

Determining the best method to correct for flux bias at 350 and 500\,$\mu$m requires
careful thought because of the way we have carried out the survey. \citet{crawford10} have
presented an elegant method for 
correcting for flux bias when a survey is carried out at more than one wavelength. However,
in this method the wavelengths are treated as equivalent, and a simple thought experiment shows
that this method will not produce the right answers for our survey.

Let us consider observations at 250 and 350\,$\mu$m.
We detect the sources at 250\,$\mu$m, selecting peaks 
in the map, and thus
the flux densities of the sources at 250\,$\mu$m will be affected by the
flux bias caused by the noise on the map (instrumental plus confusion). 
We described in \S~\ref{sect:boost250} how we estimated
the size of this bias. 

Now let us consider the 350\,$\mu$m observations. In our survey, we do 
not have to detect the source above a flux-density
threshold at 350\,$\mu$m. What we actually do is go to the position of the
250\,$\mu$m source and measure the 350\,$\mu$m flux density at this position.
Since the instrumental noise on the 350\,$\mu$m image is uncorrelated with
the instrumental noise on the 250\,$\mu$m image, the instrumental noise
on the 350\,$\mu$m image is equally likely to produce a positive or negative fluctuation 
on the map. Therefore,
in this thought experiment, there should be no flux bias at 350\,$\mu$m due to instrumental noise. 
The confusion noise, on the other hand, is highly correlated between the two bands,
because we expect the sources producing confusion to emit at both wavelengths,
250 and 350\,$\mu$m. The flux bias at 350\,$\mu$m therefore must depend on the
flux bias at 250\,$\mu$m and the covariance between the 250\,$\mu$m and 350\,$\mu$m
images.

In conclusion, using a pure intuitive approach, we would expect the differences between measured and
true fluxes at 350 and 500 \,\micron\, to be proportional to the fraction of the contribution of the 
confusion noise to the total noise at 250\,\micron, to the beam size and to the average colour of the sources. 

If we treat the problem in a more formal way, we reach very similar conclusions. We can quantify this argument in the following way.
Let us suppose that a source has measured flux densities of $F_{\rm m,250}$ and
$F_{\rm m, 350}$ at the two wavelengths. We then use the analysis of \S~\ref{sect:boost250} to estimate the
flux-bias factor $fb_{250} = F_{\rm m,250}/F_{\rm t,250}$ at 250\,$\mu$m. The source 
will therefore be sitting on a fluctuation on the 250\,$\mu$m image, the 
combination of instrumental
and confusion noise, given by:

\begin{equation}\label{eq:delta250}
\Delta_{250} = F_{\rm m,250} - F_{\rm t,250} = F_{\rm m,250} (1 - 1/fb_{250}).
\end{equation} 

\noindent We make the same assumption as with the analysis of the flux bias at
250\,$\mu$m that the fluctuations on the two images follow a Gaussian distribution,
with the simultanous probability of the fluctuations at
250 and 350\,$\mu$m following a bivariate Gaussian distribution:

\begin{equation} 
P(\Delta_{250},\Delta_{350}) = \frac{\exp(-{1 \over 2} \mathbf{r^T  C^{-1}  r})}{2 \pi \sqrt{\det {\bf (C)}}}
\end{equation}

\noindent in which $\bf r$ is the vector ($\Delta_{250}, \Delta_{350}$) and $\bf C$ is the
covariance matrix. If we define the magnitude of the fluctuation under
the source at 250\,$\mu$m using Eq.~\ref{eq:delta250}, it is fairly 
easy to show that the expectation value of the
fluctuation under the source at 350\,$\mu$m is given by

\begin{equation}
{\rm E}[\Delta_{350}] = \frac{\rm Covar}{\rm Var_{250}} \Delta_{250}
\end{equation}

\noindent in which $\rm Covar$ is the covariance between the two images
and $\rm Var_{250}$ is the variance of the image at 250\,$\mu$m. Note the important
point that this analysis does
not require any assumptions about the proportions of the variance that come from
instrumental noise and confusion. 

For a source with measured flux densities
of $F_{\rm m,250}$ and $F_{\rm m,350}$, we estimate the variance and the covariance
from the matched-filter-convolved image using the following relationships:

\begin{align} 
{\rm Var}_{250} =\frac {1}{N_{\rm pix}} \sum_i^{N_{\rm pix}} (F_{i,250} - <F_{i,250}>)^2 \nonumber\\ 
{\rm for\,all\,} F_{i,250} < F_{\rm m,250}
\end{align}

\begin{align} 
{\rm Covar} = \frac{1}{N_{\rm pix}} \sum_i^{N_{\rm pix}} (F_{i,250} - <F_{i,250}>)(F_{i,350} - <F_{i,350}>)\nonumber\\
{\rm for\,all\,} F_{i,250} < F_{\rm m,250}, F_{i,350} < F_{\rm m,350}.
\end{align}

\noindent We only include pixels with flux densities less than the measured flux densities
in the calculations 
for the same reason as in \S~\ref{sect:fluxunc}: a pixel brighter than the flux density of a source
should not be used in the calculation of the variance or covariance because otherwise
we would have measured a higher flux density for that source.
The flux bias at 350\,$\mu$m is then calculated
from:

\begin{equation}
fb_{350} = \frac{F_{\rm m,350}}{(F_{\rm m,350} - {\rm E}[\Delta_{350}])}.
\end{equation}

\noindent This analysis leads to an individual estimate of the flux bias at 350
$\mu$m for each source. To generate an average flux-bias relationship, we
calculate the mean value of the flux bias in bins of 350\,$\mu$m flux density.
This relationship is shown in Figure~\ref{fig:boost2bands} and is listed in Table~\ref{tab:boost}. The figure and
table also include the relationships between average flux bias and measured
flux density at 500\,$\mu$m, calculated in the same way.

As for the 250\,$\mu$m source counts (see \S~\ref{sect:boost250}) we can now correct 
the observed 350\,$\mu$m and 500\,$\mu$m source counts
for completeness and flux bias, and then compare the corrected counts
with the HerMES source counts \citep{bethermin12}. 
To correct the counts, we first 
estimate the incompleteness for each source using its measured
350\,$\mu$m or 500\,$\mu$m flux density and the completeness values 
listed in Table~\ref{tab:compl}.
We then correct the measured flux density of each source for flux bias using the
values listed in Table~\ref{tab:boost}. We then add up the number of
sources in each bin of corrected (true) flux density.

Figure~\ref{fig:counts2bandsfromcat} shows the results. There is 
very good agreement with the corrected counts
and the counts of \citet{bethermin12} at 350\,$\mu$m.
The agreement is less good at 500\,$\mu$m, with the HerMES counts falling below
the corrected H-ATLAS counts. We found a similar discrepancy when we corrected
the 500\,$\mu$m counts for the effect of Eddington bias (see \S~\ref{sect:compl2bands}, 
Figure~\ref{fig:counts2bands}) rather than
correcting the individual sources for flux bias and incompleteness, although the direct
inversion method produced a smaller discrepancy.
One possible explanation of the disagreement, supported
by the fact that we see a similar discrepancy from both methods for reconstructing
the source counts, is either that the
500\,$\mu$m source counts in the HerMES and H-ATLAS fields are genuinely
different or that there is some systematic error in 
the 500\,$\mu$m source counts produced by the method
used to produce either the H-ATLAS catalogue or the HerMES catalogue.

A second possibility is that the discrepancy is
caused by the Gaussian approximation we have made
in the flux-bias analysis,
since the bright sources in
the non-Gaussian tail of the distribution would increase the
flux-bias factor, 
an effect which one might expect to
be greatest at 500\,$\mu$m because of the larger beam size.

We refer to \S~\ref{sect:corrections} for a discussion about the differences between the counts derived
by the matrix inversion method and the counts obtained correcting the flux densities of single sources.

\begin{figure}
\center
\includegraphics[width=80mm,keepaspectratio,angle=0]{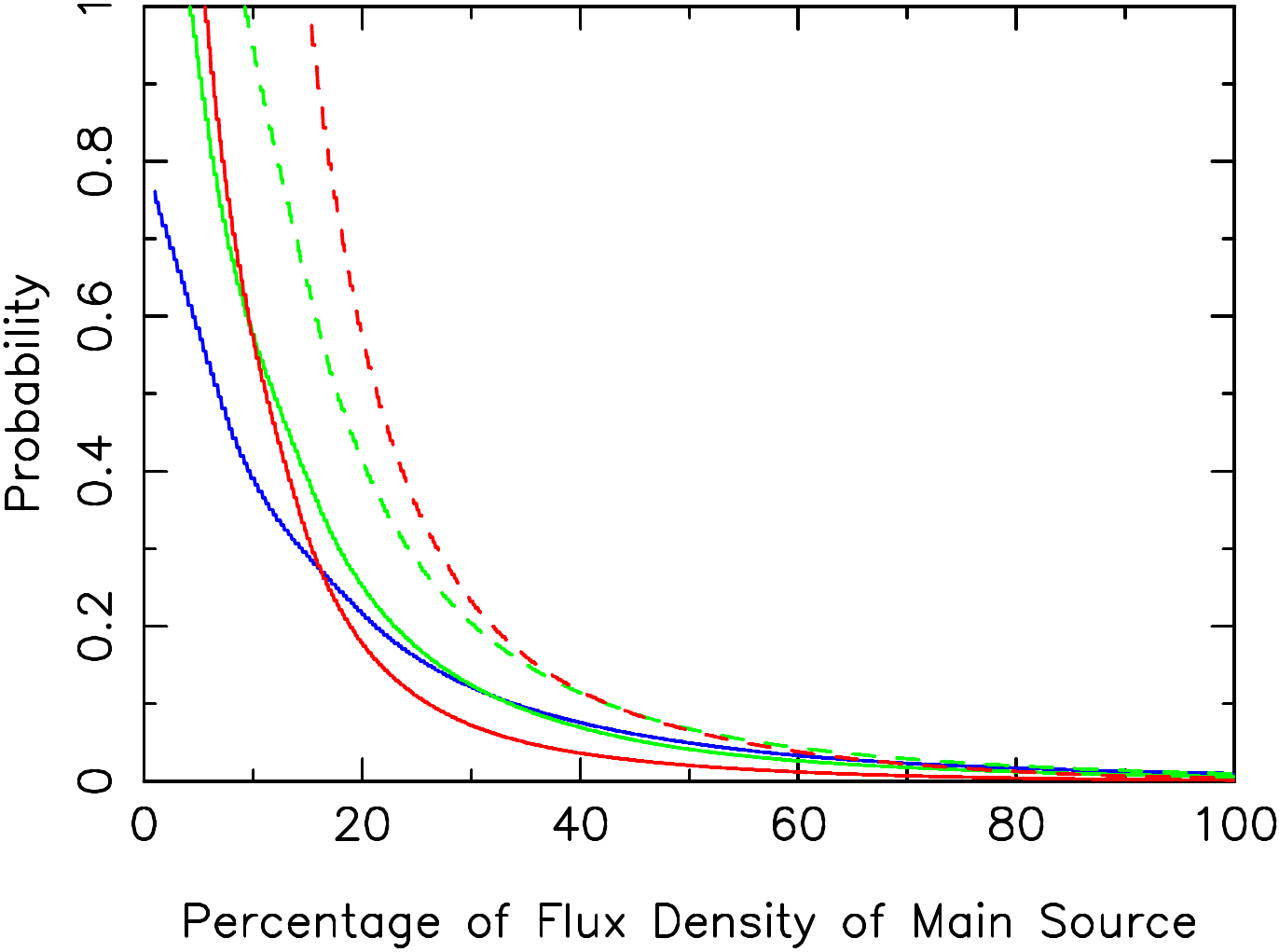}
\caption{The probability that a second source is so close to a primary
source that the two sources can not be distinguished. The x-axis gives the
flux density of the second source expressed as a percentage of the
flux density of the
primary source, and the y-axis gives the probability that
there is a second source at least this bright and close enough to
the primary source that the two cannot be distinguished.
The solid lines show the predictions at 250\,$\mu$m ({\it blue}),
350\,$\mu$m ({\it green}) and 500\,$\mu$m ({\it red}) if the effect
of source clustering is not included in the calculation.
The dashed lines show the predictions for 350\,$\mu$m ({\it green})
and 500\,$\mu$m ({\it red}) if the effect of source clustering is
included in the calculation. See \S~\ref{sect:boost2bands} for more details.
We have not including a clustering model at 250\,$\mu$m because
\citet{maddox10} did not find any significant clustering in that
waveband.
}
\label{fig:blending}
\end{figure}

In Figure~\ref{fig:blending}, we have used the counts from \citet{bethermin12} 
to predict the probability of a second source falling 
so close to a primary source that the two sources cannot
be distinguished. We have assumed that the maximum distance between
two such sources is equal to the
half-width half-maximum of the PSF, and
we have set the 
flux density of
the primary source
equal to the catalogue limit (\S~\ref{sect:catalogue}) at the chosen wavelength.
The solid lines in Figure~\ref{fig:blending} show the predictions if the effect of
source clustering is not included. There is very little difference
between the predictions for the three wavelengths.
Therefore, the larger beam size at 500\,$\mu$m makes little difference.

The dashed lines, however, show the predictions if we include the effect
of source clustering. We have used the results of \citet{maddox10}, who
found no evidence for the clustering of 250\,$\mu$m sources but strong clustering
of 350 and 500\,$\mu$m sources. We have used the values of the clustering amplitude
given by \citet{maddox10} for the samples selected at 350 and 500 $\mu$m with no
colour selection and for a value for the index of the correlation function
of 0.8. The dashed lines show the results. In this case, the probability of a
second bright confusing source is highest for 500\,$\mu$m, so it is possible
that the discrepancy in the source counts at 500\,$\mu$m is caused
by the breakdown of our Gaussian approximation caused by the increase
clustering of sources at 500\,$\mu$m.

However, whatever the cause of the discrepancy between the H-ATLAS and HerMES source
counts at 500\,$\mu$m, there is little practical effect on the catalogues.
Our estimated flux bias at 500\,$\mu$m for a source at the 
4$\sigma$ limit of the catalogue is $\simeq 4\%$. Increasing the flux-bias
factor to 10\% would be enough to make the H-ATLAS and HerMES counts agree.
Therefore, there is a 6\%  systematic uncertainty in
the 500\,$\mu$m flux density for a source at the catalogue limit,
much less than the statistical uncertainty of 25\%.

\section{Comparisons with Other Photometry}\label{sect:comparisons}

\subsection{Comparison with Science Demonstration Phase Data}\label{sect:SDPcomp}

We have compared our catalogue 
with the one we released after the SDP \citep{rigby11}. The two catalogues are based
on almost exactly the same Herschel observations,
so any differences in the flux densities
represents either
a change in the flux calibration or in the methods
used to produce the catalogues. 

We first consider the SPIRE results.

The first difference the reader can notice is in the instrumental noise of the SPIRE maps: the values measured
in the SDP release were 4.1, 4.7 and 5.7\,mJy/beam at 250, 350 and 500\,$\mu$m, which are different, in particular
at 350\,\micron, with respect to the values in Tables~\ref{tab:noisevar} and \ref{tab:noisegauss}. The differences are due to two effects.
The first effect is caused by the use of the matched filter instead of the PSF to smooth the maps: as explained in \S~\ref{sect:mapfiltering}, 
the matched filter is optimal to maximise
the signal-to-noise in presence of both instrumental {\it and} confusion noise. This means that it might not be the optimal filter to reduce
the instrumental noise alone, in fact the released filtered maps at 250 and 500\,\micron\ have a slightly larger instrumental noise than the SDP ones
had. The second effect is caused by the difference in the choice of pixel sizes: the values used in the SDP maps were 5, 5 and 10\,arcsec, while
the released maps have pixel sizes of 6, 8 and 12\,arcsec at 250, 350 and 500\,$\mu$m respectively. A larger pixel size increases the number of
samples per pixel, decreasing the instrumental noise (and vice-versa). This is particular evident at 350\,\micron, where the pixel area is now about
2.5 larger than the SDP one.

The second difference is in the confusion noise of the SPIRE maps. SDP estimates of the confusion noise were 5.3, 6.4 and 6.7\,mJy/beam at 250, 350 and 500\,$\mu$m.
In this release, we have used Eq.~\ref{eq:confnoise_cat} to calculate a ``customised'' confusion noise for each flux density and argued that the values
in Tables~\ref{tab:noisevar} and \ref{tab:noisegauss} can be considered upper and lower limits respectively (see \S~\ref{sect:conf_noise}). The SDP values lie between
the two limits, but of course any particular estimate of the noise for a single source will be different with respect to the previous release.

In this data release,
we have chosen to include in the catalogue sources detected above 
a detection limit of $4\sigma$, which
corresponds in 2-scan regions to flux-density limits
of 29.6, 37.6 and 40.8\,mJy at 250, 350
and 500\,$\mu$m, respectively. Table~\ref{tab:compl} shows
that the survey is $\simeq 90\%$ complete at these limits.
Our flux-density limits are actually very
similar to the 
flux-density limits for our SDP catalogue: 34 mJy, 38 mJy and 44\,mJy at 250, 350 and 500\,$\mu$m,
respectively
\citep{rigby11}. The SDP limits were $5\sigma$ limits, and the difference in
the signal-to-noise is because of the more rigorous analysis
of noise we have carried out for the current release.

We first compared our SDP catalogue with a new catalogue
produced from
the images smoothed with the PSF (not included in the data release), 
since this comparison allowed us to look for any differences that are
not caused by the introduction of the matched filter.
We
found the median ratio of SDP flux density to flux density in the new catalogue
is 1.02, 1.07 and 1.06 at 250, 350 and 500\,$\mu$m, respectively. The changes are well within the error of 15\%
we gave for our SDP catalogues \citep{rigby11}.
These changes are not due to changes in the overall flux calibration, since
there was no change in the flux calibration at 250 and 500\,$\mu$m and change of
$<$1\% at 350\,$\mu$m (\S~\ref{sect:dataproc}).

Apart from the introduction of the matched filter, the three
significant modifications we have made
to our method since the SDP are (a) a different way of removing the emission from the Galactic
dust, (b) the use of a measured PSF rather than a
Gaussian function (\S~\ref{sect:psf}, \citealt{pascale11}) and (c) the new method
of sequentially removing sources from an image once the flux density of the source
has been measured (\S~\ref{sect:madx}). We can not be sure of the reason for the small difference
in the flux densities, but the third modification should reduce the flux densities
of sources which are close to other bright sources and thus works in the right
direction.

If we now compare the matched-filter catalogues to the SDP catalogue, the median ratio
of SDP flux density to flux density in the new catalogue is 1.06, 1.12, and 1.05 at 250, 350
and 500\,$\mu$m, respectively. 
The fact that the SDP fluxes are on average slightly higher may
represent the improvement given by the matched-filter technique in removing the contribution
to the measured flux density of a source from a confusing nearby source. 

A last difference to notice about SPIRE measurements between SDP and this data release is about flux bias
corrections (see Table~\ref{tab:boost}). The differences are mainly due to the adopted number
counts of the source populations. The model used by \citet{rigby11} has been superseded 
at faint fluxes by the deepest {\it Herschel} surveys, so we can assume our corrections are 
more robust and reliable than the previous ones.

We now consider the differences in the PACS flux densities. We selected the sources in the SDP catalogue detected
at $>5\sigma$ at one of the PACS wavelengths and looked for a source in the
new catalogue within 5\,arcsec of each SDP source.
We found the median ratio of SDP flux density to flux density in the new catalogue is 0.96 at
both 100 and 160\,$\mu$m. The small change is reassuring, given the quite large changes 
we have made in the analysis of the dataset and the improvements in the
calibration of PACS since the publication of our SDP results.

\begin{figure}
\center
\includegraphics[width=84mm,keepaspectratio,angle=0]{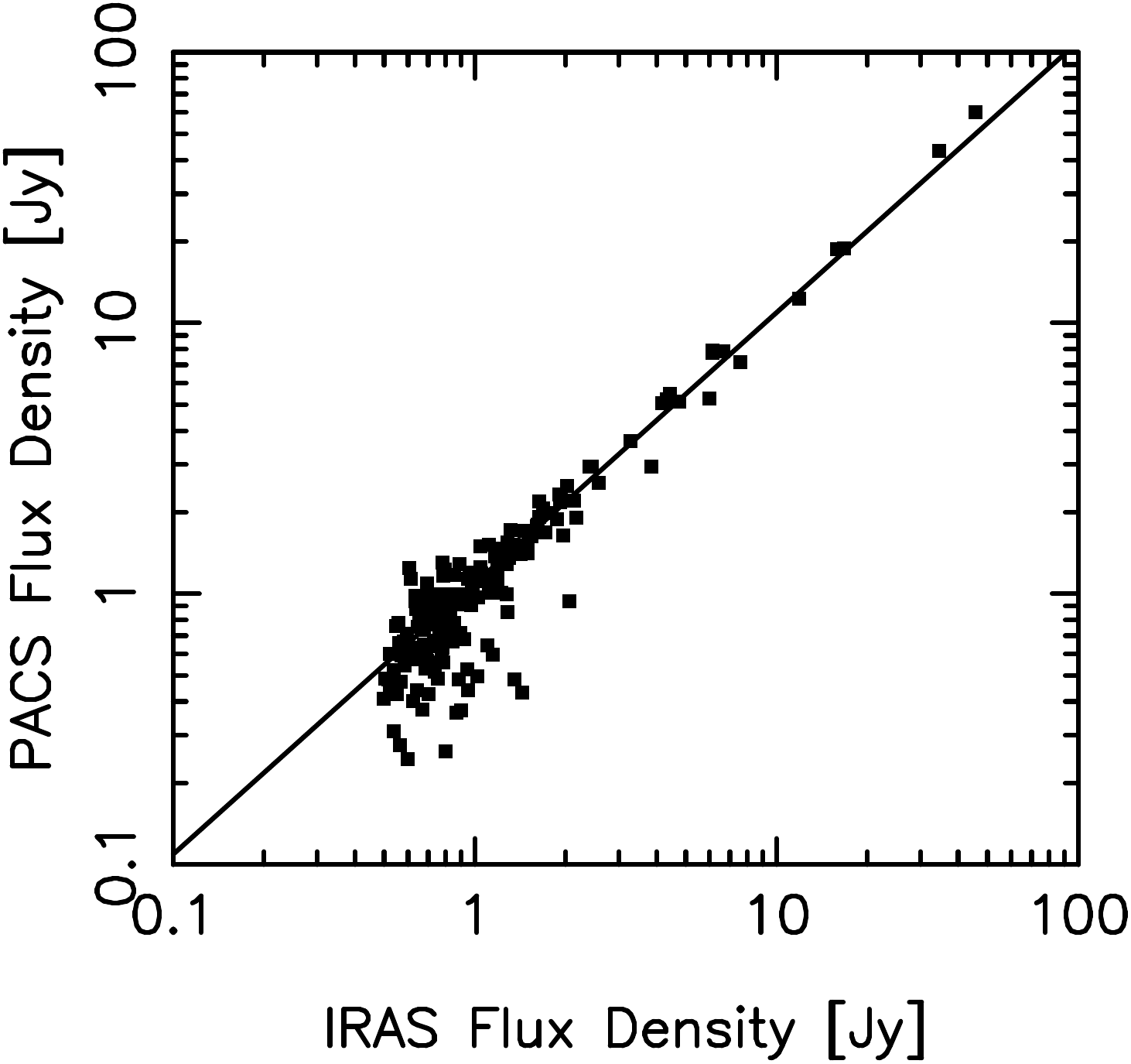}
\caption{Flux density measured by PACS at 100\,$\mu$m verses flux density measured by IRAS at
100\,$\mu$m for the galaxies in the GAMA fields that are also in the IRAS Faint Source
Catalogue \citep{wang14}. The line shows the median value
of the ratio of PACS flux density to IRAS flux density for the 75 galaxies
with IRAS flux density $>1$\,Jy.
}
\label{fig:iras}
\end{figure}

\subsection{Comparison Between PACS 100\,$\mu$m and IRAS}

We have compared our measurements of the PACS 100\,$\mu$m
flux densities with the
100\,$\mu$m flux densities for galaxies in the IRAS Faint Source
Catalogue \citep{wang14}. There are 184 galaxies for which
there are both PACS and IRAS flux densities. Figure~\ref{fig:iras} shows a comparison
of the PACS and IRAS flux densities. For galaxies with IRAS flux densities
less than 1\,Jy, there is almost no correlation between the
IRAS and PACS flux densities, which we suspect is due to the effect of cirrus emission
and flux bias on the IRAS flux densities. However, for galaxies above this
limit there is, with a few exceptions, good agreement between the two sets of
flux densities. 
Excluding the three very discrepant points, the median ratio of the PACS
flux density to the IRAS flux density for the 75 galaxies
with IRAS flux density greater than 1\,Jy is 1.09.

This result is without making a correction for the
real spectral energy distributions (SEDs), which are unlikely
to be the standard SED assumed for both PACS and IRAS: $F_\nu \propto \nu^{-1}$.
The sources detected by IRAS are mostly nearby galaxies, and therefore 
a reasonable assumption is that the real SED of a source is likely
to be a modified black-body with a dust temperature of $T_{\rm d} \simeq 20$\,K.
For this SED, the correction to the PACS 100\,$\mu$m flux densities 
is a reduction by a factor of 1.15 \citep{muller11}.
We have been unable to find any corrections listed for the
IRAS 100\,$\mu$m filter for modified blackbodies. No colour corrections are listed in 
the IRAS Explanatory Supplement for the 100\,\micron\ filter for black bodies with 
dust temperatures less than 40\,K, presumably because the long-wavelength
side of the IRAS 100\,\micron\ filter response function was not known very
accurately (IRAS Explanatory Supplement). Given these uncertainties, the systematic
9\% difference between the PACS and IRAS flux densities seems well
within the bounds of experimental uncertainty.

\section{The Catalogue}\label{sect:catalogue}

In the data release, we include a catalogue of all sources 
above the detection limit of $2.5\sigma$
at 250\,\micron\,{\it and} with a 
measured flux density $\geq 4\sigma$ at at least one 
of the three SPIRE wavelengths (250, 350 and
500\,$\mu$m). The $4\sigma$ limit is approximately
29.4, 37.4 and 40.6\,mJy at 250, 350 and 500\,$\mu$m, respectively, and
the catalogue is $\simeq$90\% complete
at all three wavelengths. These flux-density limits
are very similar to the $5\sigma$ limits  
for the catalogue we released at the end of the
Science Demonstration Phase \citep{rigby11},
the reason for the difference being the more accurate
analysis of noise for the present data release.
The catalogue contains 
113995, 46209 and 11011 sources detected at $>4\sigma$ at
250, 350 and 500\,$\mu$m, respectively. 



The catalogue contains
measurements of the
flux densities of each source in all five photometric bands, including aperture photometry
for sources for which the diameter of the optical counterpart
implies the 250\,$\mu$m source is likely to be extended and the flux from aperture photometry
is significantly larger than the flux measured by the source-detection algorithm (see \S~\ref{sect:spireext}).
The catalogues contain detections at $>3\sigma$ at
100 and 160\,$\mu$m for 4650 sources and 5685 sources, respectively,
and the typical noise at the two wavelengths is
44 and 49\,mJy, respectively (see \S~\ref{sect:pacsphot}).

Given the area of the survey and the noise, Gaussian statistics
imply that $\simeq 260$ of the
sources in our $4\sigma$ catalogue should be spurious, which is
$\simeq 0.2\%$ of the total. However, this is likely to be an overestimate.
The method we used to calculate the value of $\sigma$ for each source was designed
to provide 
a good estimate of the error in the flux density of the source.
The more commonly used method of deriving a value of
$\sigma$ for a signal-to-noise calculation is to
fit a Gaussian to the negative part of the pixel flux distribution
(\S~\ref{sect:conf_noise}, Figure~\ref{fig:noisegauss}). 
This method produces a much lower value of $\sigma$ than we have actually
used, and therefore our signal-to-noise estimates
are likely to be conservative and our catalogue more reliable than
the Gaussian statistics imply.
In practice, we suspect a more important problem than spurious sources is
likely to be single sources that are actually multiple sources, 
since our estimates of the fraction of single sources that are likely to
actually be more than one source depends sensitively on the assumptions
made about the correlation function for submillimetre sources, although
this seems unlikely to be an issue at 250\,$\mu$m where the
sources appear to be only very weakly correlated (see \S~\ref{sect:boost2bands}, Figure~\ref{fig:blending}).

The catalogue also contains information about the optical counterparts to the sources, 
with fluxes and redshifts measured by the GAMA survey. How these counterparts were found
is described in the second data-release paper \citepalias{bourne16}. 

\section{Summary}\label{sect:summary}

\begin{itemize}

\item[$\bullet$] We have described the first major data release of the {\it Herschel} Astrophysical Terahertz Large Area
Survey (H-ATLAS). The data release consists of images at 100, 160, 250, 350 and 500\,$\mu$m and catalogues
of the sources detected in the images. 

\item[$\bullet$] The survey was carried out in parallel mode with
the PACS and SPIRE cameras on Herschel. We describe the data-reduction pipelines that 
we designed
to overcome the specific challenges posed by our very large datasets.

\item[$\bullet$] The $1\sigma$ noise for source detection on the SPIRE images 
is approximately 7.4, 9.4 and 10.2\,mJy at 250, 350
and 500\,$\mu$m, respectively.
Our catalogues include 114052, 43019 and 10555 sources detected at $>4\sigma$ at
250, 350 and 500
$\mu$m, respectively. 

\item[$\bullet$] We include
measurements of the
flux densities of each source in all five photometric bands, including aperture photometry
for sources for which the diameter of the optical counterpart implies
the 250\,$\mu$m source is likely to be extended.
The catalogues contain detections at $>3\sigma$ at
100 and 160\,$\mu$m, the two PACS wavebands, for 9314 sources and 16345 sources, respectively,
and the typical noise at these wavelengths is
38 and 37\,mJy, respectively.

\item[$\bullet$] We describe In-Out simulations designed to get as close as possible to the
ground truth of how sources in the sky are converted to sources in the catalogues.
The In-Out simulations have allowed us to
determine the probability that a source with true flux density, $F_{\rm t}$ is
converted into a source in the catalogue with measured flux density $F_{\rm m}$.
We show how this information can be represented both in the form of analytic
functions representing the conditional probability, $P(F_{\rm m})|F_{\rm t})$, and
in a matrix form. We use the results to show how the completeness of the survey
depends on the {\it true} flux density at 250\,$\mu$m, both for the
survey as a whole and for the deeper part of the survey.

\item[$\bullet$] We describe a novel matrix-inversion
method that uses the results of the In-Out simulations to
correct the observed source counts for Eddington bias. 
We show that the corrected source counts for our survey agree well with the results
of the deeper HerMES survey \citep{bethermin12}, although there is
a discrepancy at 500 $\mu$m, where the corrected H-ATLAS source counts
are slightly higher than the HerMES source counts.

\item[$\bullet$] We have used the method of \citet{crawford10} to estimate
the flux bias at 250\,$\mu$m, finding that the flux-bias factor is approximately
20\% at the $4\sigma$ catalogue limit. We have then used these flux-bias estimates
and the covariance between the 250, 350 and 500\,$\mu$m images to
estimate the flux bias at 350 and 500\,$\mu$m, finding that the flux-bias factor
is $\simeq$6\% and 4\% at the catalogue limits
at 350 and 500\,$\mu$m, respectively. We have used
the flux-bias estimates to determine how the completeness of the survey 
depends on {\it measured} flux density at all three wavelengths.

\item[$\bullet$] We have used our estimates of the flux bias and the completeness
to produce estimates of the source counts at all three wavelengths, an alternative
method to the matrix-inversion method. We find similar results,
producing source counts that agree well with the source counts from the deeper
HerMES survey \citep{bethermin12} at 250 and 350\,$\mu$m and with a similar
but slightly larger discrepancy at 500\,$\mu$m.

\item[$\bullet$] A second data-release paper \citepalias{bourne16} describes the properties
of the optical counterparts to the submillimetre sources.

\end{itemize}

\section*{Acknowledgments}
We are grateful to G\"oran Pilbratt and the staff of the Herschel Science Centre for prompt and expert support during our survey.
We thank George J. Bendo, Dieter Lutz, Matt J. Griffin, Andreas Papageorgiou, H\'{e}l\`{e}n Roussel for helpful conversations about technical aspects of the survey and data reduction.
We also thank Andrew W. Blain, Asantha Cooray, Gianfranco De Zotti, Mattia Negrello, Joaquin Gonz\'{a}lez-Nuevo, Michal J. Michalowski, Douglas Scott, Daniel J.B. Smith, Pasquale Temi, Mark A. Thompson, Sebastien Viaene for useful and insightful comments.
We thank Robbie Auld and Michael Pohlen for their contribution to the project.
We express our infinite gratitude for {\sc topcat} \citep{taylor05}.
EV and SAE acknowledge funding from the UK Science and Technology Facilities Council consolidated grant ST/K000926/1.
MS and SAE have received funding from the European Union Seventh Framework Programme ([FP7/2007-2013] [FP7/2007-2011]) under grant agreement No.~607254.
SJM, LD and PJC acknowledge support from the European Research Council (ERC) in the form of the Consolidator Grant {\sc CosmicDust} (ERC-2014-CoG-647939, PI H.L.Gomez).
SJM  LD and RJI acknowledge support from the ERC in the form of the Advanced Investigator Program, {\sc COSMICISM} (ERC-2012-ADG\_20120216, PI R.J.Ivison). 
NB acknowledges the support of the EC FP7 SPACE project {\sc ASTRODEEP} (Ref.~No.~312725).
CF acknowledges funding from CAPES (proc.~12203-1).
The {\it Herschel}-ATLAS is a project with {\it Herschel}, which is an ESA space observatory with science instruments provided by European-led Principal Investigator consortia and with important participation from NASA. The {\it Herschel}-ATLAS website is \url{http://www.h-atlas.org}.

\appendix

\section{The Data Products}\label{app:data}

In this section we describe the basic data products we are releasing to the astronomical
community.
These consist of SPIRE and PACS maps of the GAMA fields and catalogues containing the sources
detected at $>4\sigma$ in at least one of the three SPIRE wavebands. 
The data products can be obtained from the website h-atlas.org.
In this section we give an overview of the different data products, describe
what data products are appropriate for different scientific projects, and discuss 
some of the limitations of the data products. More detailed technical information can be found on the
website or in the rest of the paper.

\subsection{The SPIRE maps}\label{app:spire}

We provide three classes of image:

\begin{itemize} 

\item[{\bf 1}] The raw images produced using the method of \S~\ref{sect:dataproc} but without the subtraction of any
large-scale background emission, for example
from interstellar dust (``cirrus emission'').
We also supply maps of the instrumental noise on the images obtained using the the coverage map calibrated by the 
jackknife method (\S~\ref{sect:noise_from_jn}). These images are our most basic data product and should
be a good representation of the submillimetre sky up to an angular scale of $\simeq 20$\,arcmin (\S~\ref{sect:dataproc}), but these
images should not be used without further modelling to investigate the emission on larger scales, 
for example
the extragalactic background radiation.

\item[{\bf 2}] The raw images with the large-scale background subtracted using {\it Nebuliser} (\S~\ref{sect:dataproc}).

\item[{\bf 3}] The images from {\bf 2} convolved with the matched-filter (\S~\ref{sect:mapfiltering}). These images were created to
give the highest possible signal-to-noise for point sources, and the value in each pixel
of the images is our best estimate of the flux density of a point source at that position.
We also provide a noise map, which contains our best estimate of the instrumental noise at each
position. Eq.~\ref{eq:confnoise_cat} gives our estimate of the confusion noise as a function of the flux density of the source: this value
should be added in quadrature to the instrumental noise (\S~\ref{sect:madx}).

\end{itemize}

All of these images have different scientific purposes. For the reader interested in aperture
photometry of a nearby galaxy, the appropriate images to use are the 
second set. It is possible to use the first set but we recommend the
second set because any background emission has been removed with {\it Nebuliser} 
- \S~\ref{sect:spireext}). Photometry performed using apertures larger than 1.5\,arcmin on images {\bf 2} might be
slightly underestimated due to the use of {\it Nebuliser}.
The images we 
supply are in the units often called ``Jy/beam'', which means that 
the value in any pixel is the flux density that a
point source would have if it were centered in that pixel. These images can easily be turned 
into a form suitable for aperture photometry by dividing the images by the 
SPIRE beam area. The current estimates of the beam area are
469, 831 and 1804\,arcsec$^2$ at 250, 350 and 500\,$\mu$m, respectively,
but see the SPIRE Handbook for updates.
We have calculated the errors on the flux densities by carrying out Monte-Carlo simulations in the vicinity
of the source. The reader can either use the same technique or use the following formula which describe
quite well the results of our Monte-Carlo simulations:
\begin{equation}\label{eq:aperture_noise}
\sigma_{\rm ap} = \sqrt{ \sum_i^{N_{\rm ap}} \frac{\sigma^2_{{\rm inst},i}}{C^2_{\rm conv}} + \frac{N_{\rm ap}}{N_{\rm beam}} \sigma^2_{\rm conf} }
\end{equation}
\noindent in which $\sigma_{\rm inst,i}$ is the instrumental noise
in the i'th pixel within the aperture,  $C_{\rm conv}$ is the area of the beam divided by the area of a pixel, 
$N_{\rm ap}$ and $N_{\rm beam}$ are the angular areas in pixels subtended by the aperture and the beam,
and $\sigma_{\rm conf}$ is the confusion noise given in Table~\ref{tab:noisevar}
To calculate the number of pixels in the beam, it is recommended to use the area
of the beam given by $\pi ({\rm FWHM}/2)^2$ rather than the areas given in the SPIRE handbook, since the latter
are estimated by integrating the PSF to very large radii and are thus likely to overestimate the
effect of confusion.
All flux densities measured by aperture photometry must be corrected for the fraction of the PSF
that falls outside the aperture using the provided SPIRE aperture corrections table.

If the reader is interested in running a different source-extraction programme to MADX (\S~\ref{sect:madx}),
the second set of images is also the one to use. These images
have had the large-scale emission subtracted
using a method that does not affect the flux density of point sources.

For those interested in simply measuring the flux density for an object that
should be unresolved in the SPIRE bands, the third set of images are the ones to use, since
these have been designed to provide flux-density measurements for a point source
of the highest possible signal-to-noise (\S~\ref{sect:mapfiltering}). 

If the reader is 
is interested in carrying out
a statistical ``stacking analysis'', the set of images
from {\bf 3} is also the one
to use. In this case, the reader should
be aware that the mean of the images is not zero (\S~\ref{sect:dataproc}), and so
the mean of the map should be subtracted before performing any stacking analysis.

Note that if the reader is carrying out a stacking analysis or even simply measuring 
from the H-ATLAS image the
submillimetre flux density of a previously known object,
there is no need to make a correction for flux bias (\S~\ref{sect:boost250}), because flux bias
only affects the flux densities in our catalogue.

The width of the SPIRE filters means that
both the size
of the PSF and the power detected by SPIRE depend on the spectral energy 
distribution (SED)
of the source. The SPIRE data-reduction pipeline is based on the assumption that the flux density
of the source depends on frequency$^{-1}$, and all our
images are ultimately based on this assumption. 
The recipe for aperture photometry outlined above is also based on this
assumption.
This SED is very different
from that of most sources detected by Herschel, and so the user must make a correction
to the measured flux densities to allow for this difference. 
The reader should multiply the measured flux densities by
the $K_{ColE}$ parameter, which is given in Tables~5.6 and 5.7 in the SPIRE Handbook, which is a correction
for all the effects produced by the difference in the SED. 

The noise maps we have provided do not include a contribution from the uncertainty in the
basic calibration of flux density. 
The calibration of {\it Herschel} is still improving. At the time of writing,
the error in the flux density arising from the
uncertainty in the absolute flux density of Neptune is 4\% and there is
an additional 1.5\% error that is uncorrelated between the bands (SPIRE Handbook). 
The current recommendation (SPIRE Handbook) is that these factors
should be added, and so the reader should use a calibration error
of 5.5\%.

\subsection{The PACS Images}\label{app:pacs}

We provide a single set of images. These are the images made with {\it JScanamorphos} 
which have had any residual large-scale emission subtracted with {\it Nebuliser} (\S~\ref{sect:pacs}). The 
units
of these images are Jy/pixel, so it is straightforward to carry out aperture photometry.
Because of the complicated nature of the noise on the PACS images (\S~\ref{sect:pacs}), we do not 
provide noise
images. Instead, we provide maps showing the number of observations, $N$, contributing to 
each pixel in
the final image. We have used Monte-Carlo simulations (\S~\ref{sect:pacsphot}) to show that the
error for aperture photometry, $\sigma_{\rm ap}$,
on the PACS images
through an aperture with a radius $r$ can be represented by Eq.~\ref{eq:pacs_noise}.

The PACS point spread function is not a simple Gaussian and in fast-scan parallel mode is 
significantly extended in the scan
direction \citep{lutz15}, which means that it must vary even within a single GAMA field.
For this
reason, we recommend that no attempt should be made to maximise the signal-to-noise for
point sources by convolving the images with the PSF. Instead, we recommend that all scientific
projects should use aperture photometry. If the reader prefers to filter the map, we recommend 
the use of our Gaussian fit of the empirical PSFs, which gives a FWHM of 11.4 and 13.7\,arcsec 
at 100 and 160\,\micron\ respectively.

We recommend that aperture photometry of sources that are expected to be extended, for
example nearby galaxies, should be carried out 
by adding up the flux density in a suitable aperture; there should be no need
to estimate a sky value because we have already subtracted any residual background
emission using {\it Nebulizer} (see \S~\ref{sect:pacsmaps}).
Errors in the flux density should be estimated using Eq.~\ref{eq:pacs_noise}.
As part of the data release, we have supplied a file listing the Encircled
Energy Fraction (EEF) in the two bands out to a reference radius of 1000\,arcsec.
Both the flux densities and the errors should be corrected using the EEF.
Note that the flux densities of sources with a diameter larger
than 2.5\,arcmin may be underestimated because of the use of {\it Nebuliser}.

For sources that are expected to be unresolved, we have shown that the maximumn
signal-to-noise can be obtained by aperture photometry using very small apertures (\S~\ref{sect:pacsphot}, Figure~\ref{fig:ston}),
with a diameter $<8$\,arcsec at both 100\,$\mu$m and 160\,$\mu$m.
We recommend that if the position of an object is known precisely, the reader should use a small
aperture to carry out the photometry and then use the EEF to correct the flux density and the
error to the reference radius.
However, when using a small aperture that only contains a small number of pixels, the reader should think carefully about pixelisation effects.
This is also our recommended procedure for measuring the
average flux density for a class of objects
in a stacking analysis, as long as the positions of the objects are
known very accurately.

If carrying out
a statistical ``stacking analysis'', the reader should
be aware that the mean of the images is not zero (\S~\ref{sect:pacsmaps}), and so
the mean of the map should be subtracted before proceeding.

On top of the flux density error given in Eq.~\ref{eq:pacs_noise}, there is also
a fundamental calibration error. Our conservative estimate of this
error is that it is 7\% (see \S~\ref{sect:pacsphot}).

All our measurements of flux density
are based on the assumption that the flux density, $F_{\nu}$, of a source has the spectral
dependence $F_{\nu} \propto \nu^{-1}$. 
Because of the width of the PACS spectral response and because most sources
do not have this SED, a correction must be made to the flux densities.
A table of corrections for different SEDs
is given in M\"{u}ller et al. (2011; Tables~1 and 2). 
We recommend
that the user adopts a common-sense procedure here. To give one
easy example, if the class of
sources of interest are all nearby galaxies,
we suggest that a sensible correction would
be obtained by assuming that sources have SEDs that are
modified 
black bodies with a dust temperature chosen by the reader.
Note that the corrections can be quite large e.g. 1.16 for a modified black
body with a dust temperature of 15 K \citep{muller11}.
In the case of nearby galaxies the answer is fairly obvious. In other more tricky examples,
it should be possible to use the ratios of the flux densities in the
PACS and SPIRE bands or the redshifts of the optical counterparts
to choose what correction to make.

\subsection{The Catalogues}\label{app:cat}

We provide catalogues of the sources detected above the detection limit of $2.5\sigma$
at 250\,\micron\,{\it and} at $>4\sigma$ in at least one of the
SPIRE bands. These catalogues also contain the optical counterparts to the {\it Herschel}
sources. The details of the part of the data release that consists of measurements
by telescopes other than {\it Herschel} are given in \citetalias{bourne16}.

For each detected source, we provide measurements at 100, 160, 250, 350
and 500\,$\mu$m
bands of the flux densities
and errors on the assumption that
the source is unresolved by the telescope (\S~\ref{sect:madx}, \ref{sect:pacsphot}). For the
sources that are expected to be extended, we also provide aperture photometry designed to estimate
close to total flux densities for the sources (\S~\ref{sect:spireext}, \ref{sect:pacsphot}).
The errors given in the catalogues do not contain any calibration errors, which
are described above (App.~\ref{app:spire}, \ref{app:pacs}).

There are two significant corrections that we have not made to the catalogued
flux densities that the user should be aware of. First, both the SPIRE and PACS data-reduction pipelines 
are based on the assumption
that all sources have an SED in which $F_\nu \propto \nu^{-1}$.
If the flux-density measurement in the catalogue
was made with aperture photometry, which is the case
for all PACS measurements and some SPIRE measurements,
and the reader has some knowlege of the true SED
of the source, it is possible to correct the flux
density using the prescriptions given in
App.~\ref{app:spire} and \ref{app:pacs}. If the flux-density was measured
with the MADX detection software there is no simple prescription
for correcting the flux density. However, the systematic error
on the flux density arising from
a different SED is generally much less than the statistical
error, and is thus not usually a concern. 

For any reader
interested in precision photometry and in thus 
calculating a correction to the MADX flux densities
for a different SED, we note
that there are two effects.
First, the precise PSF of a source depends on its SED, whereas
in MADX we have used a PSF derived from observations of Neptune.
Second, there is a correction that needs to be made
to the flux density arising from
the variation in the SED across the SPIRE filter. The second effect
at
least is relatively easy to calculate using the information
in the SPIRE Handbook.

The second correction we have not made to the measured flux densities
is for the effect of flux bias. Although flux bias does not affect
{\it Herschel} flux-density measurements of objects detected by
other telescopes (App.~\ref{app:spire}), it does affect the flux-density
measurement of the sources in our catalogues. 
Despite the extensive modelling described in \S~\ref{sect:corrections}, we have not made corrections
to the flux densities in the catalogue because we are aware that there may
be improvements in the modelling of this effect in the future.
We suggest that the user either uses our estimate of this effect, which
are listed in Table~\ref{tab:boost},
or in their scientific analysis make allowance
for a systematic error in the SPIRE flux density
approximately of the
size that we estimate.

\begin{table*}
\centering
\caption{Asteroids}
\begin{minipage}{140mm}
\centering
\label{tab:asteroids}
\begin{tabular}{@{}cccc@{}}
\hline
 Position (J2000.0)     & Date & Universal Time &  Asteroid \\
\hline
08 41 32.4 $+$00 56 02 & 29/05/2010 & 07 34 01 & (568) Cheruskia \\
08 41 58.4 $+$00 55 21 & 29/05/2010 & 15 35 10 & (568) Cheruskia \\
08 56 03.6 $+$02 25 26 & 23/05/2010 & 05 47 23 & (983) Gunila \\
08 56 22.9 $+$02 25 06 & 23/05/2010 & 13 34 27 & (983) Gunila \\
11 38 03.1 $-$00 47 47 & 07/07/2010 & 22 19 42 & (1167) Dubiago \\
11 38 17.9 $-$00 48 53 & 08/07/2010 & 07 06 03 & (1167) Dubiago  \\
12 07 18.9 $+$00 54 46 & 19/12/2010 & 16 53 10 & (3446) Ritina \\
12 07 35.7 $+$00 53 60 & 20/12/2010 & 01 46 51 & (3446) Ritina \\
12 12 00.5 $-$01 18 12 & 13/07/2011 & 11 37 02 & (20) Massalia \\ 
12 12 55.6 $-$00 52 56 & 15/06/2011 & 04 14 08 & (250) Bettina \\
12 15 04.0 $-$01 37 19 & 15/07/2011 & 19 43 22 & (20) Massalia \\
12 21 26.2 $-$00 06 07 & 16/07/2011 & 23 29 13 & (56) Melete \\
12 22 48.5 $+$00 34 24 & 15/06/2011 & 05 06 44 & (171) Ophelia  \\
12 26 13.2 $-$01 26 42 & 16/07/2011 & 00 22 22 & (401) Ottilia \\ 
\hline
\end{tabular}
\end{minipage}
\end{table*}

\section{Asteroids}\label{app:asteroids}

We found the asteroids by looking at the jackknife images, the images made by taking the difference
of the two images of the same region made with the different scan directions (\S~\ref{sect:observations}). The only
sources on the jackknife images
should be objects that either vary in flux density or move. In practice, the only time-varying sources
we found were asteroids. The list of source positions is given in
Table~\ref{tab:asteroids}. In the table we also list the times and dates at which the sources 
were observed, which we determined by inspecting the time-line data for individual
bolometers. 

We used these positions, times and dates to look for the asteroids corresponding
to these sources, using the database of the Minor Planet Center (www.minorplanetcenter.net).
An interesting complication in identifying the asteroids is that there 
are parallactic shifts of several arcmin corresponding to the
change in vantage point from the Earth to the location of {\it Herschel}
at the second Lagrangian point. Rather than calculating the precise
parallactic shifts, we looked for any asteroids within 15\,arcmin 
of the source position at the appropriate time and date.
There was always one asteroid that was several magnitudes brighter
than the others at optical
wavelengths, and this asteroid was usually but not always the closest
to the position of the {\it Herschel} source. 
Based on the optical brightness and the positional offsets, we are 
confident that the asteroids we have listed in the table are the
correct identifications, although we have provided enough information
for anyone interested in calculating more precise geocentric positions
in order to make the identification completely secure. 
The number in brackets in the name of the asteroid shows the
historical order in which the asteroid was discovered.
Not surprisingly,
the asteroids discovered in the first survey
of a large area of sky at submillimetre wavelengths
were also among the first to be discovered by optical
astronomers.
 
To avoid confusion,
we have not included these sources in the catalogue for this
data release.

\bsp

\label{lastpage}

\end{document}